\documentclass{article}

\usepackage{arxiv}

\usepackage[utf8]{inputenc} 
\usepackage[T1]{fontenc}    
\usepackage{hyperref}       
\usepackage{url}            
\usepackage{booktabs}       
\usepackage{amssymb}       
\usepackage{amsmath}
\usepackage{bm}
\usepackage{nicefrac}       
\usepackage{microtype}      
\usepackage{lipsum}
\usepackage{graphicx}

\usepackage{subfigure}
\usepackage{algorithm}
\usepackage{algorithmicx}
\usepackage[noend]{algpseudocode}
\usepackage{float}

\usepackage{enumerate}

\title{A Q-values Sharing Framework for Multiagent Reinforcement Learning under Budget Constraint}

\author{
 Changxi Zhu \\
  South China University of Technology \\
  \texttt{cxzhu.cn@gmail.com} \\
   \And
 Ho-fung Leung \\
  The Chinese University of Hong Kong \\
  \texttt{lhf@cuhk.edu.hk} \\
  \And
 Shuyue Hu \\
  National University of Singapore \\
  \texttt{dcshus@nus.edu.sg} \\
  \And  
 Yi Cai \\
  South China University of Technology \\
  \texttt{ycai@scut.edu.cn} \\
}

\begin{document}
\maketitle
\begin{abstract}
In teacher-student framework, a more experienced agent (teacher) helps accelerate the learning of another agent (student) by suggesting actions to take in certain states. In cooperative multiagent reinforcement learning (MARL), where agents need to cooperate with one another, a student may fail to cooperate well with others even by following the teachers' suggested actions, as the polices of all agents are ever changing before convergence. When the number of times that agents communicate with one another is limited (i.e., there is budget constraint), the advising strategy that uses actions as advices may not be good enough. We propose a partaker-sharer advising framework (PSAF) for cooperative MARL agents learning with budget constraint. In PSAF, each Q-learner can decide when to ask for Q-values and share its Q-values. We perform experiments in three typical multiagent learning problems. Evaluation results show that our approach PSAF outperforms existing advising methods under both unlimited and limited budget, and we give an analysis of the impact of advising actions and sharing Q-values on agents' learning.
\end{abstract}


\section{INTRODUCTION}
Many real-world tasks, e.g., robotic games \cite{Kitano1997RoboCupAC} and distributed power allocation \cite{HetNetQos}, involve a set of agents in a common environment. By learning from trial-and-error, an agent adapts its behavior in an unfamiliar environment, which forms a popular mechanism Reinforcement Learning (RL). However, RL methods require a long period of interactions with the environment, resulting in limited scalability and applicability in complex situations. When RL is used in a Multiagent System (MAS), which is also referred as Mutiagent Reinforcement Learning (MARL), these problems are further intensified since all agents are changing and adapting their behaviors \cite{Marinescu2017Nonstationary}, thereby increasing the amount of interactions needed. To know whether an action is optimal or not in the MARL, an agent must execute the action, often many times. When the computing resource of agents is limited, the time they spend should be used more efficiently over entire lifetime.

One natural solution to speed up the learning process of RL agents is via \emph{knowledge sharing}, relying on communicating various pieces of external knowledge, such as policies, value functions and episodes (state, action, quality triplets) among them. When an agent knows rarely about current environment, there is no doubt that it can reuse knowledge from other agents who have acquired some skills in advance to reduce the time to explore. There are three problems when agents communicate with one another: (1) deciding to share what kind of knowledge; (2) deciding when to share knowledge, especially when communication is limited due to cost; (3) deciding how to integrate received knowledge. Recently, one notable approach is the \emph{teacher-student} framework, which focuses on \emph{Action Advising} \cite{Torrey2013TeachingOA,Silva2017SimultaneouslyLA}. In this framework, a more experienced agent or human expert (teacher) aids the learning of another agent (student) by suggesting actions to take in certain states. The advising opportunities are established jointly by teacher and student, which is suitable for the case of limited communication. In order to model communication cost, the number of times that a student asks for action advice and a teacher provides advice are constrained by two budgets respectively, i.e., budget constraint. The framework has achieved promising results on both single-agent and multi-agent problems. In this paper, we consider that multiple RL agents cooperatively try to solve a task, where they learn to cooperate with each other by coherently choosing their individual actions based on their own observations, such that the resulting joint action is optimal. Each agent learns to act optimally by maximising its Q-function, which implies the cumulative discounted rewards for every state. Under the circumstance, however, a student may fail to cooperate well with others even by following a teacher's suggested actions, since all other agents are learning and adjusting their own polices as training progresses. More importantly, it is still hard for the student to learn, specifically, the Q-values corresponding to the advised actions. The problem of the teacher-student framework applied in cooperative MARL is that advising actions to a student do not inherently change the student's policy. The student needs to take a long period to adapt its behaviours for other agents in the changing environment.

Sharing knowledge is of utmost importance for cooperative agents learning in a shared environment. In such case, if they are equipped with the same learning structure, they most likely become interchangeable, i.e. they have identical optimal policies and value functions. Without knowledge mapping, an agent can directly utilize the knowledge, e.g., Q-values, from more experienced agents as its own. If a student is able to choose next action upon its teachers' Q-values in current state, then it can act without the necessary of taking more time to learn for that state. Using the Q-values from a teacher requires that the agents involved in advising relations are reinforcement learners and have similar (even the same) reward functions. Although it may sacrifice flexibility compared with action advising, nevertheless, we here argue that sharing Q-values is more effective than advising actions for a cooperative team of RL agents. The Q-values from other agents guide an agent's exploitation of the team's currently learned Q-values while not only learning the optimal Q-function by itself. Another critical consideration in our work is that sharing Q-values for all agents at every time step need time and can be costly, especially for some multiagent systems composed of a large population of agents, such as in the field of \emph{social learning} \cite{Hao2014SocialLearning}. Similar to the advising budget in teacher-student framework, we assume that each agent are constrained by two budgets for asking for and giving Q-values. It is vital to decide proper time to share Q-values.

In this paper, we present a \emph{partaker-sharer} advising framework (PSAF) for cooperative MARL agents to share Q-values under budget constraint, where only a limited number of Q-values can be shared in the whole learning process. In PSAF, a more experienced agent (sharer) can provide the other agent (partaker) with the maximum Q-value for the state of the partaker, which elucidates the sharer's best action. Each agent can play the role of partaker or sharer in different sharing processes. An agent can decide when to start asking for Q-values according to how many time it has explored for current state. That is, if it visits current state very few times, it can initiate a sharing process, take the role of partaker, and send a request to all other agents. In order to prevent from sharing even worse Q-values to the partaker, another agent will evaluate how confident it is in the partaker's state. If the agent has updated its maximum Q-value many more times than the partaker, then it can join in the partaker's sharing process, take the role of sharer, and share its maximum Q-value. There are two numeric budgets of each agent, for requesting and providing Q-values respectively.

Our contributions in this paper are threefold. First, we propose a novel Q-values sharing framework PSAF for cooperative MARL agents learning with budget constraint. Since agents only can share a limited number of Q-values, each agent in PSAF should decide when to ask for Q-values and share its Q-values, and how to integrate the shared Q-values into its own Q-function. Second, we highlight that the amount of budget is essential for agents to advise actions as well as sharing Q-values, and introduce new metric for the sharer to decide sharing opportunities, in order to use the budget more efficiently. We also explore the effect of different amount of budgets on the performance of all comparing methods. Third, we show that our proposed framework PSAF significantly accelerates the learning process of agents in three typical multiagent learning problems: Predator-Prey domain, Half Field Offense and Spread game. We try to figure out when agents share Q-values and actions. The distribution of sharing opportunities shows that in PSAF, most Q-values are shared when a partaker has updated the Q-values for corresponding state-action pairs very few times while a sharer has updated many times. Besides, we compare and contrast the effects of sharing Q-values and advising actions on agents' learning, which has not been investigated in previous works.

The remainder of this article is organized as follows. Section 2 describes related work for sharing Q-values and the teacher-student framework. In Section 3, we present the background information on multiagent reinforcement learning, temporal difference learning algorithms with eligibility traces, and a jointly-initiated framework. Section 4 formally presents our method PSAF, including when to ask for and give Q-values, and how to use the received Q-values. In section 5, we conduct empirical demonstrations in Predator-Prey domain, Half Field Offense and Spread game, showing that our approach outperforms existing advising methods under unlimited and limited budgets. In section 6, we give an analysis of advising actions and sharing Q-values, and discuss which situations are suitable for PSAF. Finally, in section 7, we draw the conclusions of this article and propose future directions.

\section{RELATED WORK}
The idea of sharing Q-values has its roots in multiagent reinforcement learning. The Q-values of reinforcement learners are the expected cumulative discounted reward gained by performing actions in states during learning. Sharing Q-values reduce the time to learn actions to be rewarded. The centralized learning method \cite{Gupta2017Cooperative} learns a shared Q-function for the actions and observations of all agents in an environment. Both the state and action space scale exponentially with the number of agents, rendering this approach infeasible for thousands of agents learning together. By contrast, we consider that agents are decentralized in both training and execution, which is convenient for agents joining and leaving \cite{Matignon2012IndependentRL,Ilhan2019TeachingOA,Hong2018ADP}. Concretely, each agent is equipped with separate processor and memory to update and store its own policy. One simple method without any communication is to use independent learners \cite{Claus1998TheDO, Matignon2012IndependentRL}, where each agent individually learns its policy or Q-function based on its own observations. However, it is nontrivial how decentralized agents benefit from sharing Q-values with one another, in order to accelerate learning process. The first line of sharing Q-values assumes that the Q-values being shared has been prepared before current learning. One typical approach is transfer learning, especially for transferring Q-functions among different tasks \cite{Taylor2007Transfer,SurveyTransfer}, whereby an agent can learn faster in a target task after training on a less complex task. Even though the method leads to a dramatic speedup in learning, it requires to build hand-coded action-value function relationship, for example, how to map different state representations and reward functions for different tasks. When several agents learn to cooperate in a multiagent system, they may need to coordinate in a few states while act independently in the other states. By using this idea, Kok and Vlassis \cite{Kok2004Sparse} propose Sparse Tabular Multiagent Q-learning that maintains a predefined list of states in which coordination is necessary. Only in these states, agents perform joint actions and a shared Q-table is updated. Then they learn their own Q-values in all remaining states. In order to automatically identify coordinating states, De Hauwere et. al propose CQ-learning \cite{CQLearning2010} and FCQ-learning \cite{FCQLearning2011}. Both methods assume that an agent has already learned its optimal policy alone before learning together with other agents. When an agent executes in the multiagent environment, a state is marked as conflict if it detects a change in the received immediate reward compared with the expected reward derived from its optimal single-agent policy. Then it incorporates its previously acquired Q-values from single-agent environment into currently joint Q-function, otherwise previous Q-values are used to select next actions. However, these works assume that agents have already learnt (or are able to learn) some kind of single-agent knowledge (e.g., local value function) before the multi-agent learning process. By comparison, we consider that multiple agents start learning with no previous knowledge in a shared environment, then they must rely on their partners who may have explored different regions of the state space. Tan \cite{Tan1993MultiAgentRL} points out that cooperative Q-learning agents can communicate several aspects of their learning process for current task, such as policies (or complete Q-tables), episodes (state, action, quality triplets), and sensation. The Q-values for every state-action pair of all agents are averaged in a predefined time interval, and then the averaged Q-values are assigned for each agent. Nevertheless, imagine that agents communicate over wireless channels such as Wi-Fi and LTE, sharing the whole Q-values may incur high-communication cost, especially when agents learn in a large-scaled multiagent system. One of the relevant works most similar to ours is Parallel Transfer Learning (PTL) \cite{PTL2014,PTL2019}, which enables multiple agents to transfer their Q-values when they are simultaneously learning in different tasks or the same task. In order for knowledge sharing to be beneficial, the Q-values to be shared are carefully selected. When all agents learn together in a common environment, a fixed number of Q-values is transferred at every time step according to the importance of associated states. In PTL, it is necessary for agents to connect with each other constantly in order to share their Q-values. In comparison, we seek to address how agents communicate with each other only if necessary, and an agent is able to decide for what situation it can share its Q-value when requested. The major merit of our framework is its flexibility and applicability for an enormous quantity of agents learning cooperatively under budget constraint.

There are many ways that address the issue of how to use received Q-values from other agents, even though they generally ignore the problems of when to shared Q-values, as emphasized in our work. These methods can be classified into: a) Average-Q; b) Weighted-Q; c) Replace-Q. The Average-Q method, also known as Policy Average \cite{Tan1993MultiAgentRL}, is the simplest way to utilize the shared Q-values. This method assumes that all agents have the same contribution to the corresponding state-action pairs. When an agent communicates with other agents, its own Q-values as well as the shared Q-values are averaged, so as to generate the new Q-values for corresponding state-action pairs. Since agents may have experienced different areas of the state-action space at a given time step, some outdated Q-values are taken into account in the Average-Q method. An agent needs to combine the shared Q-values more effectively. The Weighted-Q method allows a weighted combination of received and local Q-values. Ahmadabadi and Asadpour \cite{Ahmadabadi2002Expertness} propose several weighted strategies to generate new Q-values from an agent's current Q-values and the received Q-values. In this method, each agent measures the expertness of its teammates and assigns a weight to their knowledge and learns from them accordingly. In PTL \cite{PTL2019}, as the number of times that an agent visits a state increases, locally learned Q-values are taken more into account than the received Q-values. However, when an agent knows nothing or very few about a particular state, incorporating its Q-values may sharply decrease the new Q-values, which makes the learning process become unstable. In the Replace-Q method, an agent accepts the shared knowledge fully as it has no better information. The more complex case is that the received Q-values are potentially from multiple sources. The agent should decide to select one of them. The experience counting, inspired by the non-trivial update counting method in \cite{Torrey2012Help}, is another popular sharing strategy \cite{Cunningham2012NonreciprocatingSM}. The experience counting makes use of an experience table, in addition to a Q-function to keep track of which states and what actions an agent has actually visited during learning. The central idea behind this method is that the most experienced agent should have the most contribution to the Q-value with regard to corresponding state-action pair. At every time step, all agents use the same Q-value which is from the the agent with the most experience for every state-action pair. Considering a tabular representation of a Q-function, which is, perhaps, the most commonly applied in RL tasks, the new Q-values generated from Average-Q, Weighted-Q and Replace-Q can be directly assigned to corresponding state-action pairs. Another possible way for absorbing the shared Q-values is to update local Q-values in a Q-learning-like rule, e.g., CQ-learning and FCQ-learning. In the Q-learning-like rule, after performing actions in current states, the Q-values from other knowledge resources for the next states are used to bootstrap the Q-values for the current states. This method does not need to directly access an agent's Q-function (e.g., Q-table), which can be extended to the problem with continuous state or action space.

Recently, the teacher-student framework \cite{Torrey2013TeachingOA} has received a lot of attention \cite{Silva2017SimultaneouslyLA,Garcia2019PPR,Omidshafiei2019LearningTT,SurveyTransfer}. In this framework, a more experienced agent (teacher) accelerate the learning process of another agent (student) by providing advice on which action to take. The student updates its policy based on rewards received from the environment, while its exploration is guided by the teacher's advice. Some works, like the supervision framework designed for organizational structure \cite{Supervision2008} or supervising reinforcement learning \cite{Supervising2010}, propose to supervise a network of agents by providing rules (forbidden actions of satisfied states), suggestions (preferred actions of satisfied states) et al. However, all supervisors are fixed, and the agents who are supervised by a supervisor need to communicate with their own supervisor at every time step. By contrast, the most significant aspect of the teacher-student framework is that advising opportunities are determined only when required, considering the practical concerns regarding attention and communication. This setting is also widely accepted by many extensions of the teacher-student framework, as depicted in \cite{Zimmer2014Teacher,Fachantidis2017LearningTT,Silva2017SimultaneouslyLA}. Moreover, both the multiagent advising framework \cite{Silva2017SimultaneouslyLA} and our work assume no fixed roles of each agent, which means each agent has the opportunity to be a teacher (or sharer). There have been three modes of deciding the advising opportunities: student-initiated \cite{Clouse1996}, teacher-initiated \cite{Torrey2013TeachingOA} and jointly-initiated \cite{Amir2016Interactive}. In the student-initiated method, a student who learns to act optimally in a task is assisted by an expert teacher whenever the student's confidence in a state is low. Sharing decisions made by the student are likely to be weak since itself is still learning. Conversely, in the teacher-initiated method, a trained teacher decides when to give advice to a student. This method requires the student's current state is always communicated to the teacher, and the teacher should constantly pay attention to the learning of the student. Transmitting every state that the student has experienced to the teacher can cause a prohibitive cost of communication. The jointly-initiated approach determines advising opportunities under the agreement of both teacher and student. In this method, the teacher is not required to constantly monitor the student. The relation of a student and a teacher is established on demand, which is particularly suitable for the case of limited communication. The work of Silva et al. \cite{Silva2017SimultaneouslyLA} is the first to apply the teacher-student framework to a multiagent system composed of multiple simultaneously learning agents. In order to overcome the challenge of having no fixed roles of teacher and student, they extend the heuristics from \cite{Torrey2013TeachingOA} to measure an agent's confidence in a given state based on the number of times that it visits the state. The number of times that a student asks for advice and a teacher gives action advice are limited by two budgets respectively. With the constraint of budget, this approach achieves state-of-the-art results even when comparing with sharing the whole episode during learning. Despite having promising results, the advising strategy that uses actions as advices may not be good enough to accelerate the overall learning process for Q-learners.

\section{PRELIMINARIES}
\subsection{Multiagent Reinforcement Learning}
We are interested in a cooperative multi-agent setting where agents get local observations and learn in a decentralised fashion. All communications among agents must be clearly specified. The learning problem of multiple decentralised agents with local observations is generally modelled as a Decentralised partially observable Markov decision process (Dec-POMDP) \cite{Oliehoek2016}, which is an extension of Markov Decision Process (MDP) \cite{Sutton1998ReinforcementLA}. A Dec-POMDP is defined by a tuple $\langle \mathcal{I}, \mathcal{S}, \bm{\mathcal{A}}, \mathcal{T}, \mathcal{R}, \bm{\Omega}, \mathcal{O}, \mathcal{\gamma} \rangle$, where $\mathcal{I}$ is the set of $n$ agents, $\mathcal{S}$ is the set of environment states, $\bm{\mathcal{A}} = \times_{i \in \mathcal{I}}\mathcal{A}_{i}$ is the set of joint actions, $\mathcal{T}$ is the state transition probabilities, $\mathcal{R}$ is the reward function, $\bm{\Omega} = \times_{i \in\mathcal{I}} \Omega_{i}$ is the set of joint observations, $\mathcal{O}$ is the set of conditional observation probabilities, and $\mathcal{\gamma}\in [0,1)$ is the discount factor. At every time step $t$ in an environment state $s'$, each agent $i$ perceives its own observation $o^i_t$ from one joint observation $\textbf{o}=\langle o^1,...,o^n \rangle$ determined by $\mathcal{O}(\textbf{o}|s', \textbf{a})$, where $\textbf{a} = \langle a^1, ... , a^n \rangle$ is the joint action that causes the state transition from $s$ to $s'$ according to $\mathcal{T}(s'|\textbf{a}, s)$, and receives reward $r^i$ determined by $\mathcal{R}(s, \textbf{a})$. We focus on cooperative Multiagent Reinforcement Learning (MARL), where several reinforcement learning agents jointly affect the environment and receive the same reward ($r^1 = r^2 = ... r^n$). The observability of agents in Dec-POMDPs has been elaborately discussed in Oliehoek and Amato's work \cite{Oliehoek2016}. In our framework, we assume that agents are able to observe each other at any time and infer their local observations.\footnote{In this paper, the local observation of an agent is also referred as the agent's own state. The former one is from a multi-agent view while the latter one is from a single-agent view} That is, individual observation (partial view) for each of the agents always uniquely identifies the environment state. Since agents are distributed, it is still difficult to learn the optimal policy for each of them \cite{Oliehoek2016}.

\subsection{Temporal difference learning}
Temporal difference (TD) learning algorithms such as Q-learning \cite{Watkins1992Qlearning} and SARSA \cite{Sutton1998ReinforcementLA} learn an action-value function, $Q(s, a)$, which is an estimate of the expected cumulative discounted reward that an agent takes action $a$ in state $s$. In our work, each agent individually receives its own state from a shared environment, and learn a Q-function. Learning a policy for an agent $i$ means to better estimate its Q-function $Q^i(s^i, a^i)$ for corresponding action $a^i$ in its own state $s^i$. Since all agents have the same reward function (equivalent to a joint reward function), their Q-functions are also the same when no explicit specialization is assigned, e.g., different learning structures. We define the \emph{experience} of agent $i$ at time step $t$ as a tuple $\langle s^i_t, a^i_t, s^i_{t+1}, r^i_{t+1} \rangle$, where state $s^i_{t+1}$ is reached by the agent after it executing action $a^i_t$ in its state $s^i_t$, and reward $r^i_{t+1}$ is received. The Q-function of agent $i$ is incrementally updated based on the agent's experience gained in learning, using the weighted average of the old value and the new information. Therefore, the update of Q-values for each state-action pair are defined as follows:
\begin{equation}
 Q^i_t(s^i_t, a^i_t) \leftarrow Q^i_t(s^i_t, a^i_t) + \alpha \delta
 \label{eq:qupdate}
\end{equation}
where $\alpha \in [0, 1]$ is learning rate, and $\delta$ is TD error. In Q-learning, we have:
\begin{equation}
\delta=r^i_{t+1} + \gamma \max_a{Q^i_t(s^i_{t+1}, a)} - Q^i_t(s^i_t, a^i_t)
\end{equation}
where $\gamma \in [0,1)$ is discount factor. RL agents face to choose whether to focus on high reward actions or taking actions with the intent of exploring the environment. We consider agents adopt $\epsilon$-greedy for action selection. As such, at every time step, with a large probability $1-\epsilon$, an agent takes the action with the highest $Q$-value in the current state, and with a small probability $\epsilon$, the agent takes a random action. Then in normal learning (without asking for advice), a Q-learning agent can choose the action to be executed at every learning step according to $\epsilon$-greedy. SARSA, which is another popular RL algorithm, defines the TD error as follows:
\begin{equation}
\delta=r^i_{t+1} + \gamma Q^i_t(s^i_{t+1}, a^i_{t+1}) - Q^i_t(s^i_t, a^i_t)
\end{equation}
where $a^i_{t+1}$ is the next action that agent $i$ will execute in $s^i_{t+1}$ according to a defined exploration strategy like $\epsilon$-greedy. These algorithms are guaranteed to converge to the optimal Q-function $Q^*_i$, from which the optimal policy $\pi^*_i$ of agent $i$ can be derived:
\begin{equation}
 \pi^*_i(s^i) = \arg\max_a{Q^*_i(s^i, a)}
\end{equation}

In order to speed up reinforcement learning, instead
of updating one Q-value at every time step, n-steps can be used for making a backup. We adopt Eligibility traces \cite{Sutton1998ReinforcementLA} to further enhance the performance of TD algorithms. In this strategy, each agent records the state-action pairs which have been recently visited. The TD error at current time step is used to update the Q-values for these state-action pairs. Eligibility traces aims to assign the credit or blame to the eligible states or actions. At time step $t$ in an episode, agent $i$ updates its accumulating traces for the states that have been visited and the actions that have been taken so far by following rules:
\begin{equation}
e^i_t(s^i,a^i)=
\begin{cases}
\gamma\lambda e^i_{t-1}(s^i, a^i), & \text{if } s^i \neq s^i_t \\
\gamma\lambda e^i_{t-1}(s^i, a^i) + 1, & \text{if } s^i = s^i_t
\end{cases}
\end{equation}
where $\lambda \in$[0,1] refers to traces decay rate. Q($\lambda$) and SARSA($\lambda$) are the extension of Q-learning and SARSA, respectively. In Q($\lambda$) or SARSA($\lambda$), an agent uses weighted TD errors by using eligibility traces as weights to update Q-values for every experienced state-action pair in current episode. Then Eq. \ref{eq:qupdate} is modified as follows:
\begin{equation}
 Q^i(s^i, a^i) \leftarrow Q^i(s^i, a^i) + \alpha \delta e^i_t(s^i,a^i)
\end{equation}

\subsection{Jointly-Initiated Framework}
Recent works on teacher-student framework mainly adopt jointly-initiated method to build advising relations, which is more appropriate for learning with budget constraint. Here we introduce a multi-agent advising framework AdhocTD\cite{Silva2017SimultaneouslyLA}, in order to illustrate how to construct advising relations (for action advice) only when necessary. This framework has shown promising results in complex stochastic environment Half Field Offense\cite{Stone2016HalfFO}, and is considered as the state-of-the-art approach. In AdhocTD, at each time step, agent $a_i$ asks for advice with an asking probability $P_{ask}$ in current state s. The asking probability is calculated as follows:
\begin{equation}
    P_{ask}(s)=(1+v_a)^{-\sqrt{n_{visit}(s)}}
\end{equation}
where $v_a$ is a predetermined parameter, $n_{visit}(s)$ is the number of times that the agent visits state $s$. Agent $a_j$ gives its best action with a giving probability $P_{give}$ for the state of agent $a_i$. The giving probability is calculated as follows:
\begin{equation}
P_{give}(s)=1-(1+v_b)^{-\sqrt{n_{visit}(s)}\times I(s)}
\end{equation}
where $v_b$ is a predetermined parameter, and $I(s)$ encodes the difference between agent $a_j$'s maximum Q-value and its minimum Q-value in the requested state $s$:
\begin{equation}
I(s)=\max_a Q(s, a)-\min_a Q(s, a)
\end{equation}
All agents are limited by a budget $b_{ask}$ to ask for advice and a budget $b_{give}$ to give advice.

\section{PARTAKER-SHARER ADVISING FRAMEWORK}
Our approach aims to accelerate learning for a cooperative team of reinforcement learning agents in a multiagent environment. These agents are assumed to be homogeneous, meaning that they are interchangeable, i.e. they have identical optimal policies and value functions. During learning process, each agent may have unique experience or local knowledge (i.e., Q-functions) of how to perform effectively in current task. They may not be willing to share the whole Q-functions with each other due to budget constraint. However, we assume that the team of agents would like to share a limited number of Q-values during entire lifetime, which is modelled as the limited number of times that an agent asks for and shares Q-values. Specifically, each agent can decide to share its maximum Q-value for the current state of the agent who asks for Q-values. The maximum Q-values from other agents guide an agent's exploitation of the team's currently learned Q-values. The goal of our method is to explore when to ask for Q-values and share the maximum Q-values, as well as how to use the received maximum Q-values.

We here propose a \emph{partaker-sharer} advising framework (PSAF) for multiple decentralized Q-learners to share Q-values under budget constraint. A sharing process is initiated by a \emph{partaker}, which is a role of an agent who asks for Q-value in its own state. The other agents can take the role of \emph{sharer}, and share their maximum Q-values for the state of the partaker. Then the partaker will choose its current action based on its own Q-values as well as the sharers' Q-values. Each agent can take different roles in different sharing processes, which depends on whether they decide to ask for and share Q-values for, to be specific, particular states. In this way, an agent may ask for Q-values in current situation while share its maximum Q-value for the requested state of another agent. The maximum number of times that an agent takes the role of partaker and sharer are two numeric budgets $b_{ask}$ and $b_{give}$ respectively.

In PSAF, we consider that an agent is more likely to ask for Q-values if it knows very few about some states, and it is more beneficial for the agent to be guided by other more experienced agents for those states. Several heuristics from \cite{Amir2016Interactive,Silva2017SimultaneouslyLA} can be used to decide when an agent takes the role of partaker. They rely on the range of Q-values for the requested state or how many times an agent visit the state. The range of Q-values can vary very much at the beginning of training so that it may mislead an agent who tries to ask for Q-values, where we expect the agent would like to get more help. AdhocTD \cite{Silva2017SimultaneouslyLA} has defined an asking function $P_{ask}(s)=(1+v_a)^{-\sqrt{n_{visit}(s)}}$ by considering the number of times an agent visits a state $s$. The function outputs a higher probability for requesting advice when the agent visits state $s$ very few times, while the value of the function declines rapidly when the agent gains more experience in that state. In this paper, we adopt the same function $P^i_{ask}(s^i)$ to allow an agent $i$ to determine when to initiate a sharing process and take the role of partaker in current state $s^i$. After successfully initiating a sharing process, another agent $j$ (from all the agents except partaker $i$) needs to decide whether it would share its maximum Q-values to the partaker. Previous works on teacher-student framework have proposed several methods (e.g., function $P_{give}$ defined by AdhocTD) to help a teacher to decide when to advise a student without accessing the student's current learning. However, if both student and teacher learn from scratch, i.e., their policies are generally non-optimal during learning, the teacher may be less helpful or even suggest worse advice if it is not able to evaluate how well the student have learned. In PSAF, during a sharing process, partaker $i$ will tell other agents how confident it is in its Q-values for current state $s^i$. Sharing Q-values with low confidence may interfere the learning of agents who take the role of partaker, and waste communications. Therefore, the other agents can joint in the sharing process (of partaker $i$) and share their Q-values if they have higher confidence in the Q-values than the partaker. Intuitively, as training progresses, if an agent updates its Q-value for a state-action pair more times, the Q-value generally becomes more reliable. We firstly propose a confidence function $\Phi^i:S\times A \rightarrow \mathbf{R}$, which encodes how confident partaker $i$ is in currently learned Q-values for state $s^i$. Formally we have:
\begin{gather}
    \Phi^i(s^i, a) = m^i_{visit}(s^i, a)
\end{gather}
where $m^i_{visit}(s^i, a)$ is the number of times that partaker $i$ updates its Q-value for a state-action pair $(s^i, a)$. In order to share much better Q-values than the partaker and then use the budget more efficiently, another agent $j$ who intends to share its own Q-values only if it has updated the Q-values many more times compared with the partaker. Here we propose another confidence function $\Psi^j:S\times A \rightarrow \mathbf{R}$, which represents the confidence of agent $j$ for state-action pair $(s^i, a)$. $\Psi^j$ is expected to scale down the updated times of the Q-value for $(s^i, a)$ to a proper value. Then we define $\Psi^j$ as follows:
\begin{gather}
    \Psi^j(s^i, a) = m^j_{visit}(s^i, a) \times \xi^j
\end{gather}
where $m^j_{visit}(s^i, a)$ is the number of times that agent $j$ updates its Q-value for state-action pair $(s^i, a)$, and $\xi^j \in [0,1]$. $\xi^j$ is used to determine how many times agent $j$ should have updated the Q-value for $(s^i,a)$ when comparing with partaker $i$, so that agent $j$ can take the role of sharer. When $\xi^j$ is 0, agent $j$ can not share its Q-values. When $\xi^j$ is 1, the agent share its Q-values as long as it has updated the Q-values more times than partaker $i$. Now we consider how to construct $\xi^j$. In a given state, if all Q-values are nearly the same, it does no matter which one is shared. If the maximum Q-value of agent $j$ in a state is much higher than other Q-values, sharing the maximum Q-value is more meaningful for partaker $i$ since the Q-value may lead to a fine-gained action. Then agent $j$ is more likely to be as a sharer even it does not update the maximum Q-value many times. For the sake of convenience, we define $\xi^j$ of agent $j$ to be the difference between the maximum Q-value and minimum Q-value normalized to $[0,1]$ in state $s^i$ as follows:
\begin{gather}
    \xi^j(s^i) = \frac{\max_aQ^j(s^i,a)-\min_aQ^j(s^i,a)}{\max_aQ^j(s^i,a)-\min_aQ^j(s^i,a) + 1}
\end{gather}
where $\max_aQ^j(s^i,a)$ and $\min_aQ^j(s^i,a)$ are the maximum and the minimum Q-value of agent $j$ in state $s^i$ respectively. If the difference between $\max_aQ^j(s^i,a)$ and $\min_aQ^j(s^i,a)$ is close to 0, then $\xi^j$ is low and agent $j$ should update its maximum Q-value in state $s^i$ quite a number of times to make sure the Q-value is reliable. If the difference between $\max_aQ^j(s^i,a)$ and $\min_aQ^j(s^i,a)$ is large, $\xi^j$ is high, so that agent $j$ will have more opportunities to share its maximum Q-value.

\begin{algorithm}[t]
\caption{Ask for Q-values and use the shared maximum Q-values in a given state}
\label{alg1}
\begin{algorithmic}[1]
\Require
agent $i$, budget $b^i_{ask}$, asking function $P^i_{ask}$, and function $\Phi^i$.
\For{each time step}
\State let current state be as $s^i$
\If{$b^i_{ask}>0$}
\State $p \leftarrow getRandomValue(0,1)$
\If{$p<P^i_{ask}(s^i)$}
\State broadcast state $s^i$ and the confidence of Q-values derived by $\Phi^i$
\For{each of the other agents}
\State add the shared Q-value to collection $\Pi$
\EndFor
\If{$\Pi \neq \emptyset$}
\State $b^i_{ask} \leftarrow b^i_{ask} -1$
\State $\Pi$ = $\Gamma(\Pi)$ \Comment{each action is associated with one Q-value}
\For{each state-action pair of $\Pi$}
\State denote $Q^j(s^i, a)$ as the Q-value for state-action pair $(s^i, a)$
\State $Q^i{(s^i, a)} \leftarrow Q^j(s^i, a) $
\EndFor
\State execute greedy exploration strategy
\EndIf
\EndIf
\EndIf
\If{no action is executed}
\State perform usual exploration strategy (i.e., $\epsilon$-greedy)
\EndIf
\EndFor
\end{algorithmic}
\end{algorithm}

Algorithm 1 describes when to ask for Q-values and how to use the shared maximum Q-values for agent $i$. At every time step, the agent in current state $s^i$ can take the role of partaker as long as budget $b^i_{ask}$ has not been used up (lines 1-3). With the probability calculated by asking function $P^i_{ask}(s^i)$, partaker $i$ initiates a sharing process and broadcasts to all other agents for requesting Q-values (lines 4-6). The broadcast message contains the partaker's state $s^i$, and its confidence of each Q-value under current situation. Then partaker $i$ waits until a predefined timeout to collect answers. We denote $\Pi$ as the collection of Q-values from all sharers for the state of the partaker (lines 7-8). If $\Pi$ is not empty, $b^i_{ask}$ is decremented by 1 (lines 9-10). Partaker $i$ may receive several Q-values for the same state-action pair in the sharing process, since some sharers may have the same best action according to their own Q-values. Besides, each sharer can not access other sharers' Q-functions to avoid additional communication before joining in the partaker's sharing process. In PSAF, the partaker can either randomly select one of the shared Q-values or choose the Q-value with maximum confidence among them, which is defined by function $\Gamma$ (line 11). Now, partaker $i$ needs to integrate the Q-value selected from function $\Gamma$ to its local Q-function in current state $s^i$. In this paper, we assume a tabular representation of the Q function. For each corresponding state-action pair, partaker $i$ replaces the original Q-value with the shared Q-value from $\Gamma$ (lines 12-14). Then, in order to exploit the shared Q-values from the whole team, the partaker executes its currently best action corresponding to the currently maximum Q-value in state $s^i$ (line 15). If agent $i$ does not ask for Q-values or no Q-value is received, the agent performs usual exploration strategy, i.e., $\epsilon$-greedy (lines 16-17).

Algorithm 2 describes when agent $j$ provides its maximum Q-values to partaker $i$ for state $s^i$. As long as budget $b^j_{give}$ has not been used up, agent $j$ compares its confidence of the maximum Q-value in $s^i$ with the partaker (lines 1-3). Agent $j$ shares its maximum Q-value if it has higher confidence in the Q-value than partaker $i$. If agent $j$ takes the role of sharer, budget $b^j_{give}$ is decremented by 1 (lines 4-5). Then sharer $j$ sends its maximum Q-value in state $s^i$, the associated action, and the confidence of the Q-value to partaker $i$ (line 6). After that, the partaker's action selection in state $s^i$ will be guided by the sharer.

\begin{algorithm}[t]
\caption{Share the maximum Q-value for a given state}
\label{alg:algorithm1}
\begin{algorithmic}[1]
\Require
agent $j$, function $\Psi^j$, budget $b^j_{give}$, state $s^i$, and function $\Phi^i$.
\State switch from current state $s^j$ to state $s^i$
\If{$b^j_{give}>0$}
\State $a^*_j \leftarrow \arg\max_{a}Q^j(s^i, a)$
\If{$\Psi^j(s^i, a^*_j) > \Phi^i(s^i, a^*_j)$}
\State $b^j_{give} \leftarrow b^j_{give} -1$
\State return $a^*_j, Q^j(s^i, a^*_j)$ and $\Psi^j(s^i, a^*_j)$
\EndIf
\EndIf
\State switch from state $s^i$ to state $s^j$
\end{algorithmic}
\end{algorithm}

\section{EXPERIMENTS}
We are interested in a multiagent system where several agents cooperatively solve a task. Each agent independently observes the environment and learns its own Q-values. The advising framework \cite{Silva2017SimultaneouslyLA} using actions as advices achieves excellent results in a similar setting as our work. And this framework has shown to surpass sharing a successful episode under budget constraint. We compare PSAF with the following approaches in our experiments.

\begin{enumerate}[a.]
\item \textbf{Multi-IQL} \cite{Claus1998TheDO}: Each agent is independent learner and learns individual Q-function. There is no communication among all agents. Multi-IQL serves as a baseline method to validate the benefit of advising actions and sharing Q-values.

\item \textbf{AdhocTD} \cite{Silva2017SimultaneouslyLA}: AdhocTD is a state-of-the-art method on action advising for multiagent learning, and the detail of this framework is illustrated in Section 3.3. We compare PSAF with AdhocTD since we argue that sharing Q-values is the most effective way to promote the learning of Q-learners.

\item \textbf{AdhocTD-Q}: We adapt AdhocTD in a straightforward way so that agents can share Q-values with one another. In AdhocTD-Q, agents use asking function $P_{ask}$ and giving function $P_{give}$ defined by AdhocTD, respectively, to ask for Q-values and share their maximum Q-values.
\end{enumerate}

We evaluate PSAF, AdhocTD, AdhocTD-Q, and Multi-IQL in three cooperative games. Predator-Prey domian is a popular benchmark for multiagent learning. Half Field Offense is a more complex robot soccer game. Spread game is a quickly deployable domain. Since current experiments only contain few agents to interact with each other, we adopt random selection strategy (function $\Gamma$) for each partaker in AdhocTD-Q and PSAF to utilize the shared Q-values. In all methods (except Multi-IQL), the number of times that an agent asks for and provides Q-values (or actions) are limited by budgets $b_{ask}$ and $b_{give}$ respectively. When we report that the difference between two curves is significant, it means we have at least 95\% confidence that one curve has larger area by using t-tests on their areas with $\alpha$ = 0.05.

\subsection{Predator-Prey domain}
\begin{figure}[htb]
  \centering
  \subfigure[Predator-Prey domain]{
    \includegraphics[width=1.4in, height=1.4in]{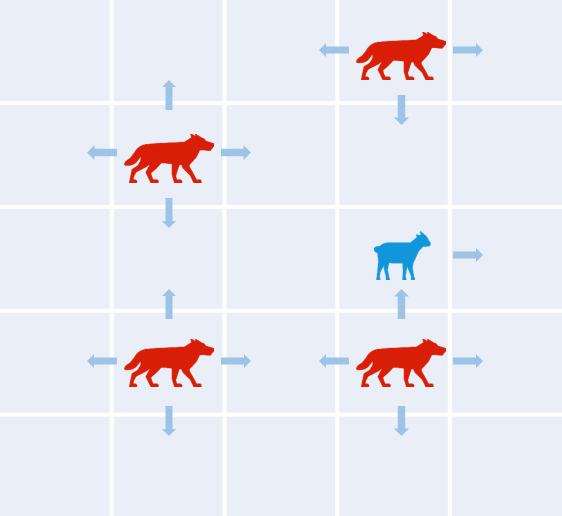}
  }
  \subfigure[The prey is caught.]{
    \includegraphics[width=1.4in, height=1.4in]{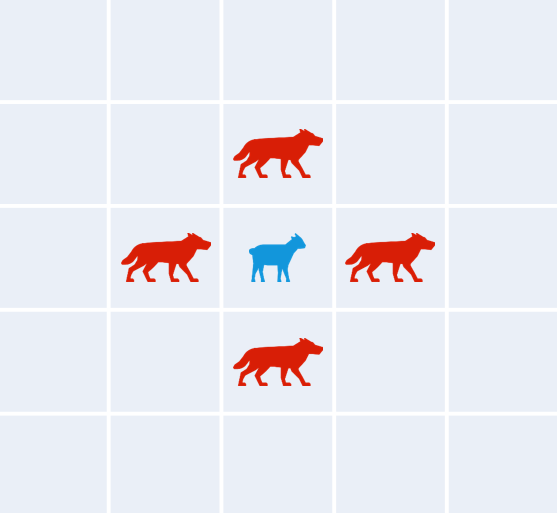}
  }
  \caption{Left: Predator-Prey domain with four predators and one prey. The red one is a predator, and the blue one is a prey. Right: Four predators are next to the prey.}
\end{figure}

Predator-Prey (PP) domain is easy to implement, customize, and the results are easy to interpret. Therefore, it has been extensively used to evaluate multiagent learning algorithms \cite{RewardShaping2014,Rosenfeld2017SpeedingUT,Le2017CoordinatedMI}.\footnote{The source code is available at http://www.biu-ai.com/RL} Our implementation includes an $N \times N$ grid world, where $N$ is the number of cells in $x$ $(y)$ direction. As shown in Figure 1a, there are four predators and one prey in the grid world. Each of them occupies one cell, and one cell is allowed to be occupied by only one agent to avoid the case of deadlock. The four predators and the prey can choose between five actions \emph{Stay}, \emph{Go Up}, \emph{Go Down}, \emph{Go Left}, and \emph{Go Right}. By executing an action, each agent moves cell by cell in corresponding direction. In this game, the prey takes a random action $20\%$ of the time, with rest of the time moving away from all predators, making the task harder than with a fully random prey. Four predators are reinforcement learning agents. They learn to cooperatively catch the prey as soon as possible. Each predator fully observes the relative $x$ and $y$ coordinates of other predators and the prey. All values of states are normalised to $[-1, 1]$ by dividing by the number of cells $N$. The prey is caught only when four predators are next to the prey in four cardinal directions, as shown in Figure 1b. If the predators catch the prey, all predators receive a reward of 1, otherwise 0. For predators not sharing actions or Q-values, each of them is equipped with $Q(\lambda)$, where $\lambda$=0.9, $\gamma$=0.9, and $\alpha$=0.1. The $\epsilon$-greedy with $\epsilon$=0.1 is used as exploration strategy for four predators. Since all predators are learning and changing their policies, it is very hard for them learning with $Q(\lambda)$. Tile coding \cite{Sherstov2005Function, Sutton1998ReinforcementLA} is used to force a generalization over the state space, with 8 tilings and tile-width 0.5.

The PP domain has one popular metric for performance evaluation. \emph{Time to Goal} (TG) is the number of steps that predators take to catch the prey. Lower TG values means that the predators catch the prey more quickly. One episode starts when four predators and the prey are initialized with random positions in the grid world. The episode ends when either predators catch the prey, or a time limit is exceeded. In each episode, the maximum number of steps that predators and the prey can play is 2,500. Four predators are trained for 10,000 episodes. After every 100 training episodes, the TG values for each of these episodes are gathered and averaged to get more stable values. The evaluation process is computationally efficient, and it is able to clearly present the trend in TG values for different methods. Then we obtain one run containing all evaluated TG values for 10,000 training episodes. The process is repeated 100 times. For the parameters $v_a$ and $v_b$ of AdhocTD and AdhocTD-Q, and $v_a$ of PSAF, we firstly tune $v_a$ and $v_b$ for AdhocTD with unlimited budget to get significant lower TG values than Multi-IQL and to spend not very high budget. Then we apply these parameters to all other experiments. Finally, we choose $v_a$=0.2 and $v_b$=1 for AdhocTD, AdhocTD-Q and PSAF.

\begin{figure}[htbp]
  \centering
  \subfigure[TG of 10,000 episodes with b=+$\infty$]{
    \includegraphics[width=2.5in, height=2.1in]{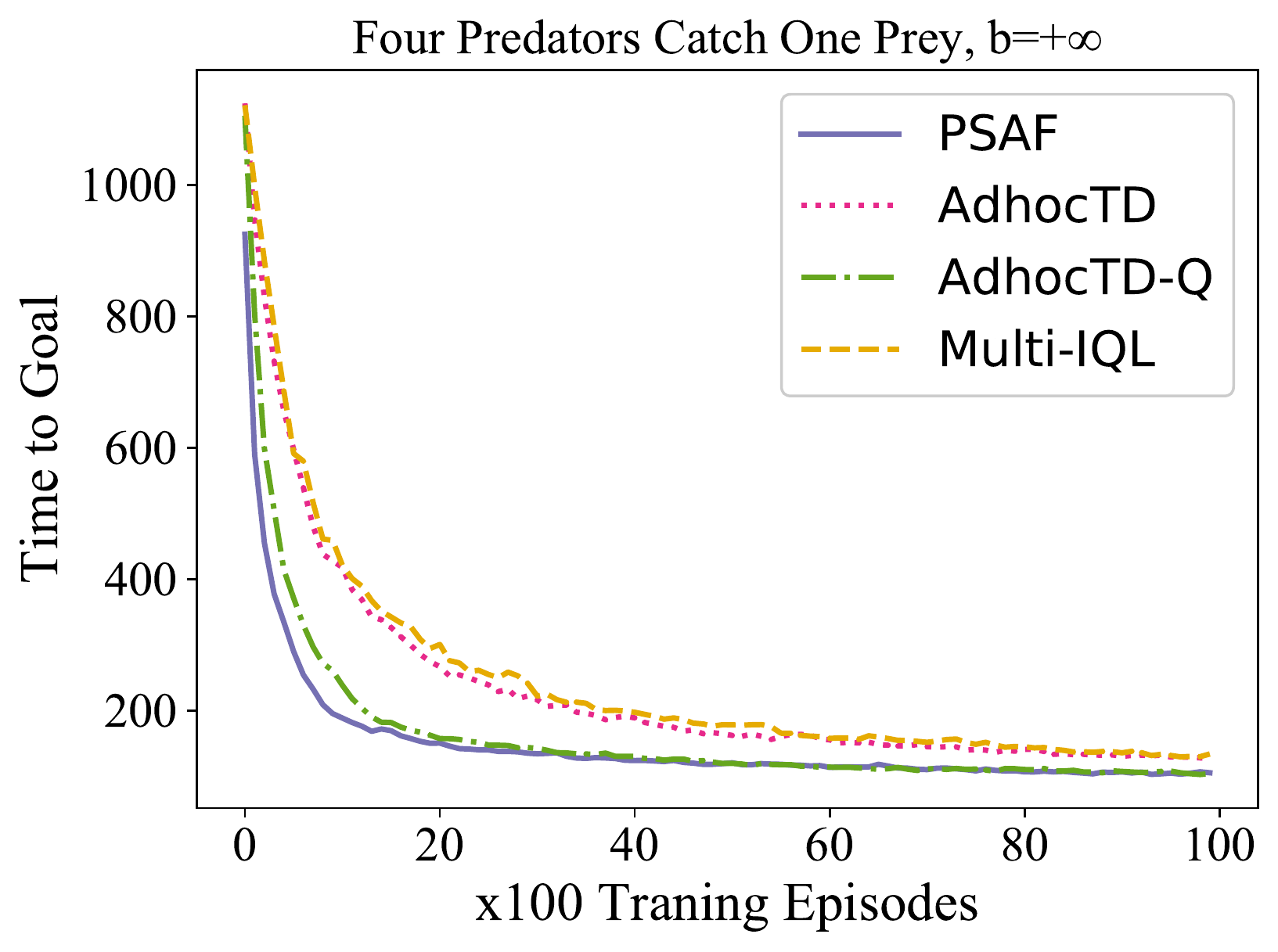}
  }
    \subfigure[TG between 50 and 200 with b=+$\infty$]{
    \includegraphics[width=2.5in, height=2.1in]{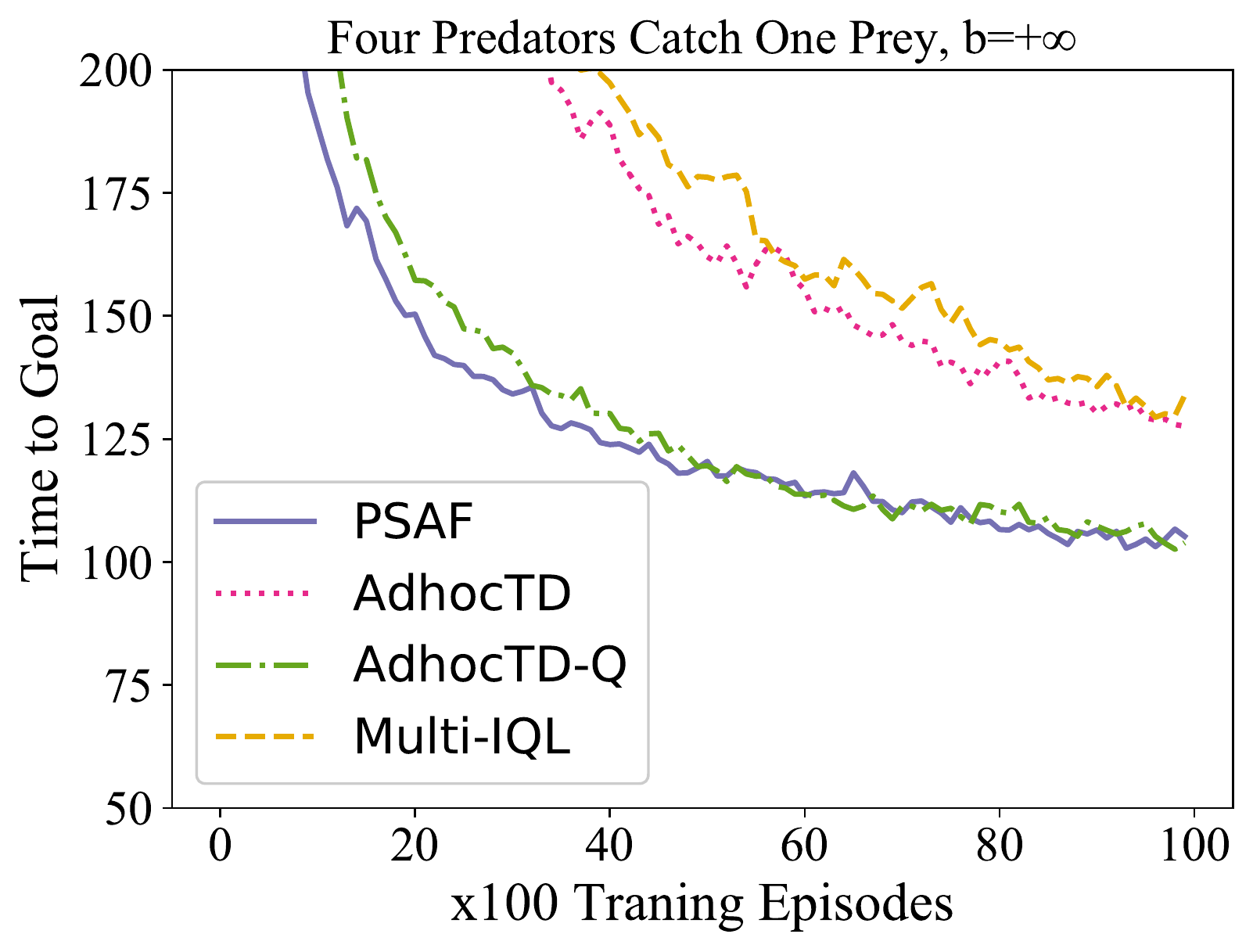}
  }

  \subfigure[TG of 10,000 episodes with b=2,500]{
    \includegraphics[width=2.5in, height=2.1in]{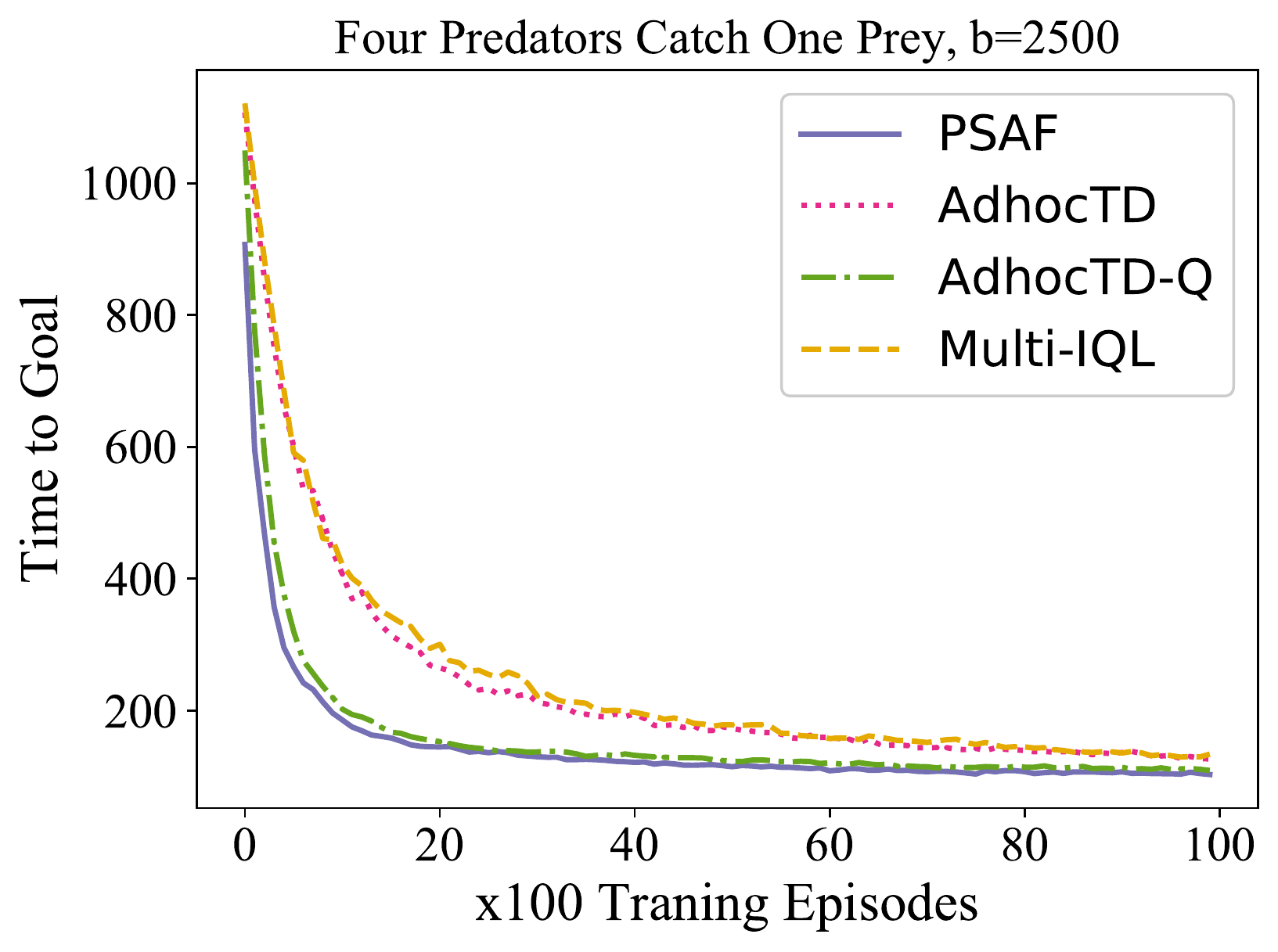}
  }
    \subfigure[TG between 50 and 200 with b=2,500]{
    \includegraphics[width=2.5in, height=2.1in]{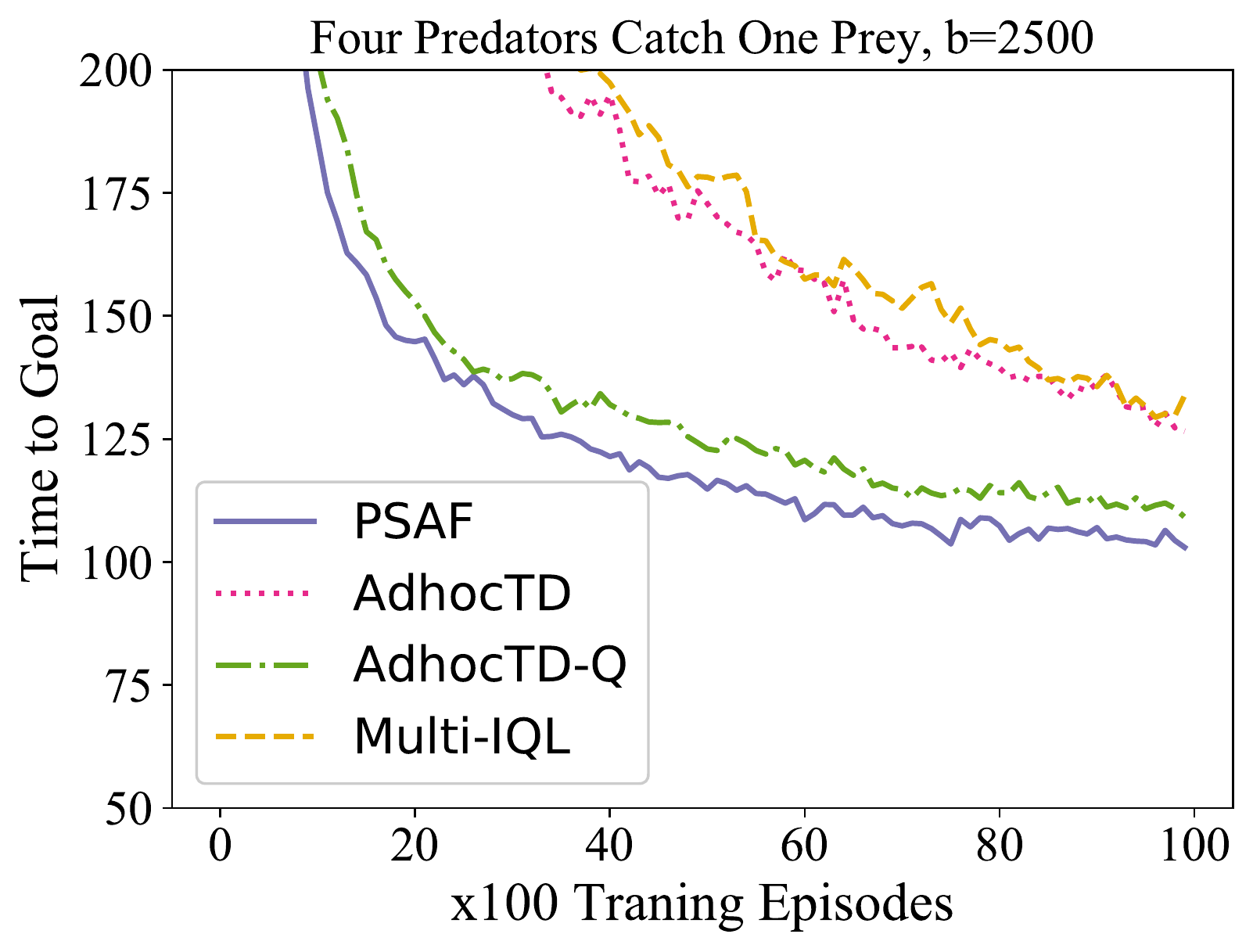}
  }

  \subfigure[TG of 10,000 episodes with b=1,800]{
    \includegraphics[width=2.5in, height=2.1in]{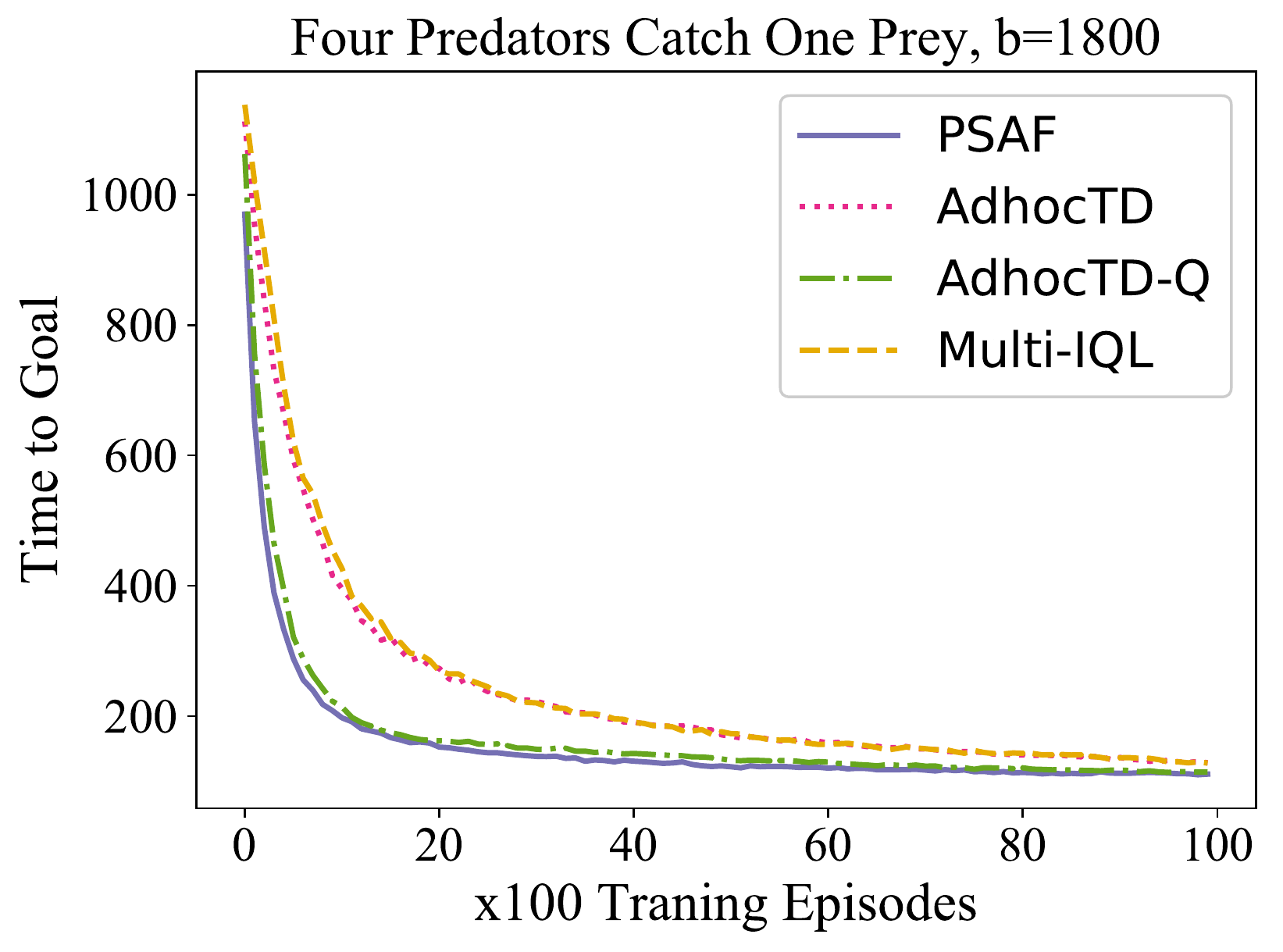}
  }
  \subfigure[TG between 50 and 200 with b=1,800]{
    \includegraphics[width=2.5in, height=2.1in]{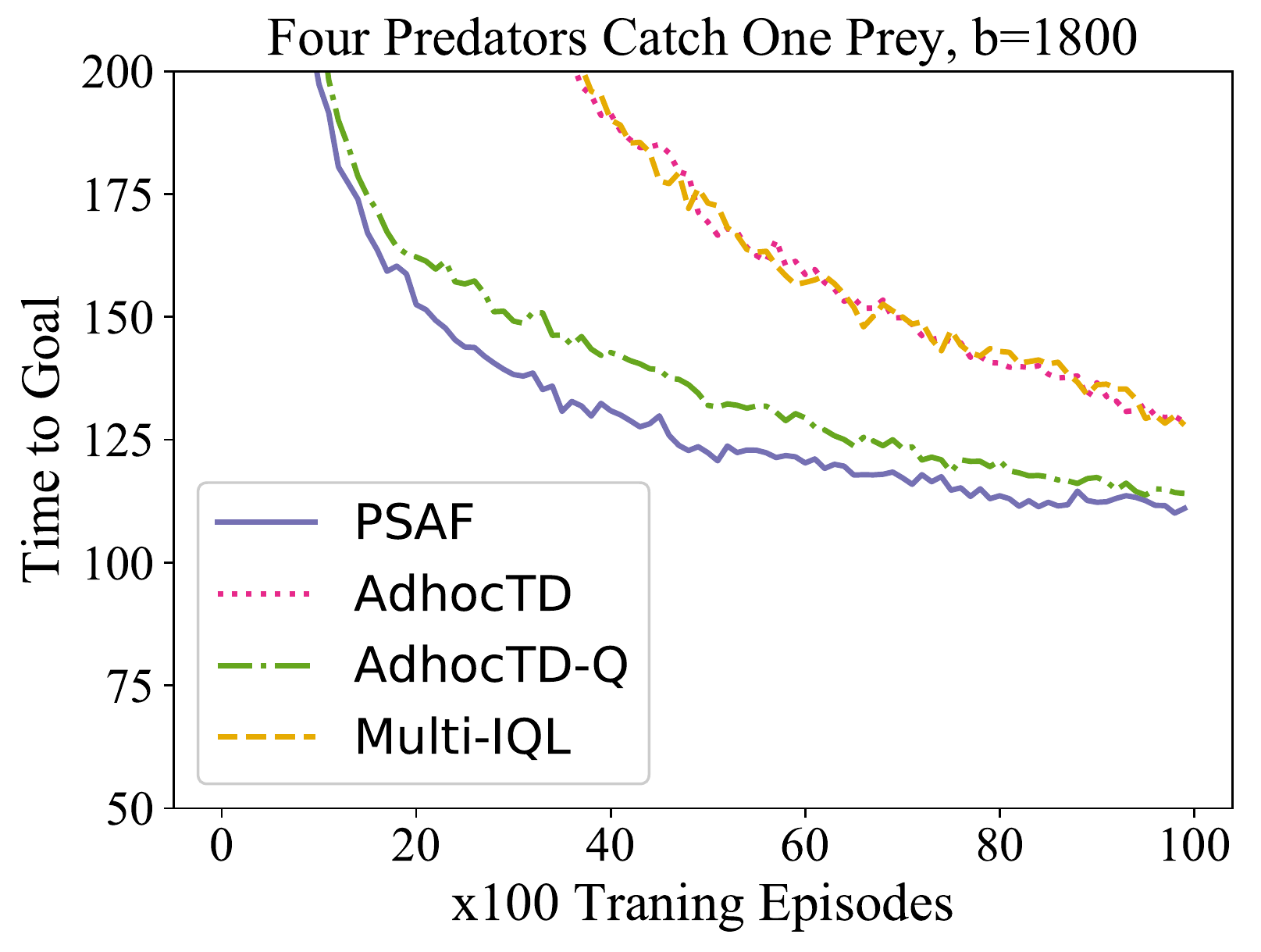}
  }
  \caption{TG of PSAF, AdhocTD, AdhocTD-Q and Multi-IQL in PP domain for cases 1, 2 and 3.}
\end{figure}

\begin{figure}[htbp]
\centering
    \subfigure[The consumption of $b_{give}$ with b=+$\infty$]{
    \includegraphics[width=2.5in, height=2in]{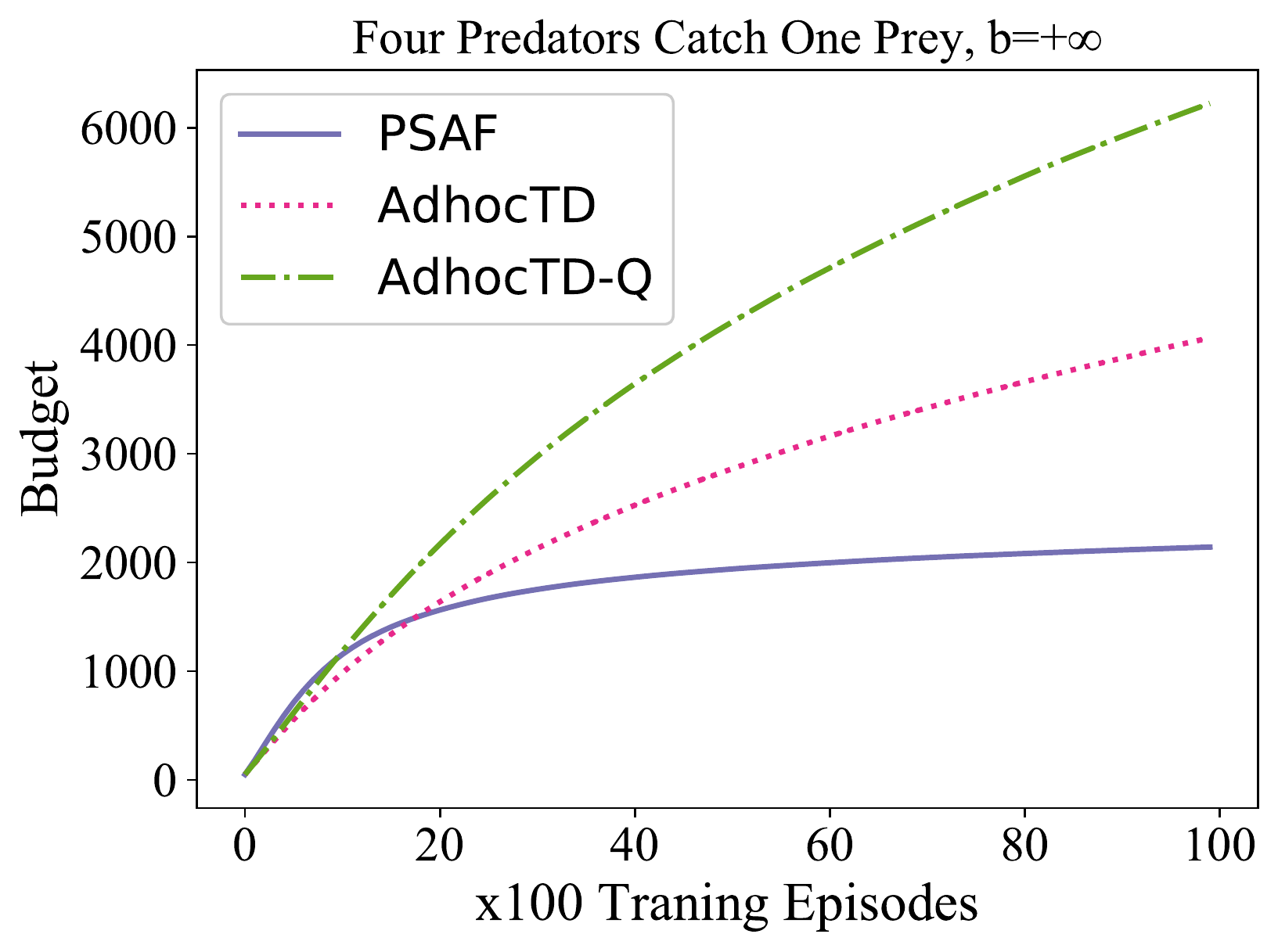}
  }
    \subfigure[The consumption of $b_{give}$ with b=2,500]{
    \includegraphics[width=2.5in, height=2in]{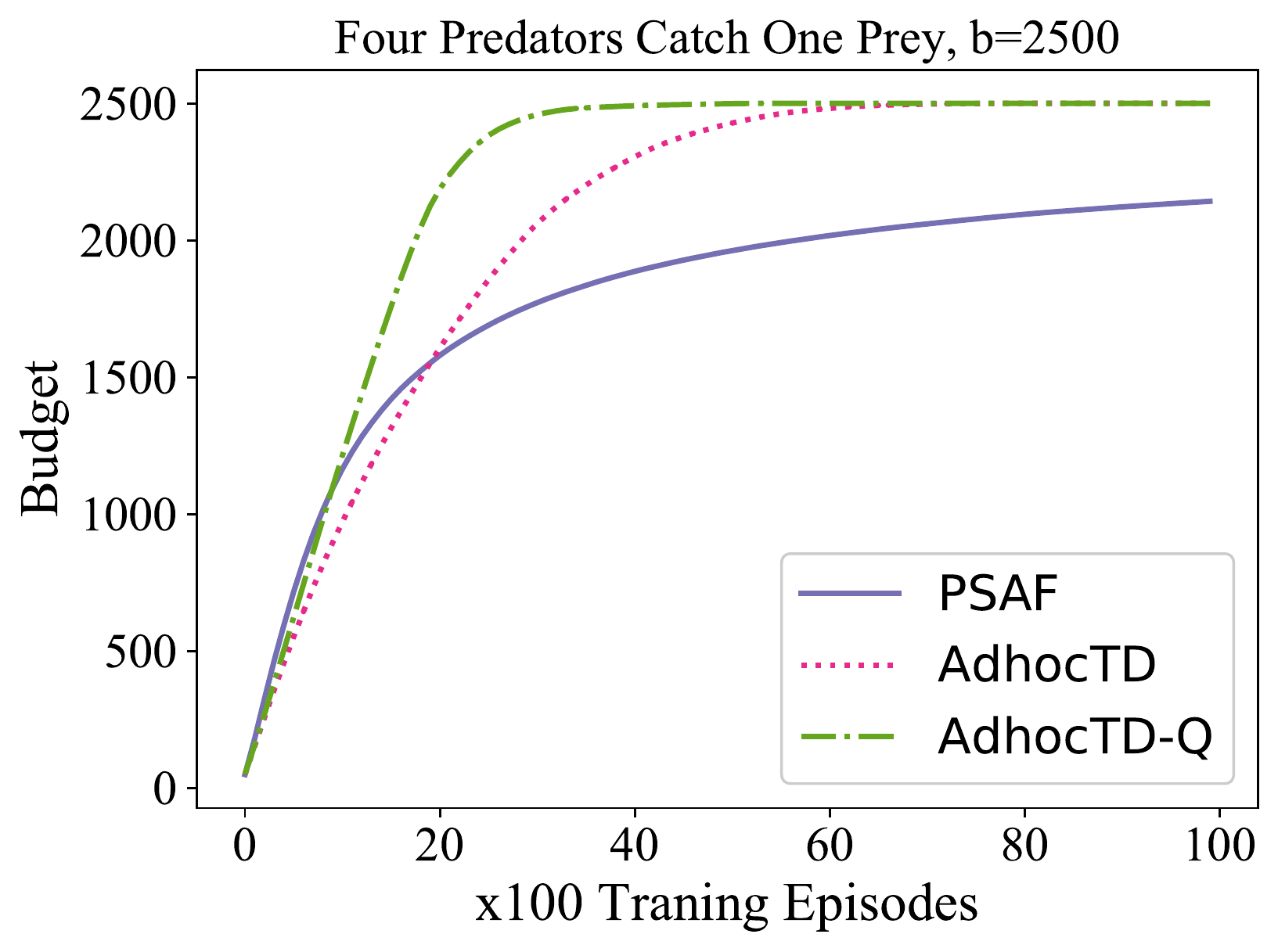}
  }

    \subfigure[The consumption of $b_{give}$ with b=1,800]{
    \includegraphics[width=2.5in, height=2in]{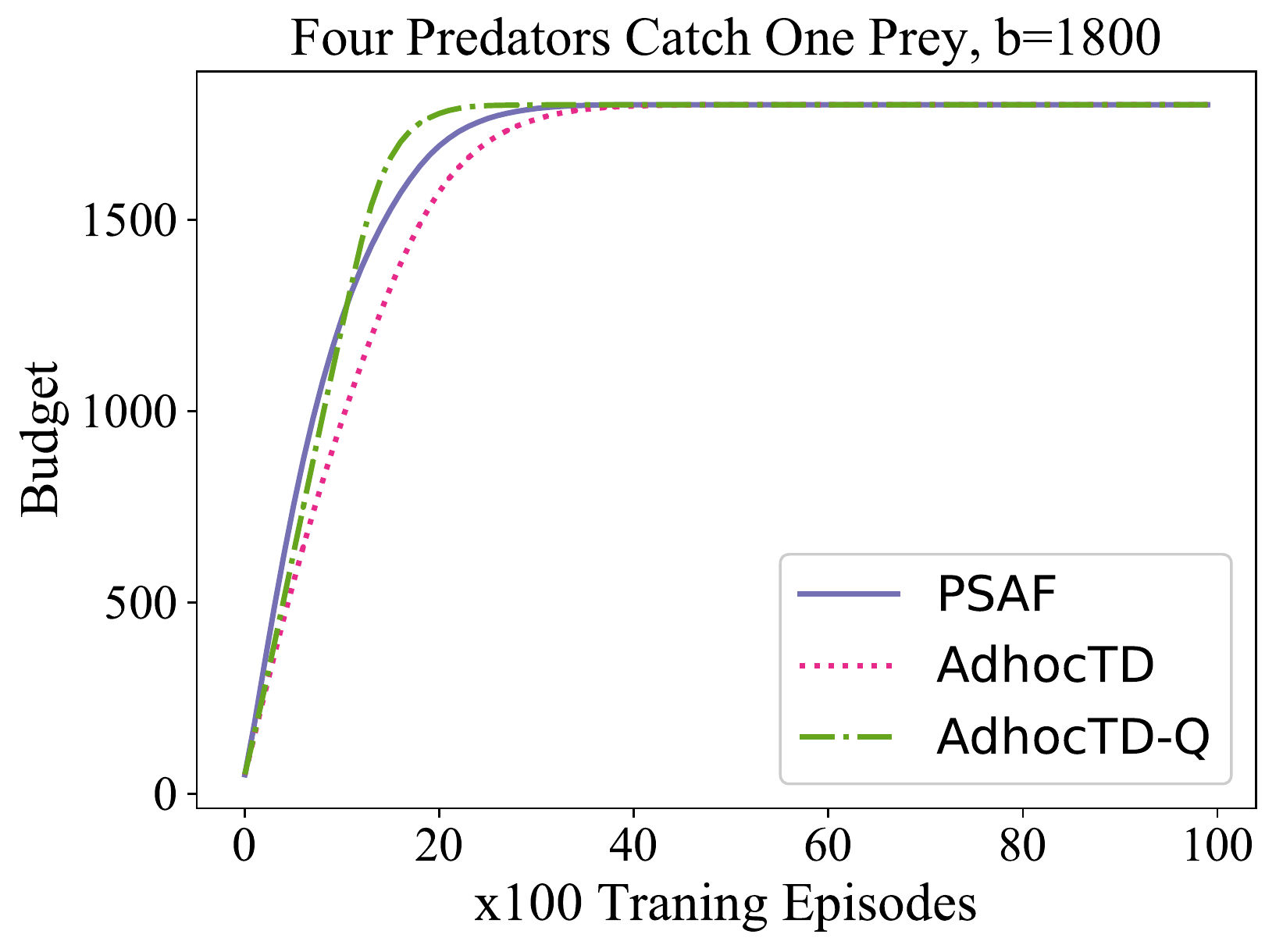}
  }
  \caption{The used budget $b_{give}$ of PSAF, AdhocTD and AdhocTD-Q in PP domain for cases 1, 2 and 3.}
\end{figure}

\begin{figure}[htbp]
  \centering
  \subfigure[The visited times of states]{
    \includegraphics[width=2.5in, height=2in]{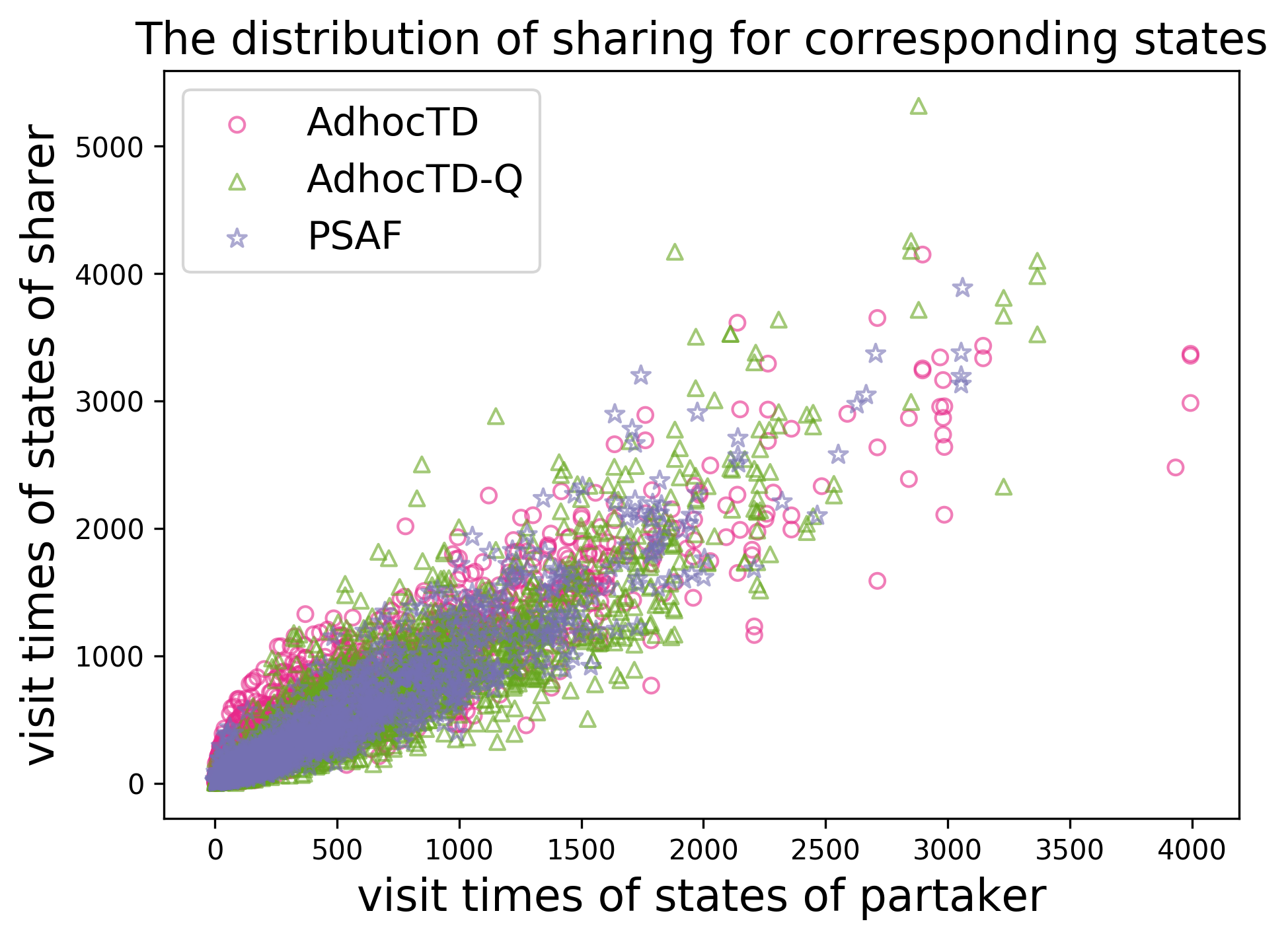}
  }
  \subfigure[The visited times of state-action pairs]{
    \includegraphics[width=2.5in, height=2in]{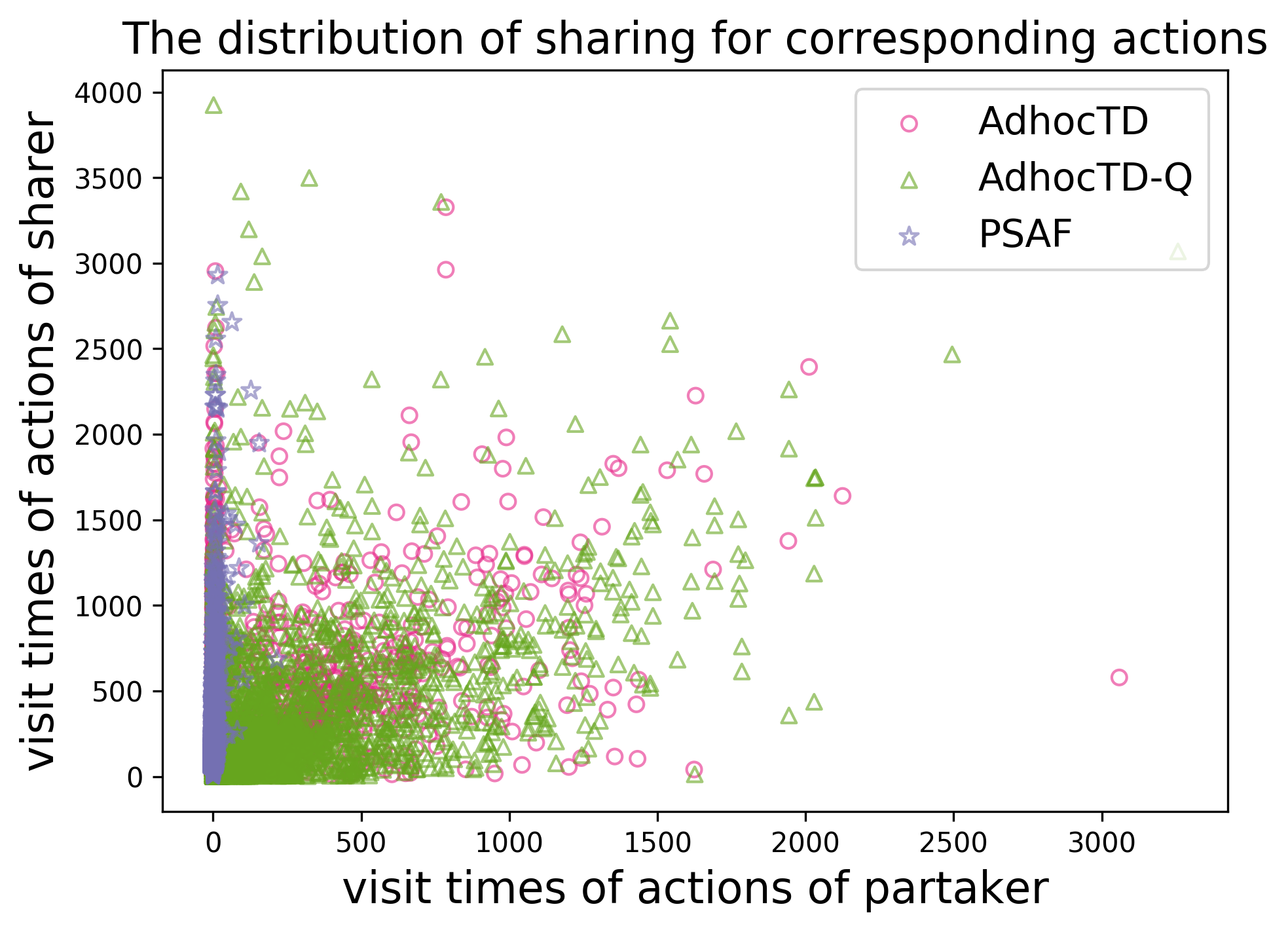}
  }
  \caption{The number of times that a partaker (student) and a sharer (teacher) visit the advised state (or state-action pair) for AdhocTD, AdhocTD-Q and PSAF in PP domain.}
\end{figure}

In order to explore the effect of different amount of budgets on the performance of each method, we consider three cases: (1) there are unlimited budgets, and we set $b_{ask}=b_{give}=+\infty$; (2) there are limited budgets, and we set $b_{ask}=b_{give}=$2,500; (3) there are even lower budgets for all methods, and we set $b_{ask}=b_{give}=$1,800. The average results of the 100 runs for three cases are shown in Figure 2. In Figure 2a, when the budget is unlimited, we can see that AdhocTD-Q and PSAF has significantly lower TG values than both AdhocTD and Multi-IQL. Sharing Q-values greatly improve the learning speed of predators comparing with advising actions, as well as disabling communication. Similar results can be seen in Figure 2c and 2e, where the budget is limited to 2,500 and 1,800 respectively. Figure 2b shows the detail of TG values under unlimited budgets. The TG of AdhocTD-Q is slightly higher than PSAF, even though AdhocTD-Q shares many more Q-values than PSAF, as shown in Figure 3a. After about 5,000 episodes, The TG of AdhocTD-Q gradually approaches PSAF. In Figure 2f, where all methods spend all budget before 4,000 episodes, PSAF still has significant improvements in learning compared with AdhocTD-Q. Besides, as available budget decreases, the difference between AdhocTD and Multi-IQL becomes smaller. Figure 3a shows that PSAF gradually consume less budgets as training time increases, and it ends up spending almost no budget, since predators are likely to explore the state space many times. However, AdhocTD-Q consumes many more budgets than other two methods, which means predators in AdhocTD-Q have more opportunities to share their Q-values. When the budget is limited to 2,500 in Figure 3b, we can see that AdhocTD-Q and AdhocTD spend all budgets completely at about 2,000 and 6,000 episode respectively, resulting in higher TG values than the case of unlimited budgets. In Figure 3c, although all methods quickly run out of all budget, PSAF still significantly outperforms other methods. All experiments in PP domain show that sharing Q-values achieves significant improvements compared with advising actions regarding TG evaluation. The performance of AdhocTD-Q and AdhocTD heavily depends on available budget. In case 1 and 2, PSAF has the best results while only needs to share about 2,000 Q-values for each predator during the whole learning process.

We now attempt to figure out when predators share Q-values or actions in AdhocTD, AdhocTD-Q and PSAF during learning. Since these methods encode the number of times an agent visits a state (or a state-action pair), we investigate how many times the advised states are visited by partakers (students) and sharers (teachers) in all sharing processes for a single run. Note that the budget is unlimited. As shown in Figure 4a, we can see that AdhocTD, AdhocTD-Q and PSAF have similar distribution of sharing opportunities regarding the visited times of states in which Q-values (or actions) are shared. Most sharing processes are initiated in states that a partaker (student) visits very few times. Due to the probabilistic functions both in asking and giving, an agent equipped with AdhocTD and AdhocTD-Q still have opportunities to be suggested even when it has visited current state many times. In Figure 4b, when a Q-value (or an action) is shared in a sharing process, we record the number of times that the partaker (student) and the sharer (teacher) visit the corresponding state-action pair. We can see that in PSAF, most Q-values are shared in state-action pairs that partakers visit very few times, while sharers visit many more times. However, in AdhocTD-Q and AdhocTD, Q-values or actions are shared even when partakers (students) visit corresponding state-action pairs very often. Figure 4 shows that sharing Q-values to a partaker is more helpful when it rarely updates the Q-values, and a sharer should make sure that the Q-value to be shared has been sufficiently learned.

\subsection{Half Field Offense}

\begin{figure}[htbp]
\centering
  \includegraphics[width=2.1in, height=1.3in]{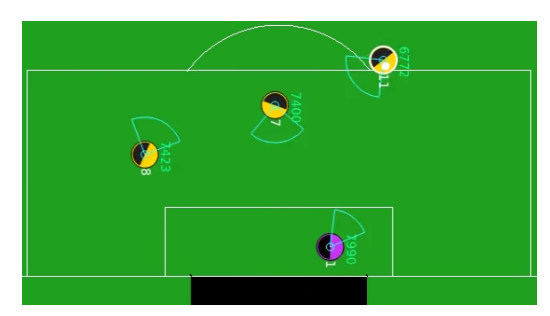}
  \caption{Half Field Offense. Three players against one goalkeeper.}
\end{figure}

Half Field Offense (HFO) \cite{Stone2016HalfFO} is a subtask of simulated robot soccer game. Our implementation is based on the work of Silva et al. \cite{Silva2017SimultaneouslyLA}, where HFO has been tested by AdhocTD to achieve promising results.\footnote{The source code is available at https://github.com/f-leno/AdHoc\_AAMAS-17.} As shown in Figure 5, there are three RL players learning to score a goal against one goalkeeper in the field. The goalkeeper is an automated agent using a policy derived from Helios, the 2012 RoboCup 2D champion team \cite{HFOPlatform}. When a player dose not have the ball, its only option is \emph{Move}, which is an automated action guided by Helios. When the player has the ball, it learns to choose between four actions \emph{Shoot}, \emph{PassNear}, \emph{PassFar}, and \emph{Dribble}. The three players benefit from cooperative behaviours, e.g., one player passes the ball to another player for better shot. In our experiments, the state of a player is composed of following observations.

\begin{enumerate}[(1)]
  \item \textbf{Able to Kick}: Boolean indicating if the player can kick the ball.
  \item \textbf{Goal Center Proximity}: The player's proximity to the center of the goal.
  \item \textbf{Goal Center Angle}: Angle from the player to the center of the goal.
  \item \textbf{Goal Opening Angle}: The largest open angle of the player to the goal with no other blocking players.
  \item \textbf{Teammate 1's Goal Opening Angle}: The goal opening angle of the player's nearest teammate.
  \item \textbf{Teammate 2's Goal Opening Angle}: the goal opening angle of the player's farthest teammate.
\end{enumerate}

These state features are normalized in the range $[-1, 1]$. They are discretized by Tile Coding \cite{Sherstov2005Function, Sutton1998ReinforcementLA}, where the number of tiles is 10, and tile width is 0.5. One episode starts when the players and the ball initialized with random positions on the field. The episode ends when either a player scores a goal, the goalkeeper catches the ball, the ball leaves the field, or a time limit is exceeded. In each episode, the maximum number of steps that players can perform is 200. We consider that three players cooperatively learn to score the ball. All players receive a common reward of $1$ when a player scores a goal, $-1$ when the ball is caught by the goalkeeper or the ball is out of bounds, and $0$ when players reach the maximum time step in current episode. We train the three players for 10,000 episodes. At every 20 training episodes, 100 testing episodes are played. During testing episodes, all players are unable to learn and share Q-values. They only execute best actions for evaluating currently learned policies. The whole process is repeated 50 runs. We use two standard evaluation metrics in HFO. The primary metric, \emph{Goal Percentage} (GP), is the percentage of all testing episodes which end with a goal being scored. The second metric, \emph{Time to Goal} (TG), is the average number of steps required to score in all testing episodes that culminates with a goal. When players do not share Q-values or actions during learning process, they use SARSA($\lambda$), where $\lambda$=0.9, $\alpha$=0.1, and $\gamma$=0.9. The exploration strategy for all players is $\epsilon$-greedy with $\epsilon$= 0.1. For $v_a$ and $v_b$ in AdhocTD, we use the same value in \cite{Silva2017SimultaneouslyLA}, which has shown satisfied results. Then we set $v_a$=0.5 and $v_b$=1.5 for AdhocTD, AdhocTD-Q and PSAF. Similar to the PP domain, we test three cases for all methods: (1) $b_{ask}=b_{give}=+\infty$; (2) $b_{ask}=b_{give}=$150; (3) $b_{ask}=b_{give}=$50.

The average results of 50 runs are shown in Figures 6-8. In Figures 6a and 6b, where the budget is unlimited, both PSAF and AdhocTD-Q have significantly higher GP than AdhocTD and Multi-IQL. The GP of AdhocTD-Q is slightly higher than PSAF between 4,000 and 6,000 episodes, while AdhocTD-Q gradually achieves similar GP as PSAF. Nevertheless, in Figure 8a, we can see that PSAF consumes much less budget than AdhocTD-Q. When the budget is limited to 150, both AdhocTD and AdhocTD-Q spend all budget before 2,000 episodes, resulting in significantly lower GP than PSAF, as shown in Figures 6c and 6d. When the available budget drops further, we can see that in Figures 6e and 6f, PSAF still achieves higher GP than other methods, which means PSAF more applicable to scenarios with a small budget. The results of all methods evaluated by TG are shown in Figure 7. We can see that when the budget is unlimited, the TG of AhocTD-Q is significantly lower than other methods. However, as expected, when the available budget decreases, AdhocTD and AdhocTD-Q has similar TG values to Multi-IQL, while PSAF significantly outperforms all other methods. Figure 8 shows that PSAF consumes much less budget than other methods. Specifically, for case 1 and 2, each player equipped with PSAF only needs to share 125 Q-values in the 10,000 training episodes, while achieves similar (even better) performance comparing with AdhocTD-Q.

\begin{figure}[htbp]
  \centering
  \subfigure[GP with $b_{ask}=b_{give}=+\infty$]{
    \includegraphics[width=2.5in, height=2.1in]{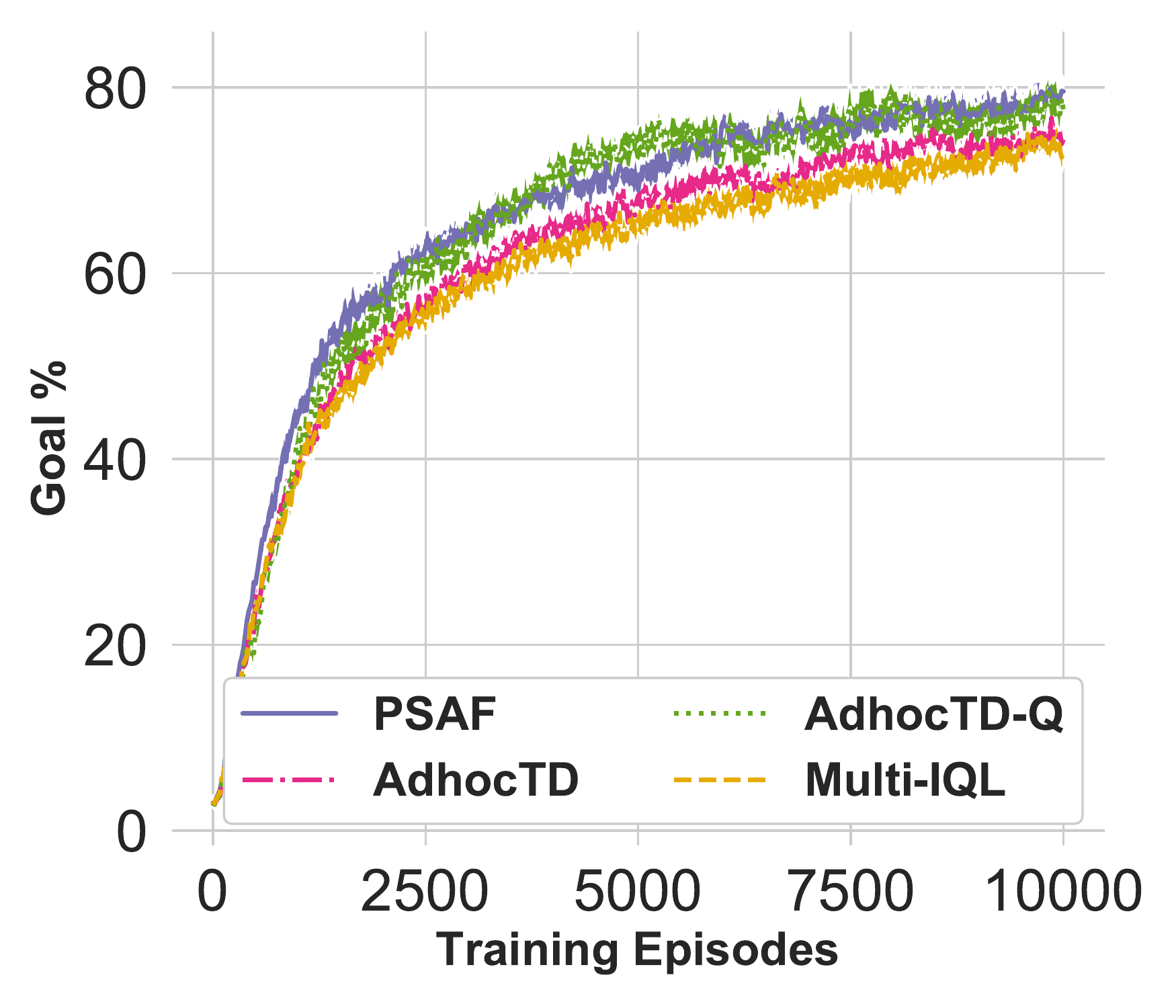}
  }
  \subfigure[GP between 7,000 and 10,000 episodes in Figure (a)]{
    \includegraphics[width=2.5in, height=2.1in]{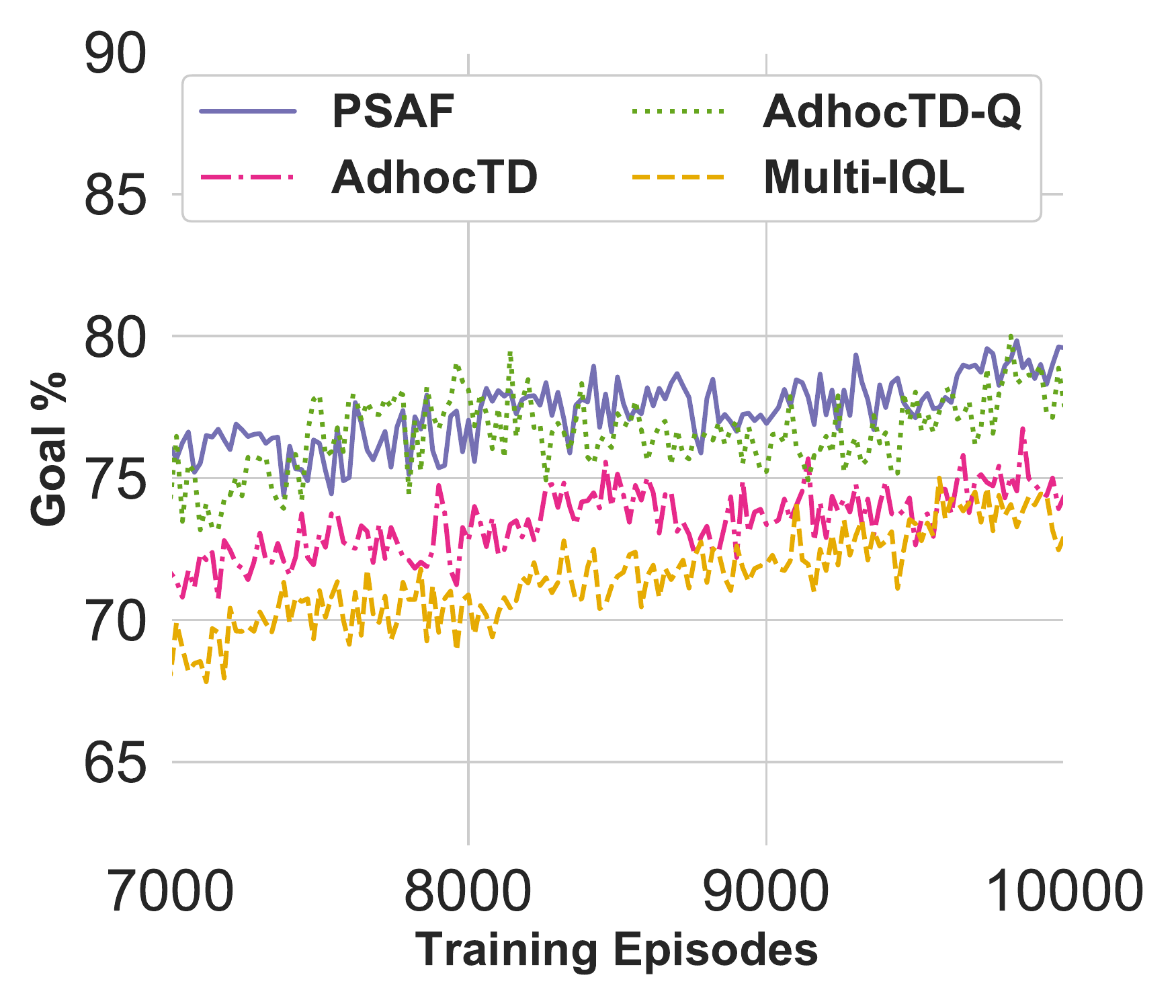}
  }

    \subfigure[GP with $b_{ask}=b_{give}=$150]{
    \includegraphics[width=2.5in, height=2.1in]{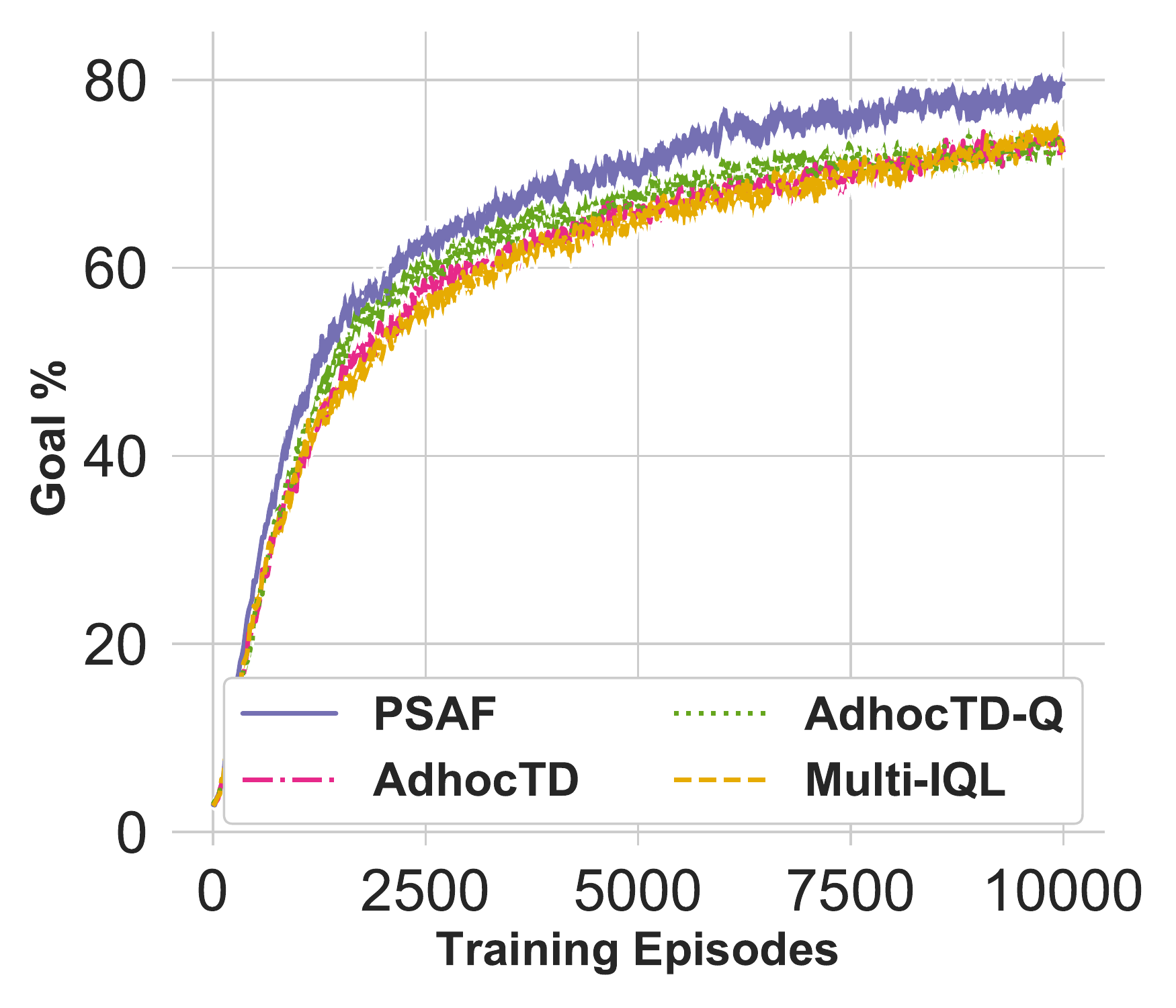}
  }
  \subfigure[GP between 7,000 and 10,000 episodes in Figure (c)]{
    \includegraphics[width=2.5in, height=2.1in]{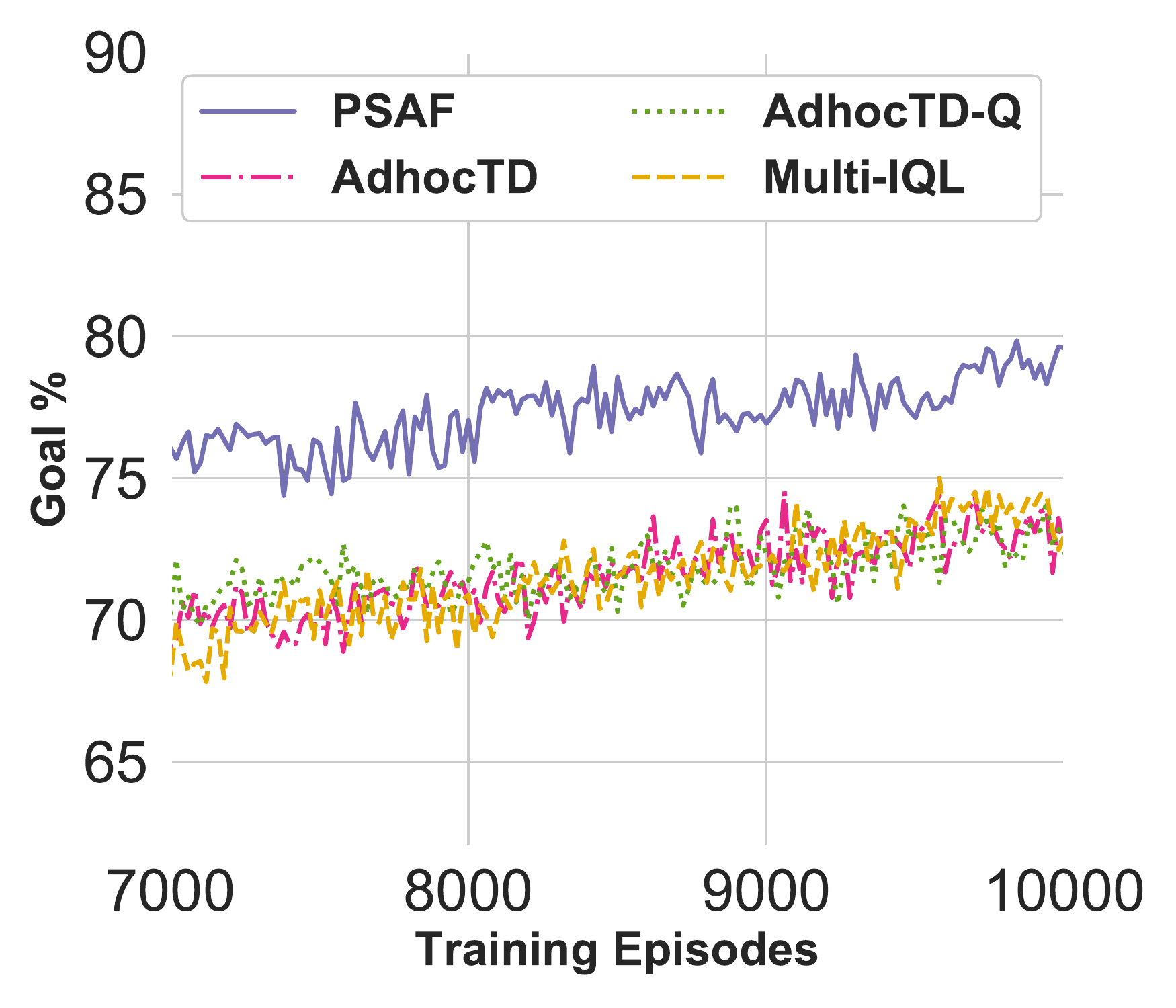}
  }

    \subfigure[GP with $b_{ask}=b_{give}=$50]{
    \includegraphics[width=2.5in, height=2.1in]{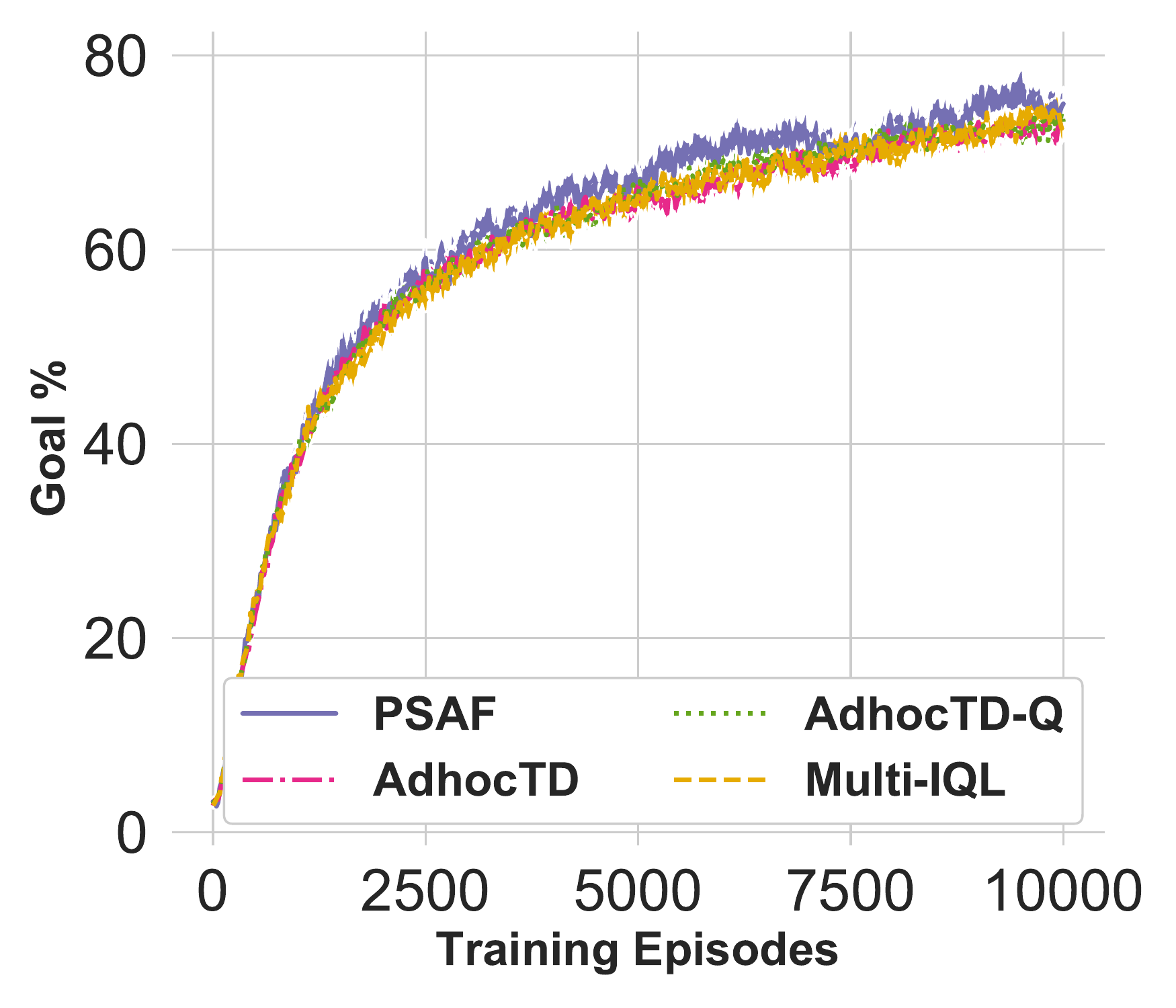}
  }
  \subfigure[GP between 7,000 and 10,000 episodes in Figure (e)]{
    \includegraphics[width=2.5in, height=2.1in]{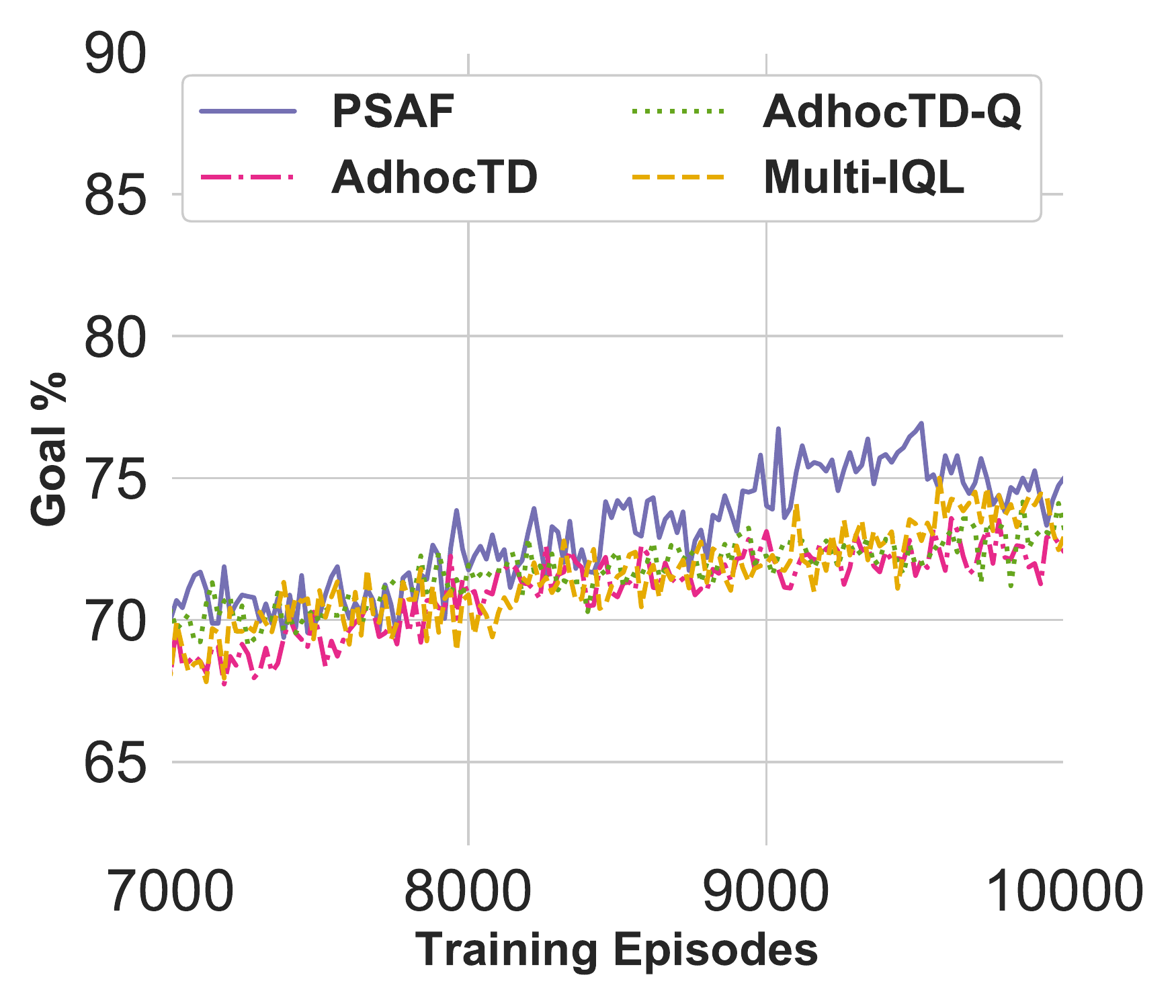}
  }
  \caption{GP of PSAF, AdhocTD, AdhocTD-Q and Multi-IQL in HFO for cases 1, 2 and 3.}
\end{figure}

\begin{figure}[htbp]
  \centering
  \subfigure[TG with $b_{ask}=b_{give}=+\infty$]{
    \includegraphics[width=2.5in, height=2.1in]{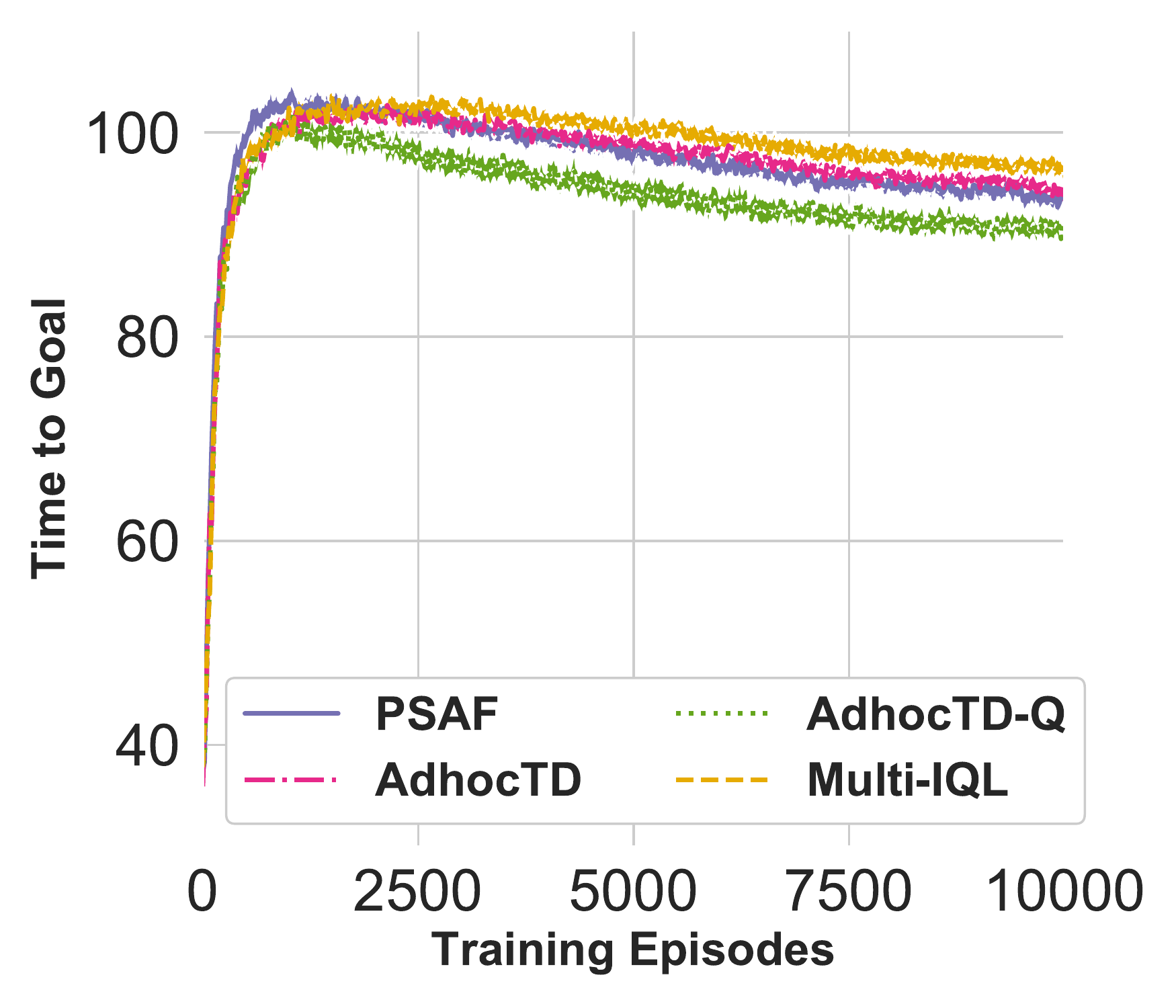}
  }
  \subfigure[TG between 7,000 and 10,000 episodes in Figure (a)]{
    \includegraphics[width=2.5in, height=2.1in]{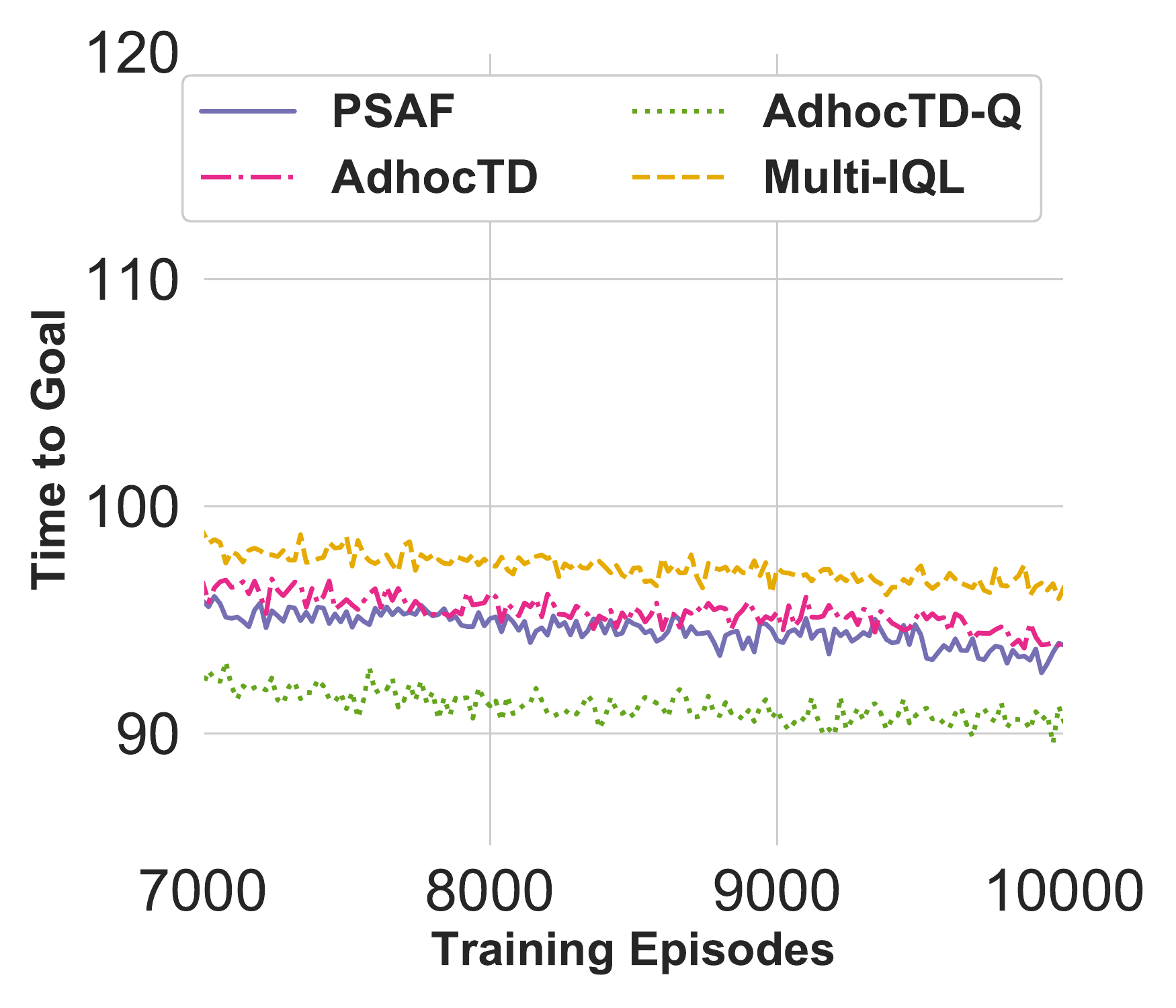}
  }

    \subfigure[TG with $b_{ask}=b_{give}=$150]{
    \includegraphics[width=2.5in, height=2.1in]{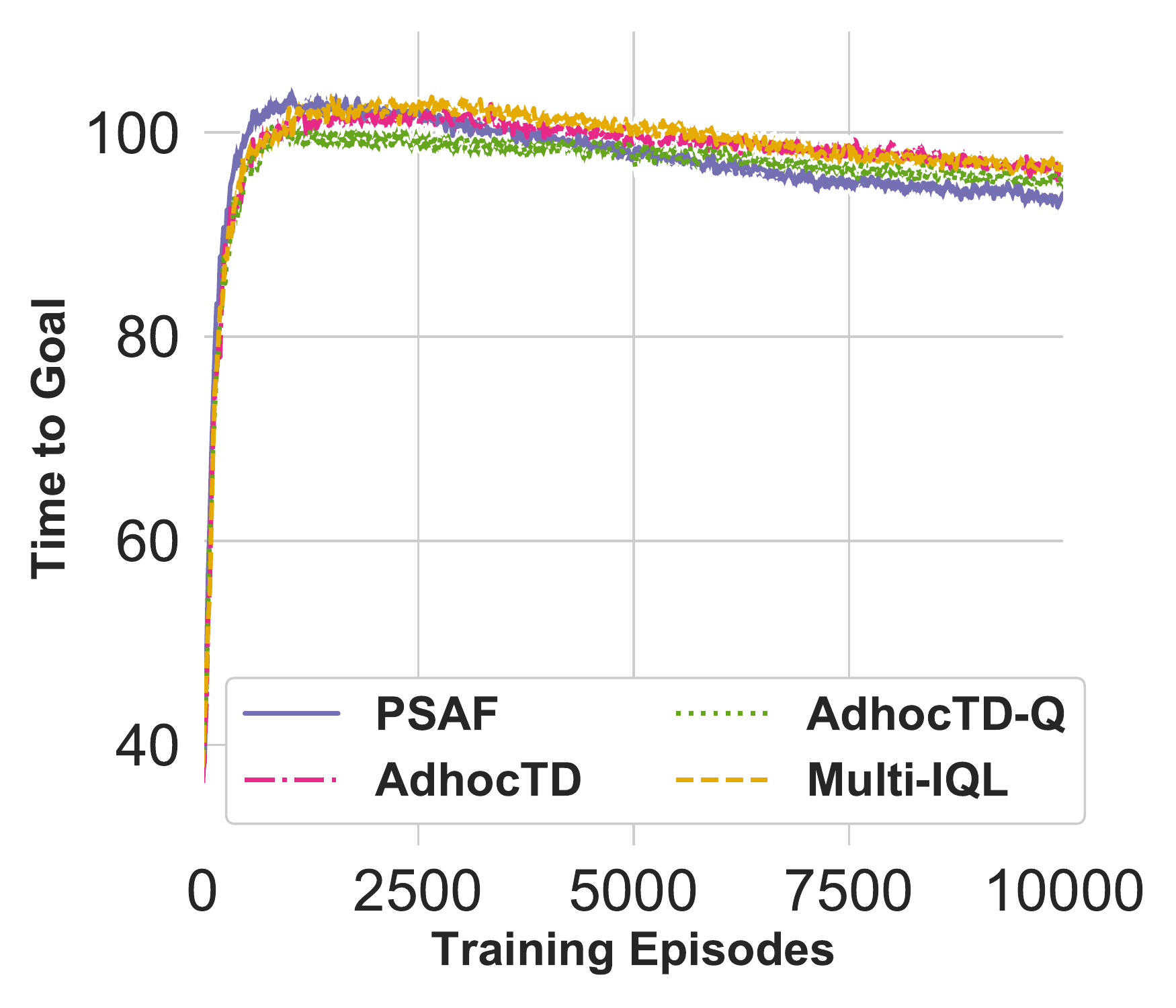}
  }
  \subfigure[TG between 7,000 and 10,000 episodes in Figure (c)]{
    \includegraphics[width=2.5in, height=2.1in]{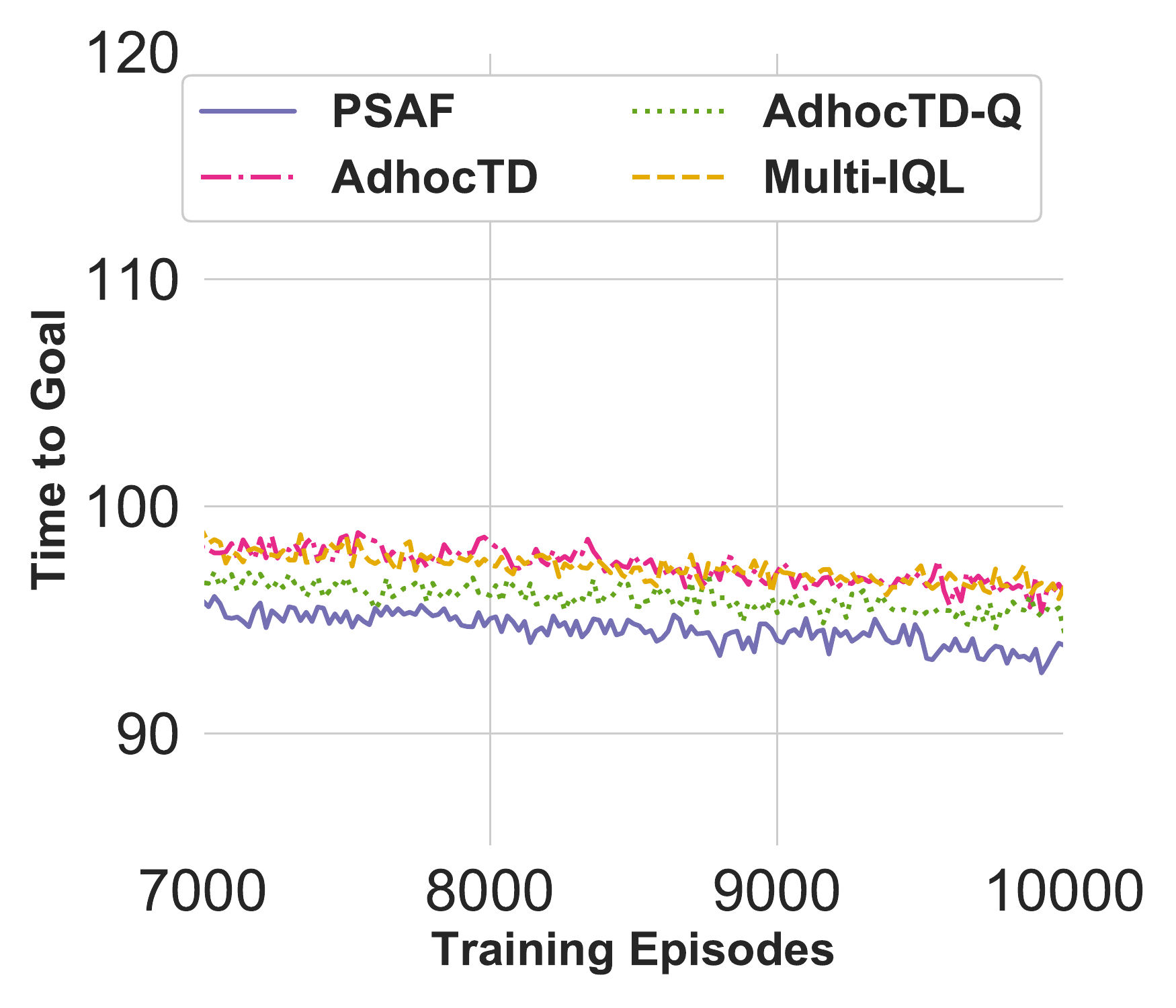}
  }

    \subfigure[TG with $b_{ask}=b_{give}=$50]{
    \includegraphics[width=2.5in, height=2.1in]{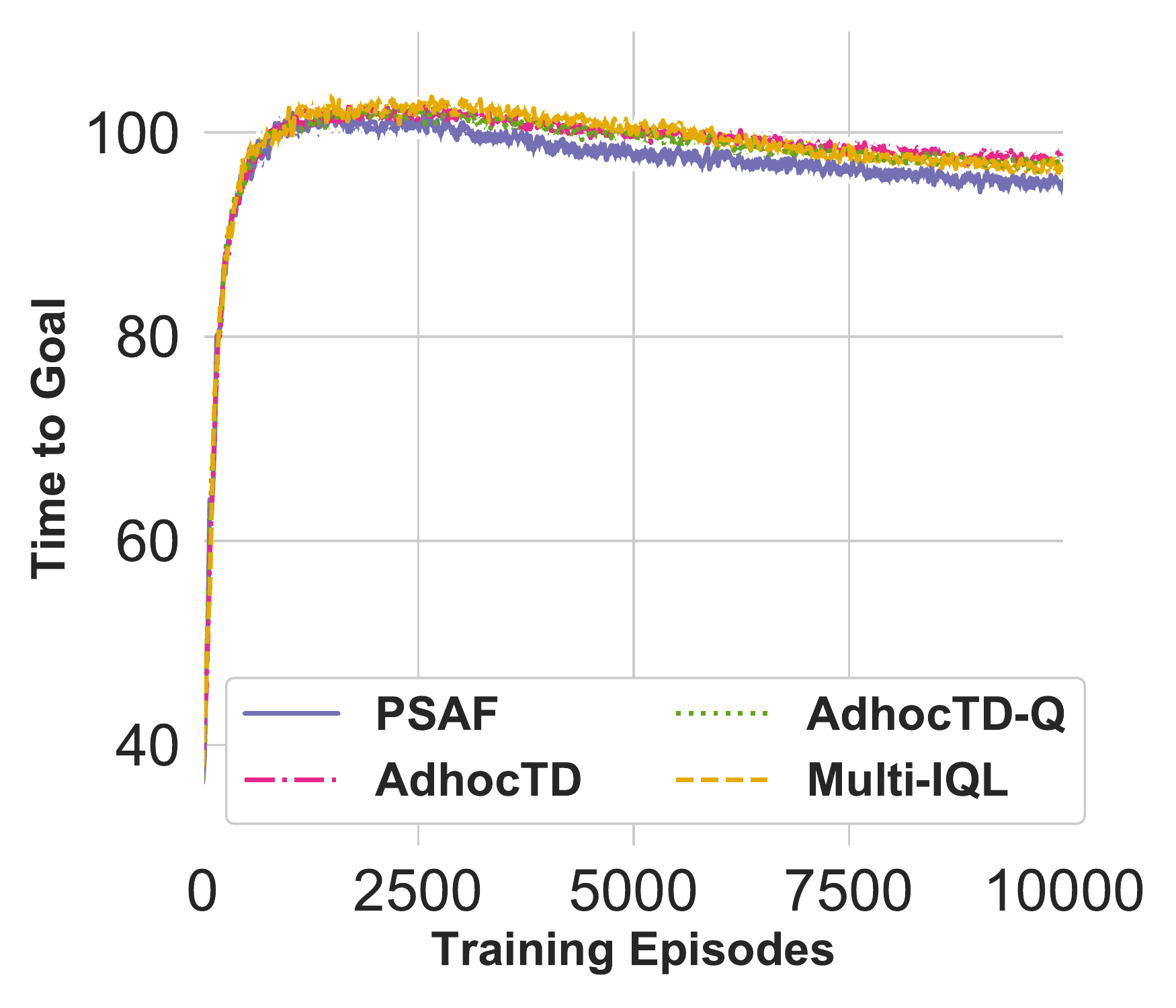}
  }
  \subfigure[TG between 7,000 and 10,000 episodes in Figure (e)]{
    \includegraphics[width=2.5in, height=2.1in]{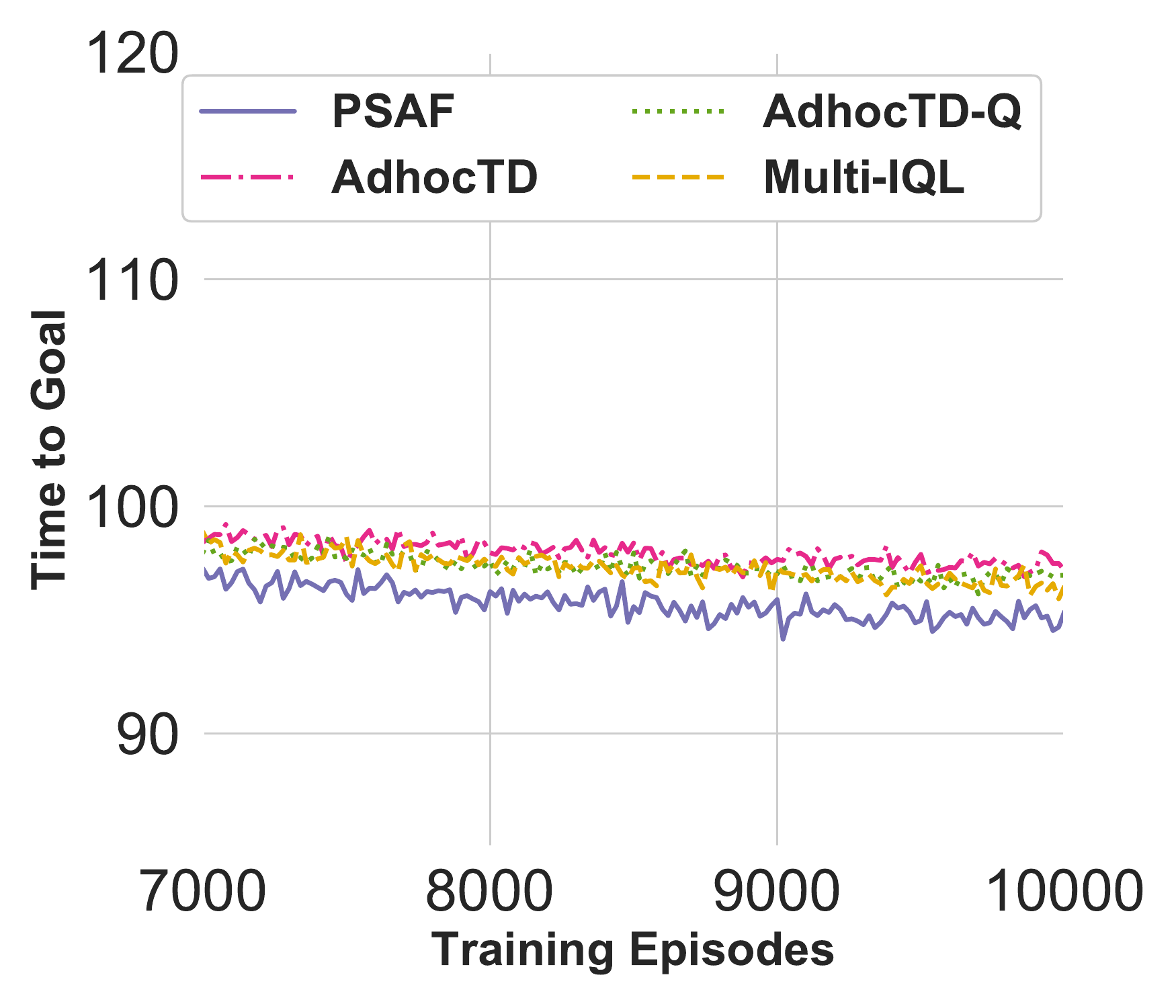}
  }
  \caption{TG of PSAF, AdhocTD, AdhocTD-Q and Multi-IQL in HFO for cases 1, 2 and 3.}
\end{figure}

\begin{figure}[htbp]
  \centering
  \subfigure[Unlimited budget]{
    \includegraphics[width=2.5in, height=2in]{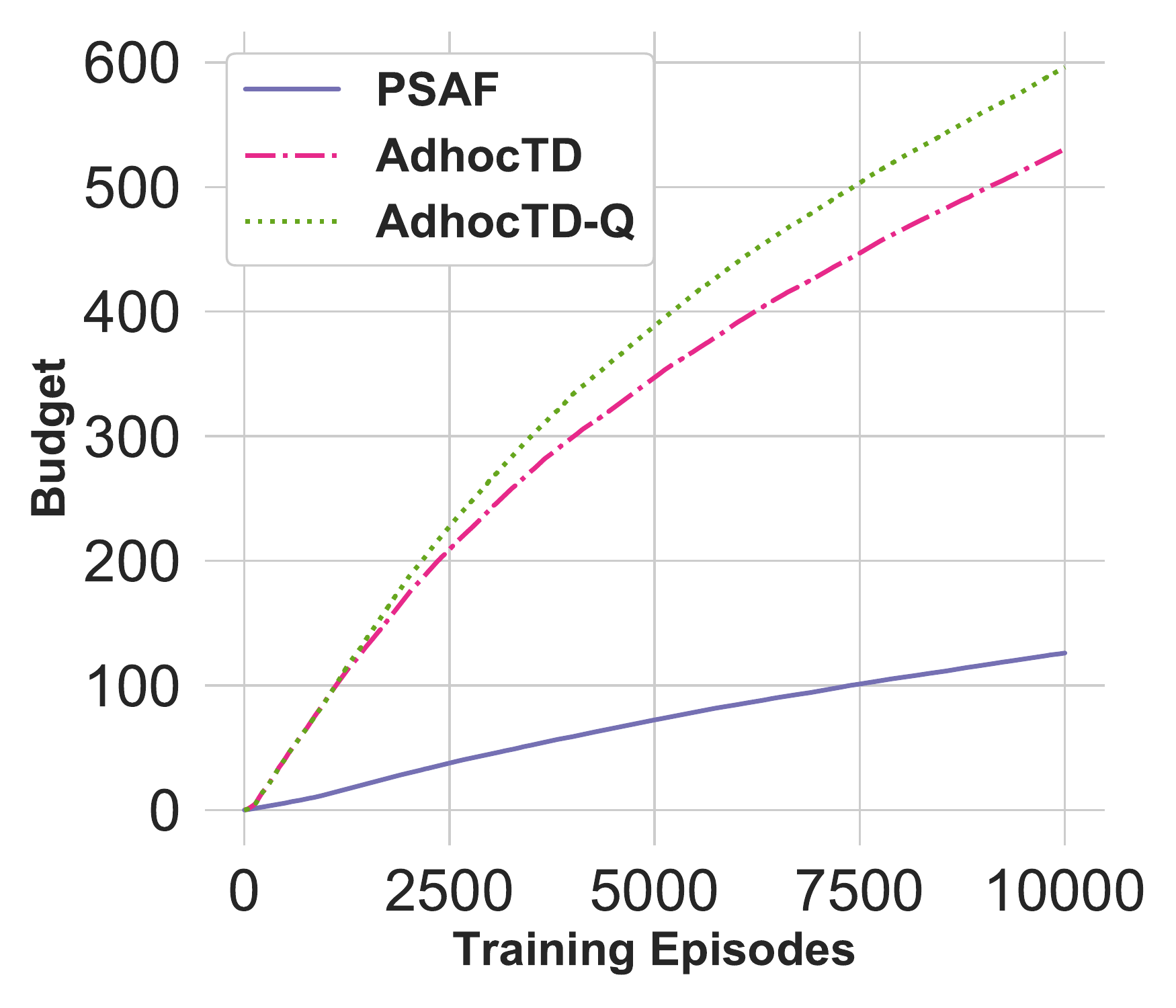}
  }
  \subfigure[The budget is 150]{
    \includegraphics[width=2.5in, height=2in]{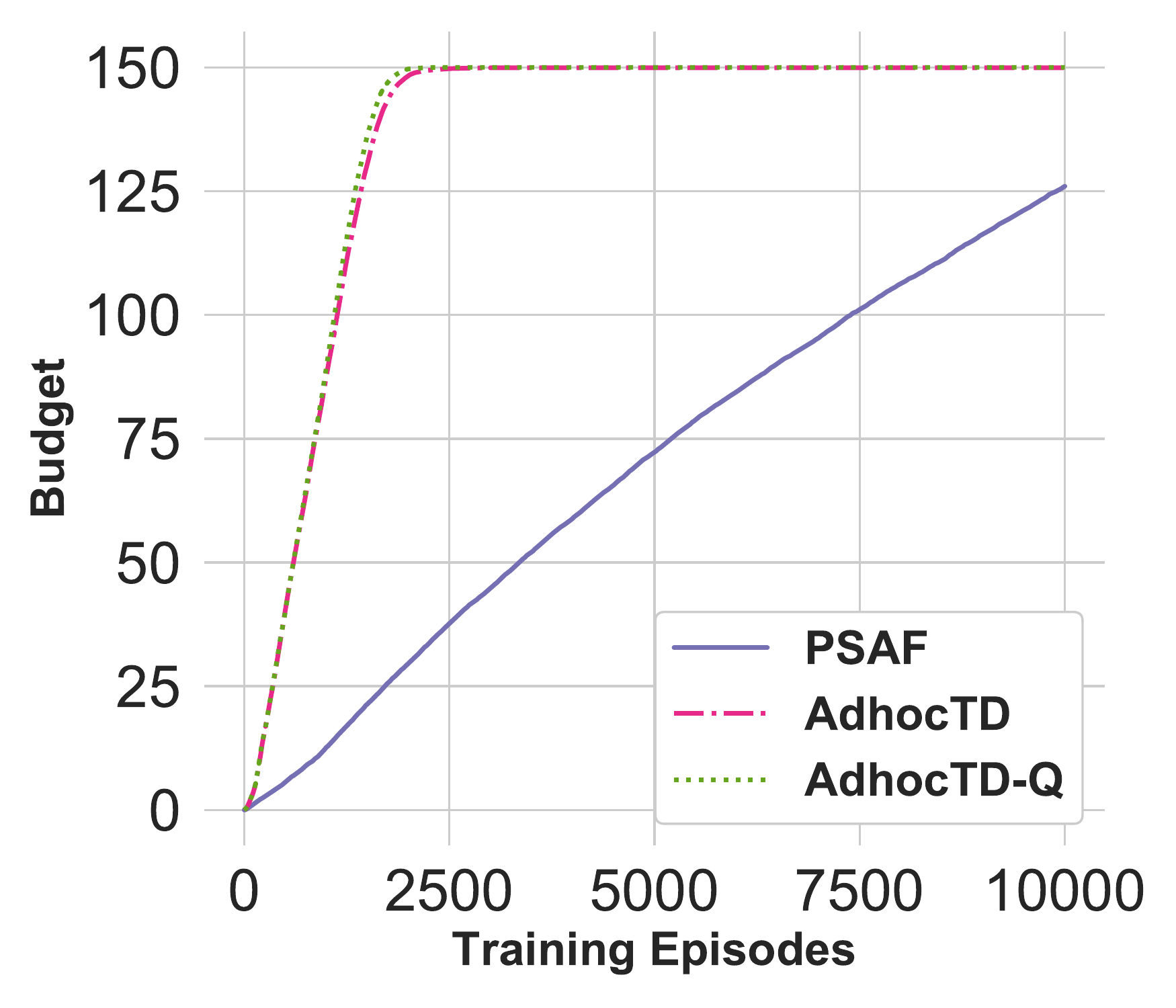}
  }

    \subfigure[The budget is 50]{
    \includegraphics[width=2.5in, height=2in]{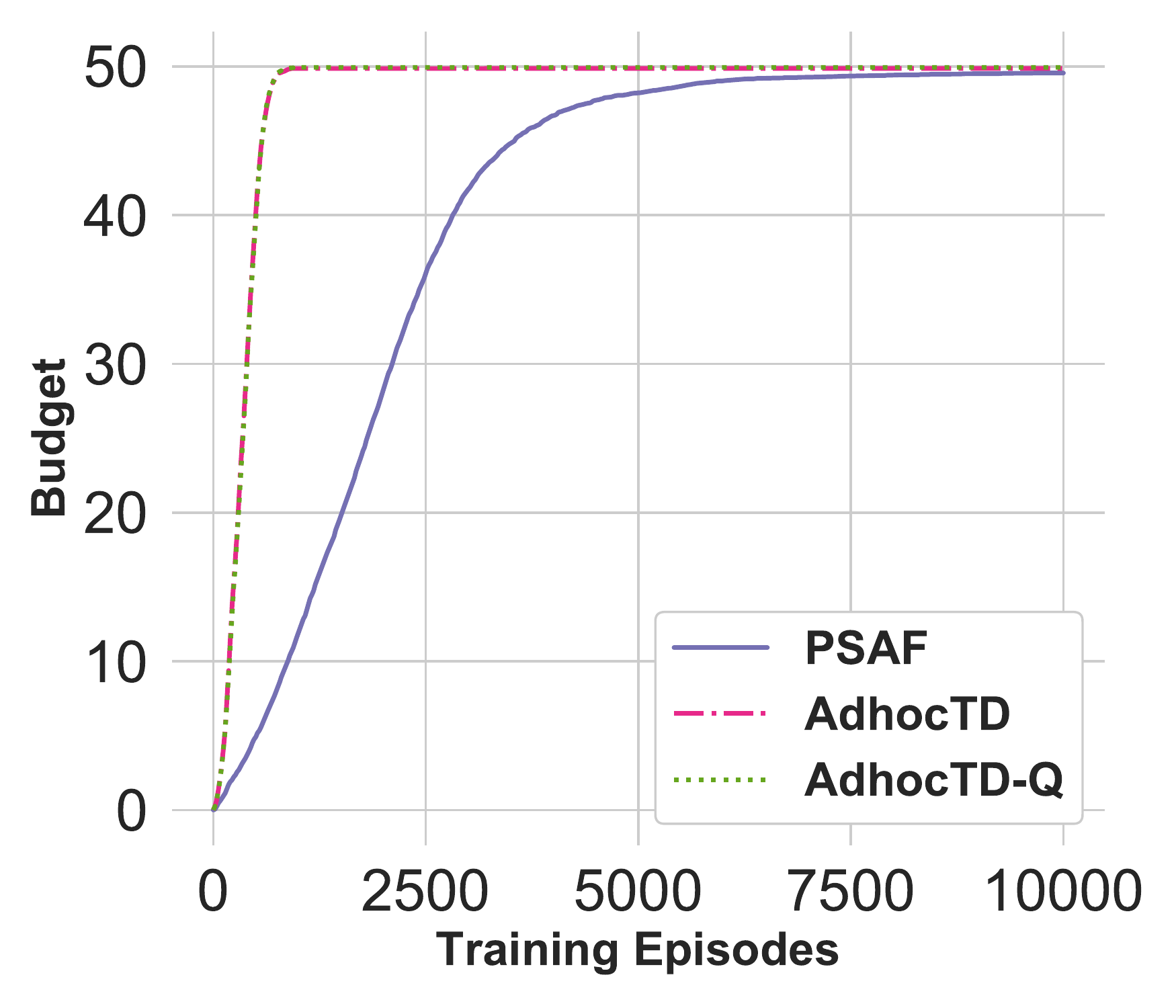}
  }
  \caption{The consumed budget $b_{give}$ of PSAF, AdhocTD and AdhocTD-Q in HFO for cases 1, 2 and 3.}
\end{figure}

\begin{figure}[htbp]
  \centering
  \subfigure[The visited times of states]{
    \includegraphics[width=2.5in, height=2in]{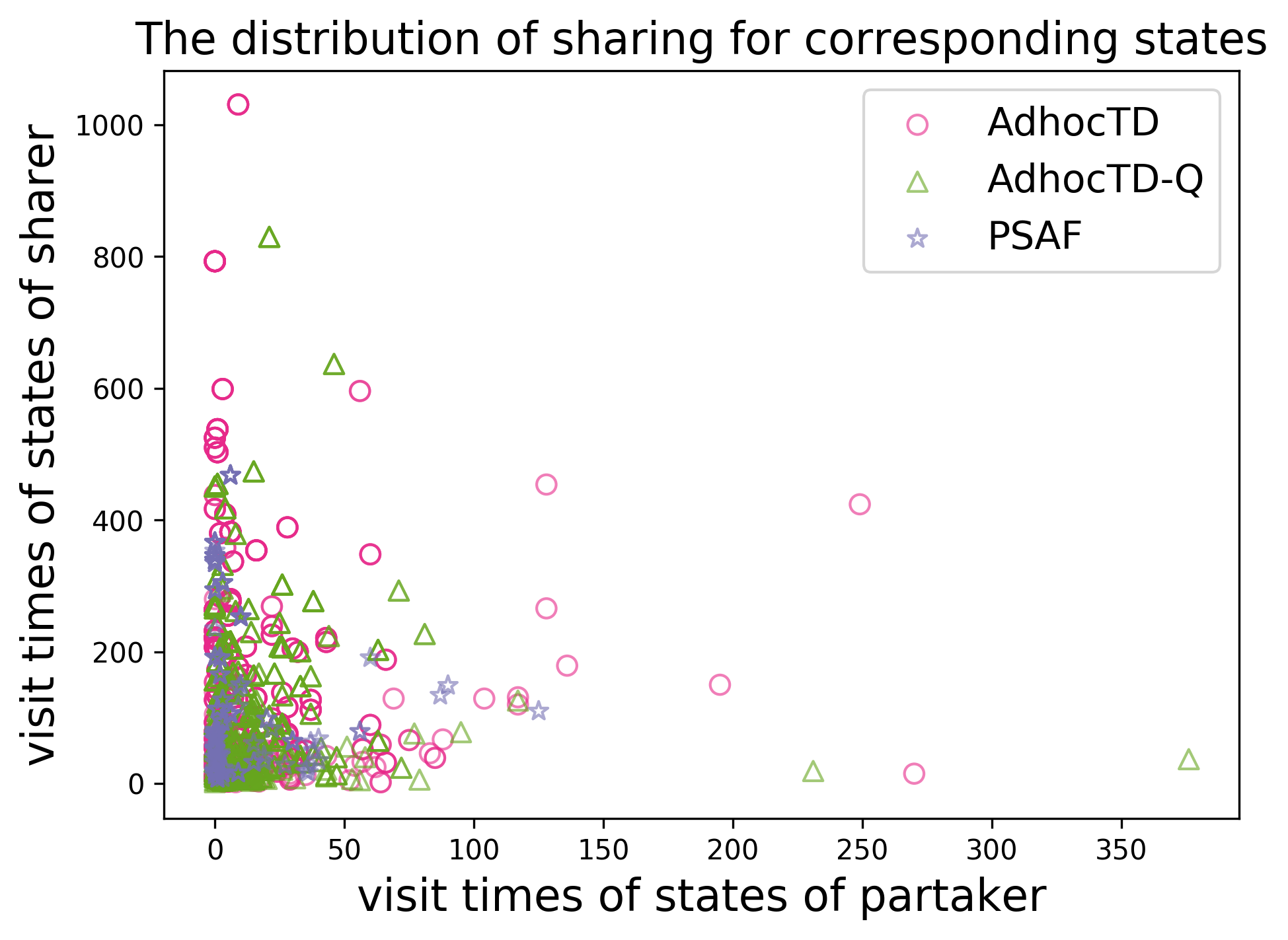}
  }
  \subfigure[The visited times of state-action pairs]{
    \includegraphics[width=2.5in, height=2in]{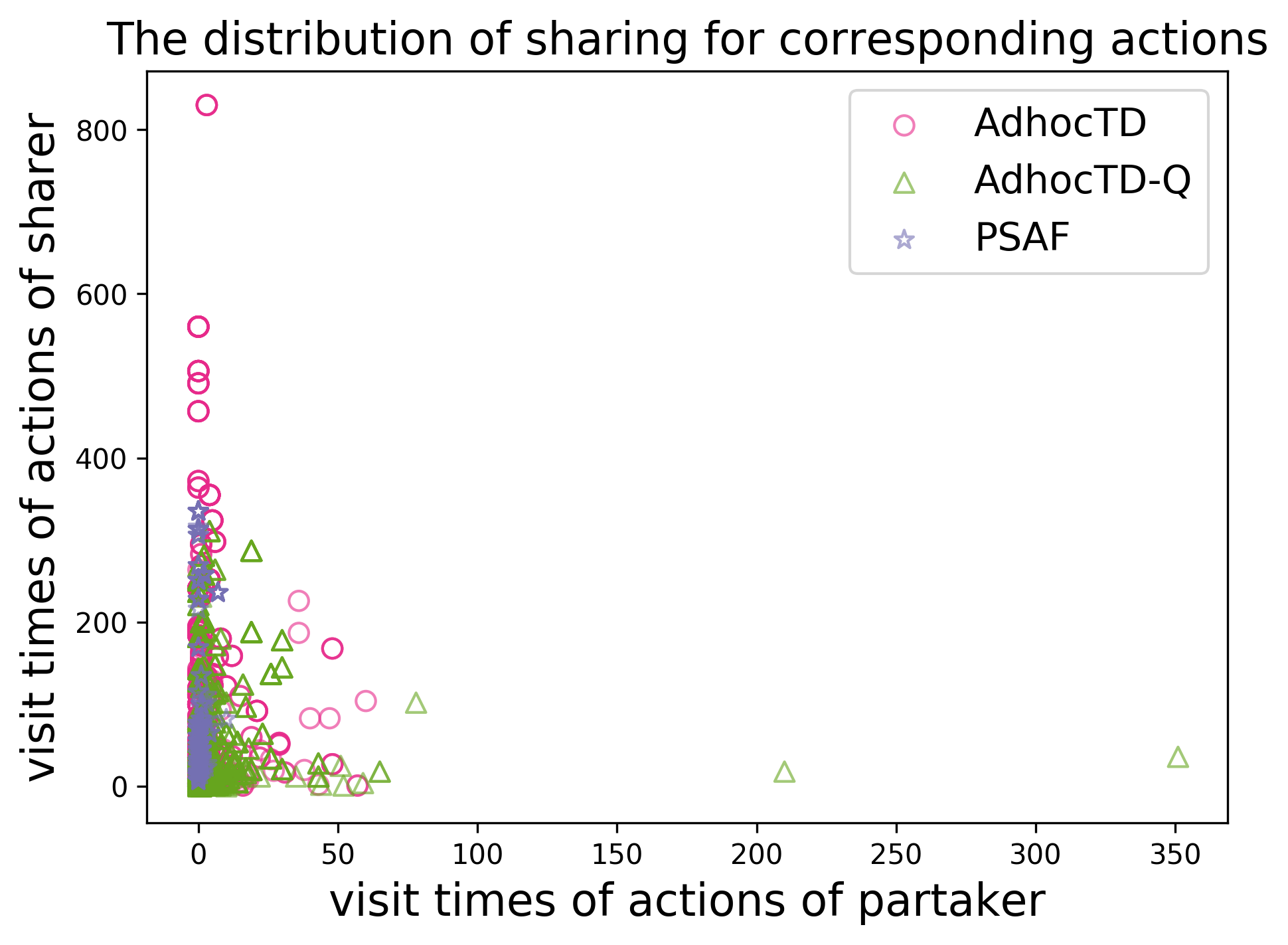}
  }
  \caption{The number of times that a partaker (student) and a sharer (teacher) visit the advised state (or state-action pair) for AdhocTD, AdhocTD-Q and PSAF in HFO.}
\end{figure}

The distribution of sharing opportunities in HFO is very different to the PP domain. As shown in Figure 9, we record the visited times of advised states and corresponding state-action pairs for AdhocTD, AdhocTD-Q and PSAF with unlimited budget. We can see that most Q-values (or actions) are shared when a partaker (or a student) visits current state very few times and a sharer (or a teacher) visits the state many times. In Figure 9b, it is interesting to note that the teacher in AdhocTD and AdhocTD-Q can give advice when it explores corresponding state-action pairs many more times than the student. This might explain why AdhocTD-Q achieves similar results as PSAF, and AdhocTD is much better than Multi-IQL when the budget is unlimited in HFO, while not in the PP domain. Nevertheless, PSAF consumes much less budget than other methods since sharing opportunities in PSAF concentrate around the states (or state-action pairs) that a partaker rarely visits.

\subsection{Spread Game}
\begin{figure}[htbp]
  \centering
  \subfigure[Spread game]{
    \includegraphics[width=1.4in, height=1.4in]{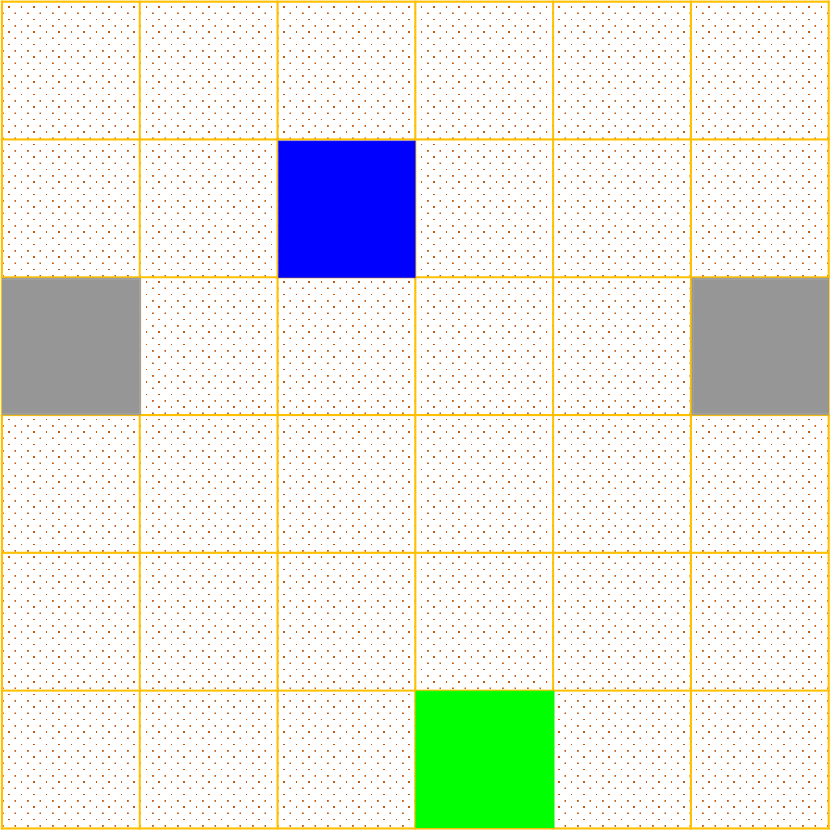}
  }
  \subfigure[Two agents cover two landmarks]{
    \includegraphics[width=1.4in, height=1.4in]{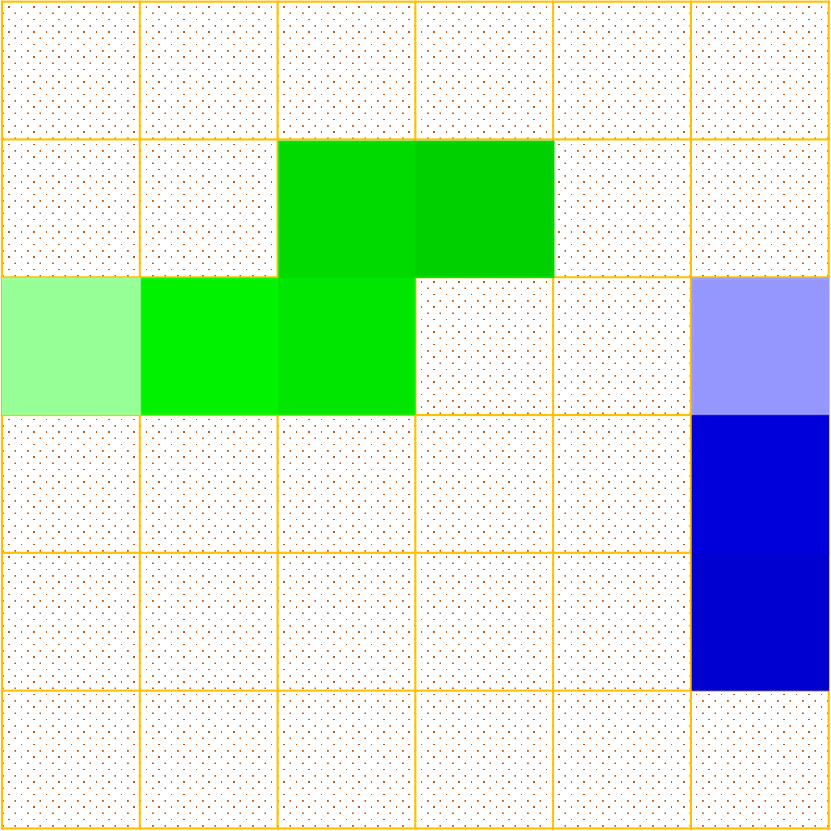}
  }
  \caption{Spread game with two agents and two landmarks. The grey one is a landmark, and the blue (or green) one is an agent.}
\end{figure}

Spread is a grid world game with discrete space and time, in which cooperative agents must spread and cover a number of landmarks, in order to maximise their shared rewards. Despite this environment's representational and mechanical simplicity, it still is capable of presenting complex behavioural challenges for MARL, for example, agents need to cover landmarks without overlap. We conduct experiments in spread domain to focus on behavioural learning while keeping the computational expense at minimum. Our implementation is based on \cite{Ilhan2019TeachingOA}.\footnote{The source code is available at https://github.com/chauncyzhu/spreadgame.} As shown in Figure 10, we consider that the grid world is sized $6 \times 6$ and consists of 2 agents and 2 landmarks. The positions of the landmarks are fixed. In this game, the agents can overlap, and ignore each other when occupying the same cell, while finally they should learn to cover the landmarks as many as possible. The agents navigate around the grid world by using a discrete set of actions \emph{Stay}, \emph{Move Up}, \emph{Move Down}, \emph{Move Left}, and \emph{Move Right}. Movement actions executed by an agent change its position on the grid world by 1 cell in the corresponding direction. The agents learn to cover all landmarks as quickly as possible. At every time step, each agent observes the relative positions of the other agent and landmarks in form of \emph{x}-axis and \emph{y}-axis values. All values of states are normalised to $[-1,1]$ by dividing by the size of the grid world. One episode begins when the agents and landmarks are initialized with random positions in the grid world. The episode ends when agents cover all landmarks or the maximum time step 20 of each episode is reached. When two agents cover the two landmarks, they receive a common reward 1. When only one agent cover one landmark, both of the two agents receive reward 0. When no landmark is covered, then the agents receive punishment -1. In order to act optimally and maximise the total reward, the agents must move to the appropriate landmark following the shortest path.

\begin{figure}[htbp]
  \centering
  \subfigure[ARS with $b_{ask}=b_{give}=+\infty$]{
    \includegraphics[width=2.5in, height=2.1in]{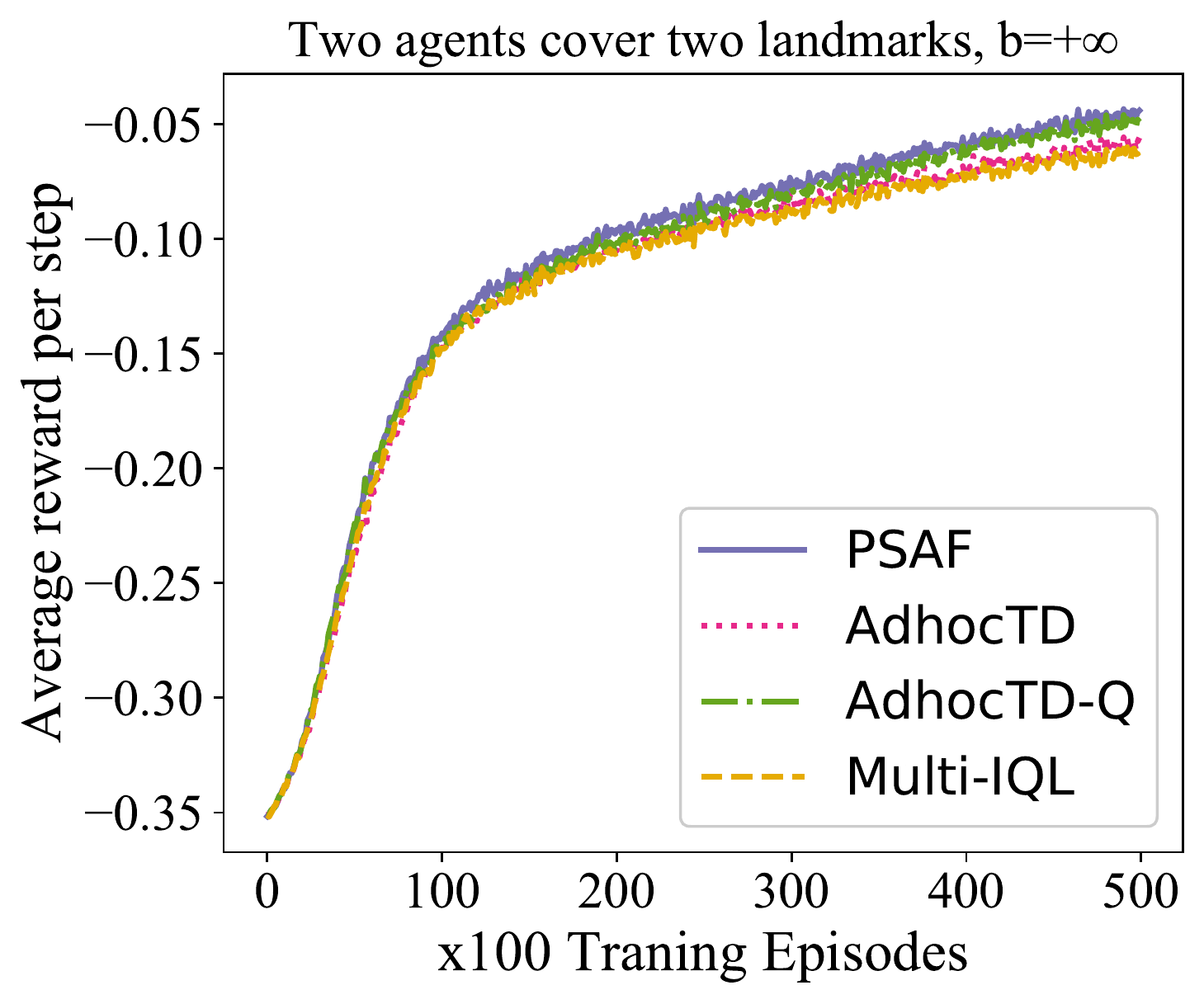}
  }
  \subfigure[AUC of ARS with $b_{ask}=b_{give}=+\infty$]{
    \includegraphics[width=2.5in, height=2.1in]{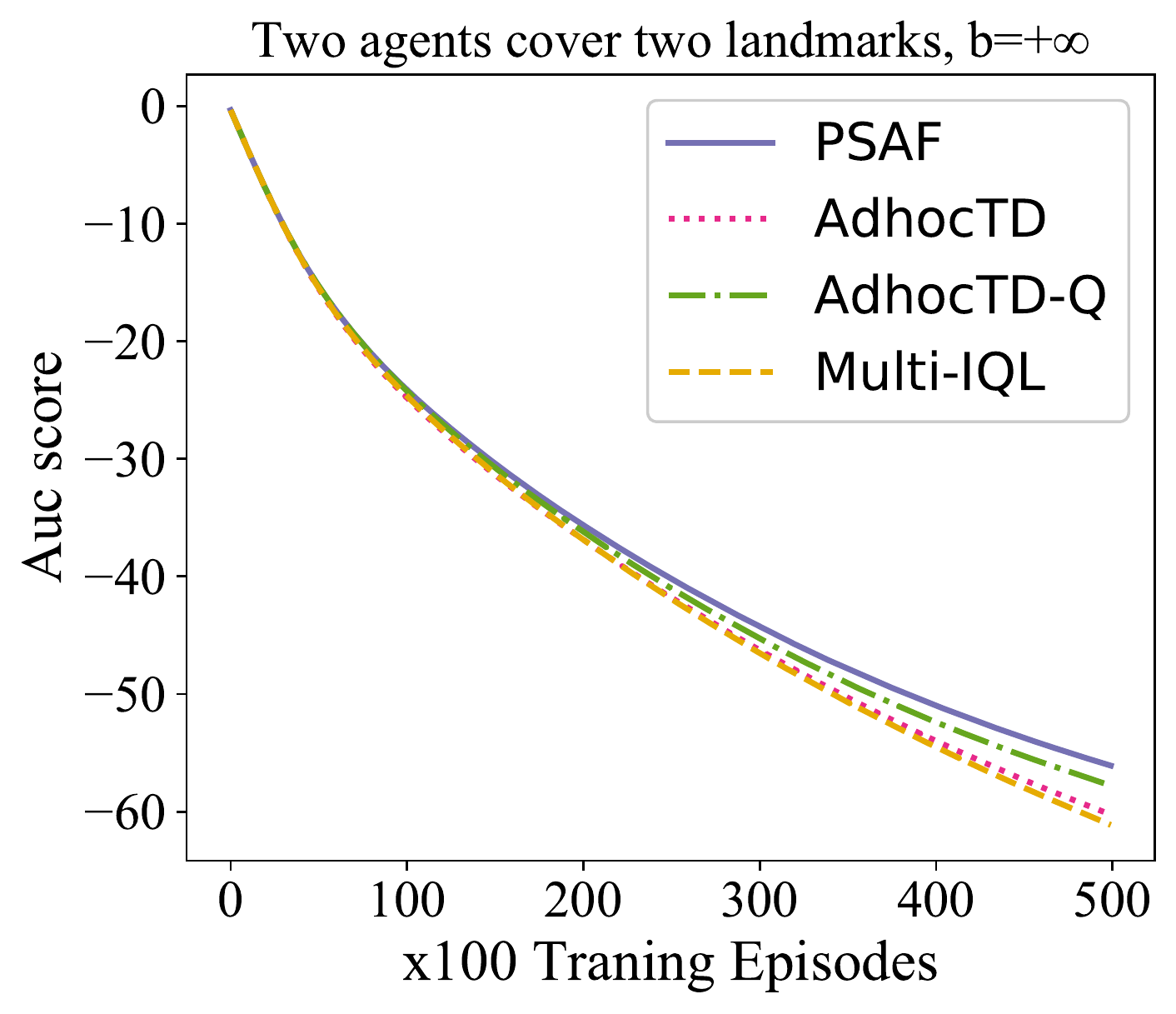}
  }

  \subfigure[ARS with $b_{ask}=b_{give}$=3,000]{
    \includegraphics[width=2.5in, height=2.1in]{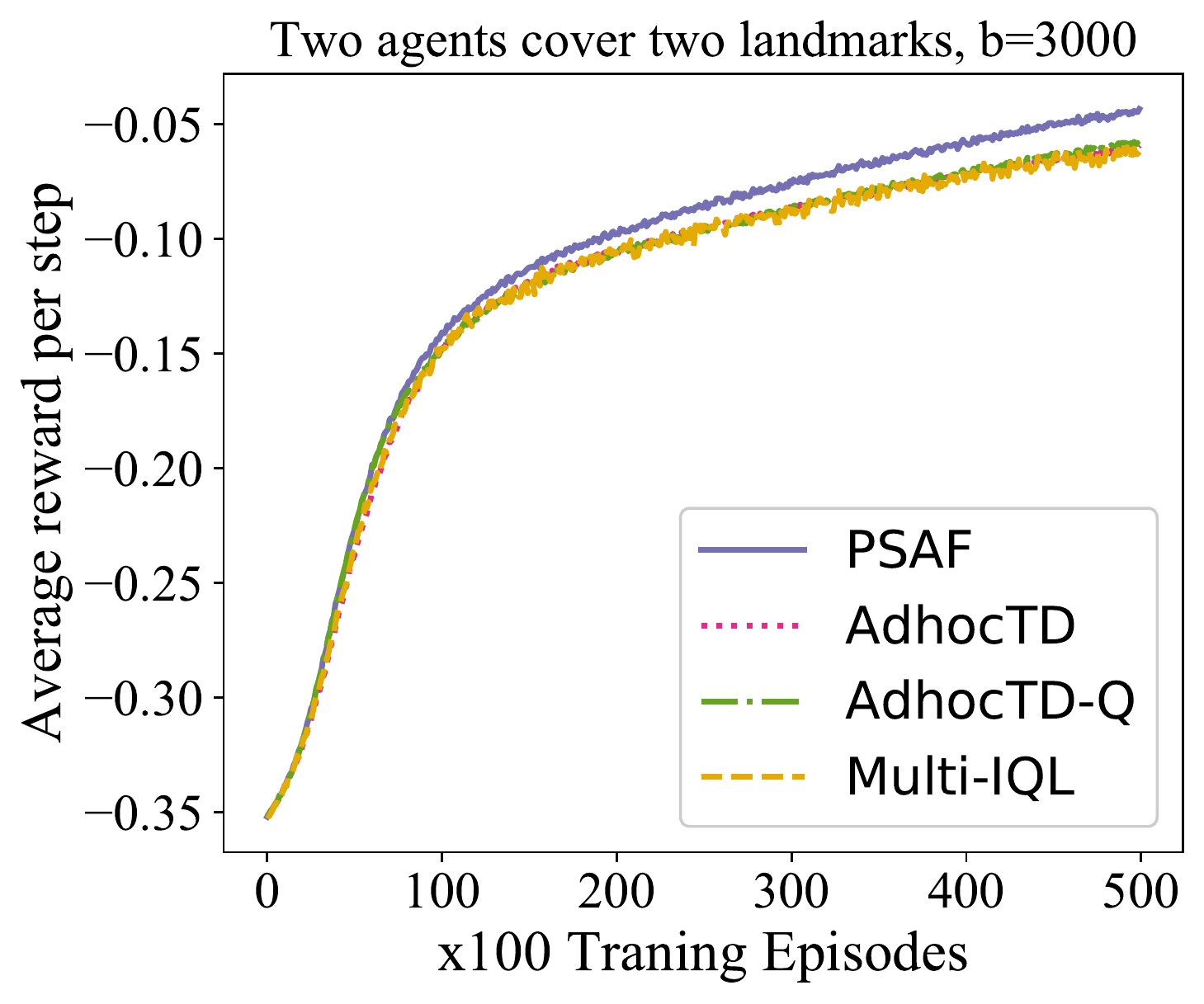}
  }
  \subfigure[AUC of ARS with $b_{ask}=b_{give}$=3,000]{
    \includegraphics[width=2.5in, height=2.1in]{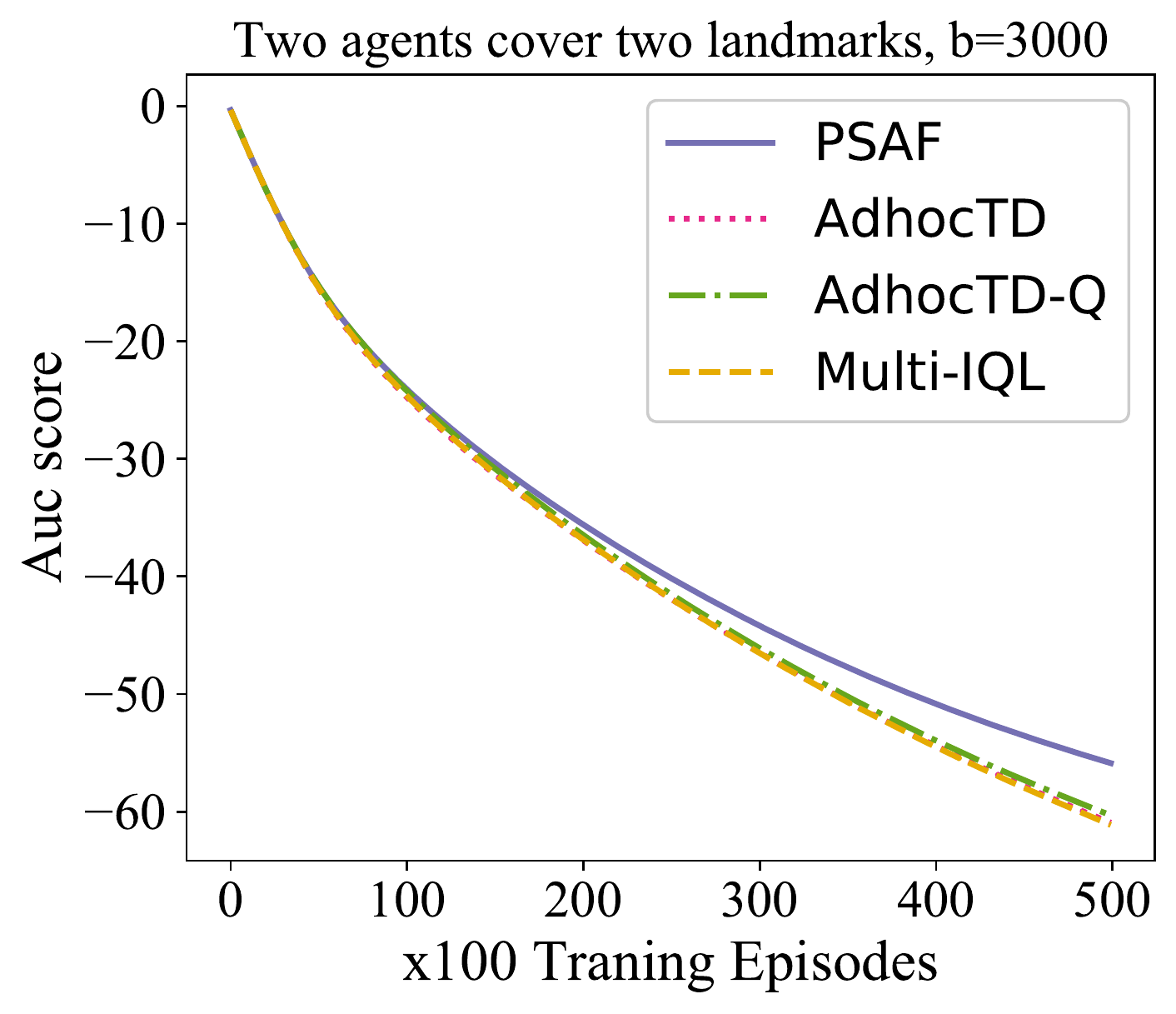}
  }

  \subfigure[AUC of ARS with $b_{ask}=b_{give}=$1,500]{
    \includegraphics[width=2.5in, height=2.1in]{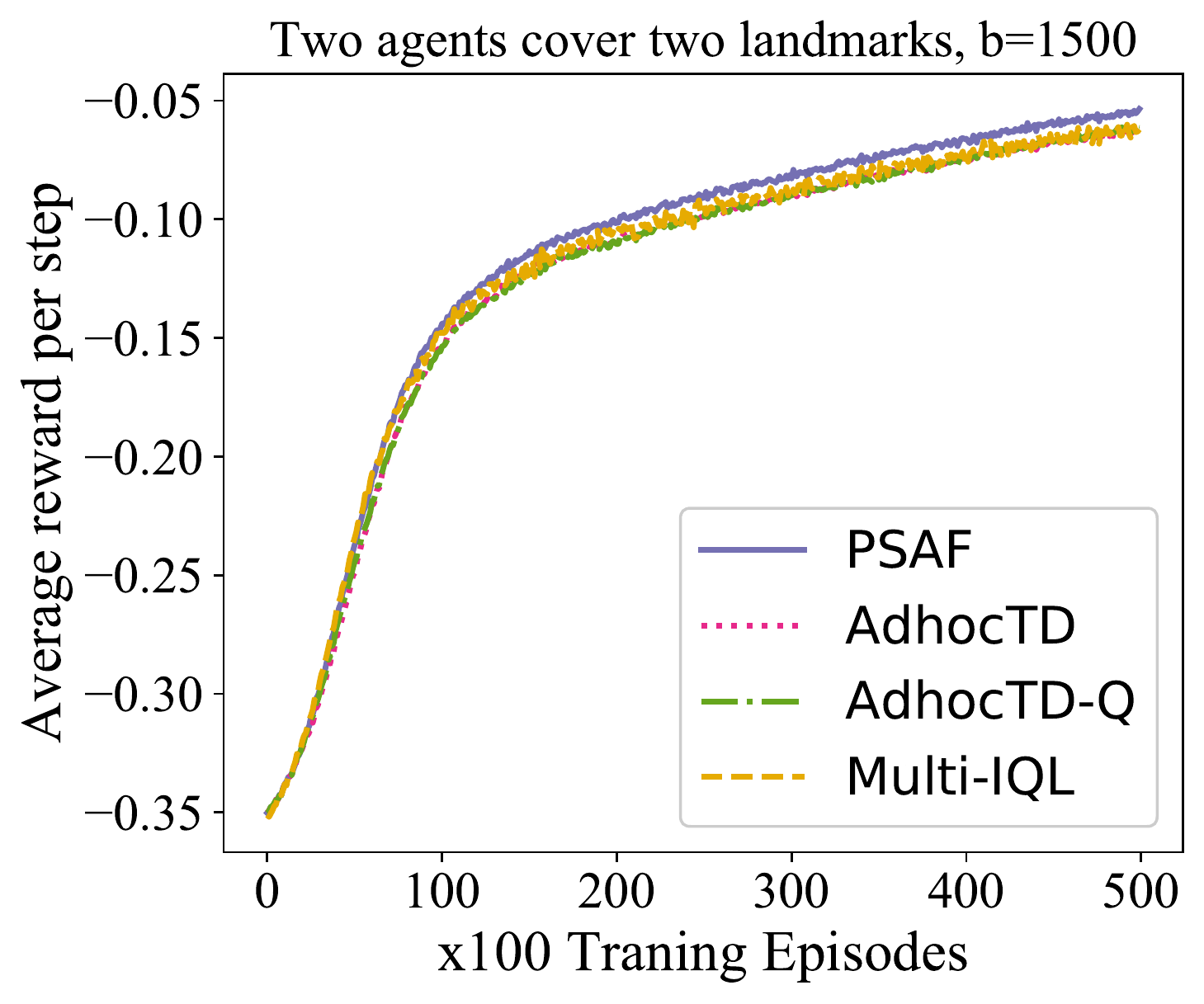}
  }
  \subfigure[AUC of ARS with $b_{ask}=b_{give}=$1,500]{
    \includegraphics[width=2.5in, height=2.1in]{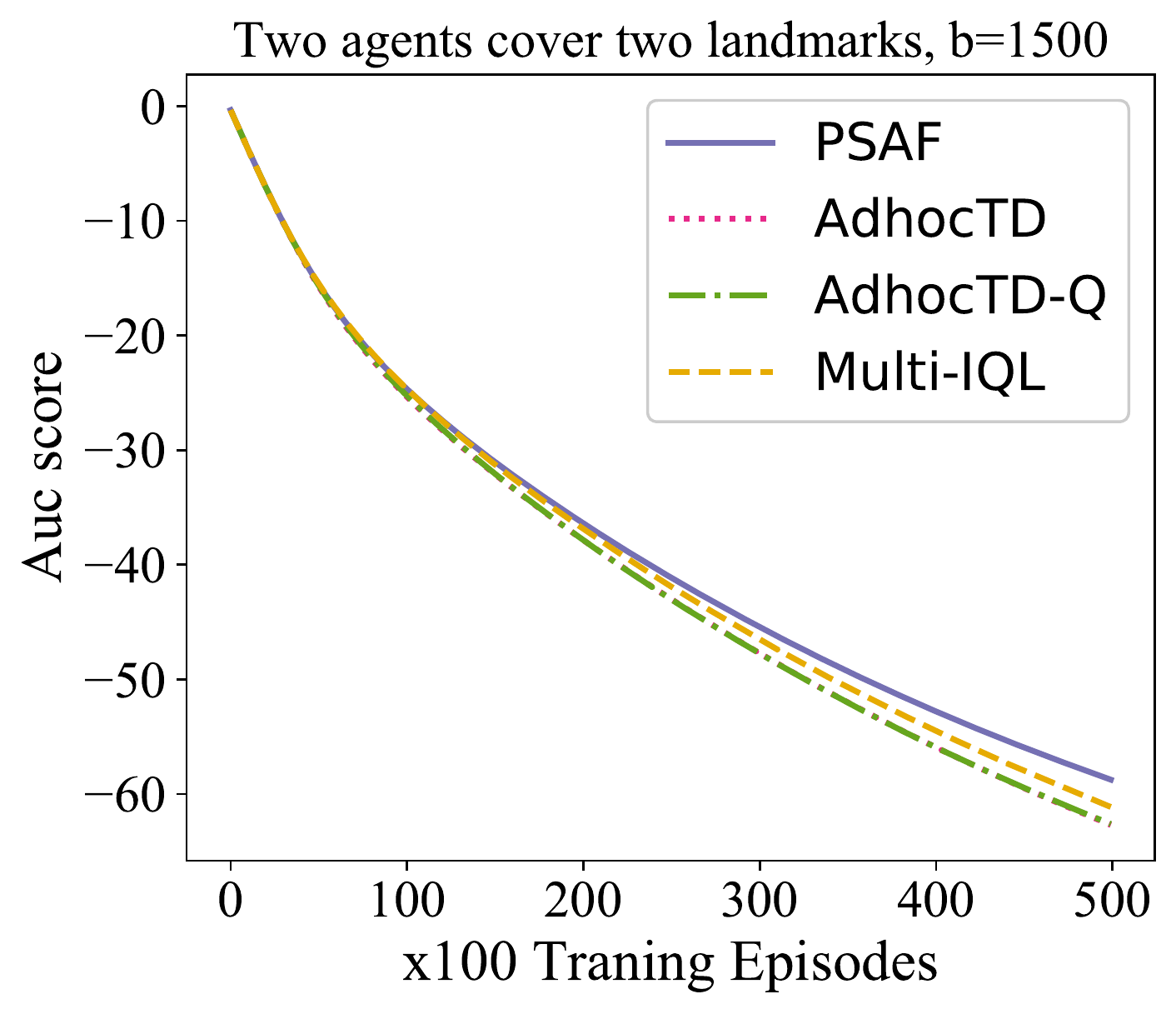}
  }
  \caption{ARS and AUC value of PSAF, AdhocTD, AdhocTD-Q and Multi-IQL in Spread for cases 1, 2 and 3.}
\end{figure}

\begin{figure}[htbp]
  \centering
  \subfigure[Unlimited budget]{
    \includegraphics[width=2.5in, height=2in]{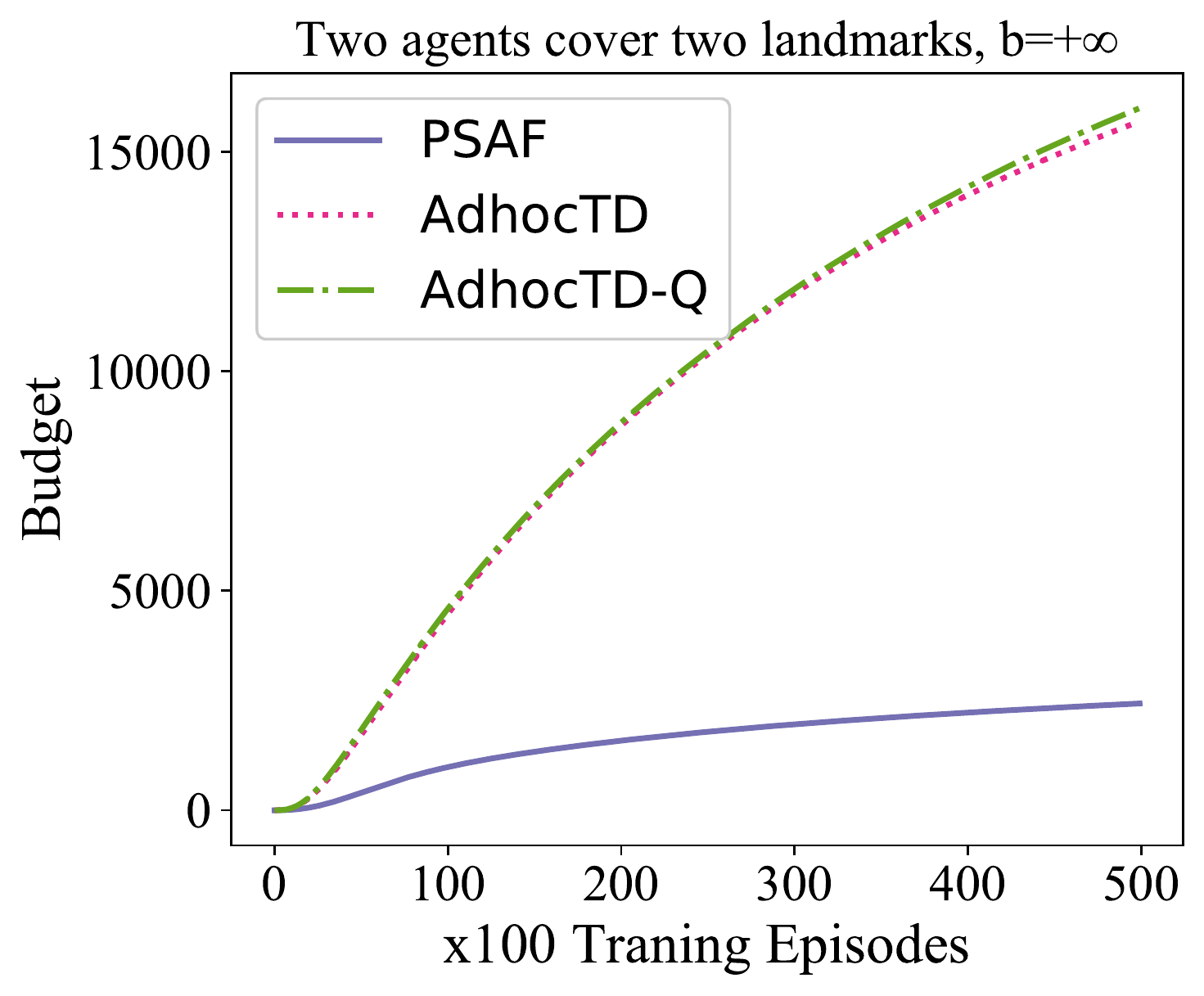}
  }
  \subfigure[$b_{ask}=b_{give}$=3,000]{
    \includegraphics[width=2.5in, height=2in]{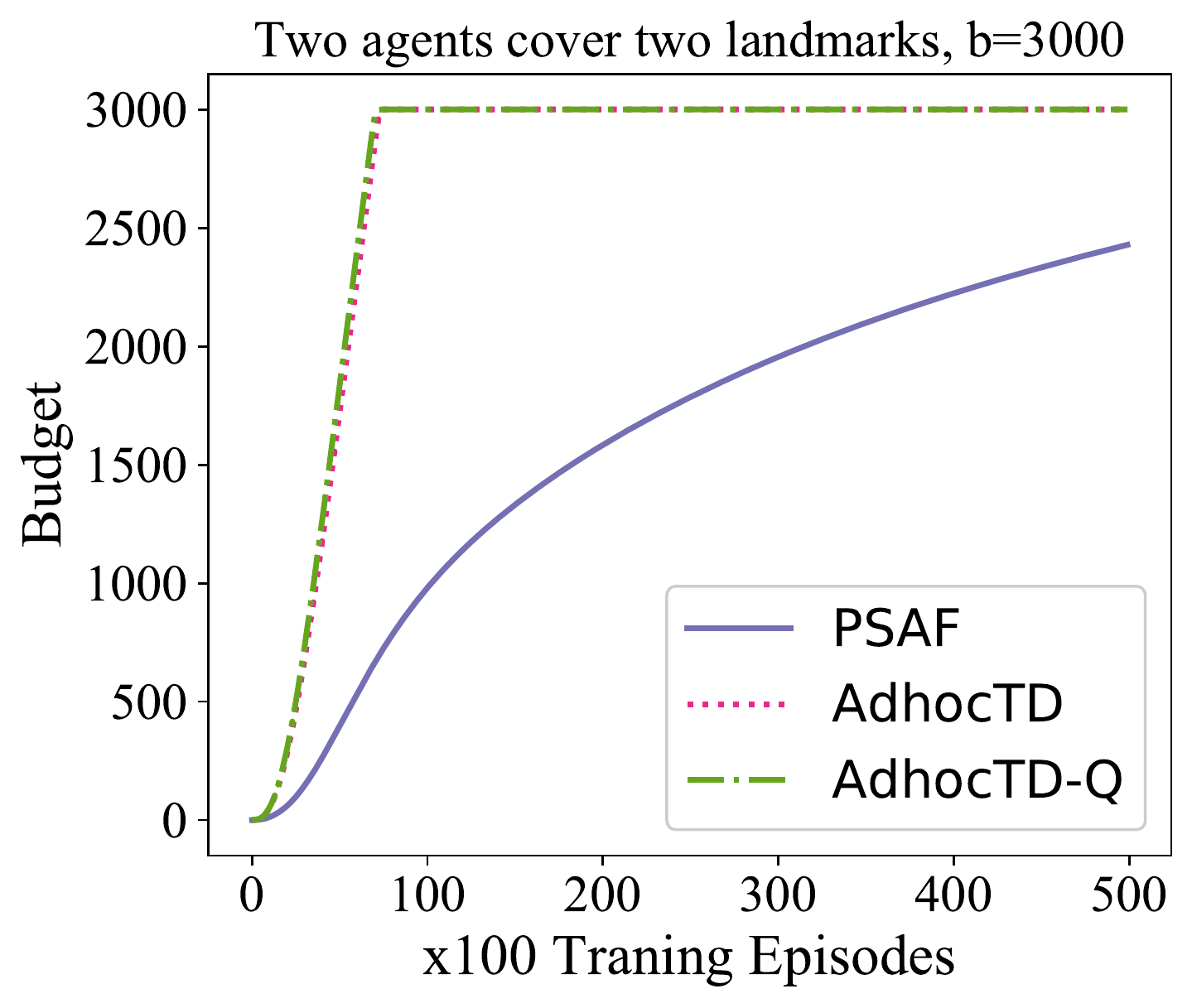}
  }

    \subfigure[$b_{ask}=b_{give}$=1,500]{
    \includegraphics[width=2.5in, height=2in]{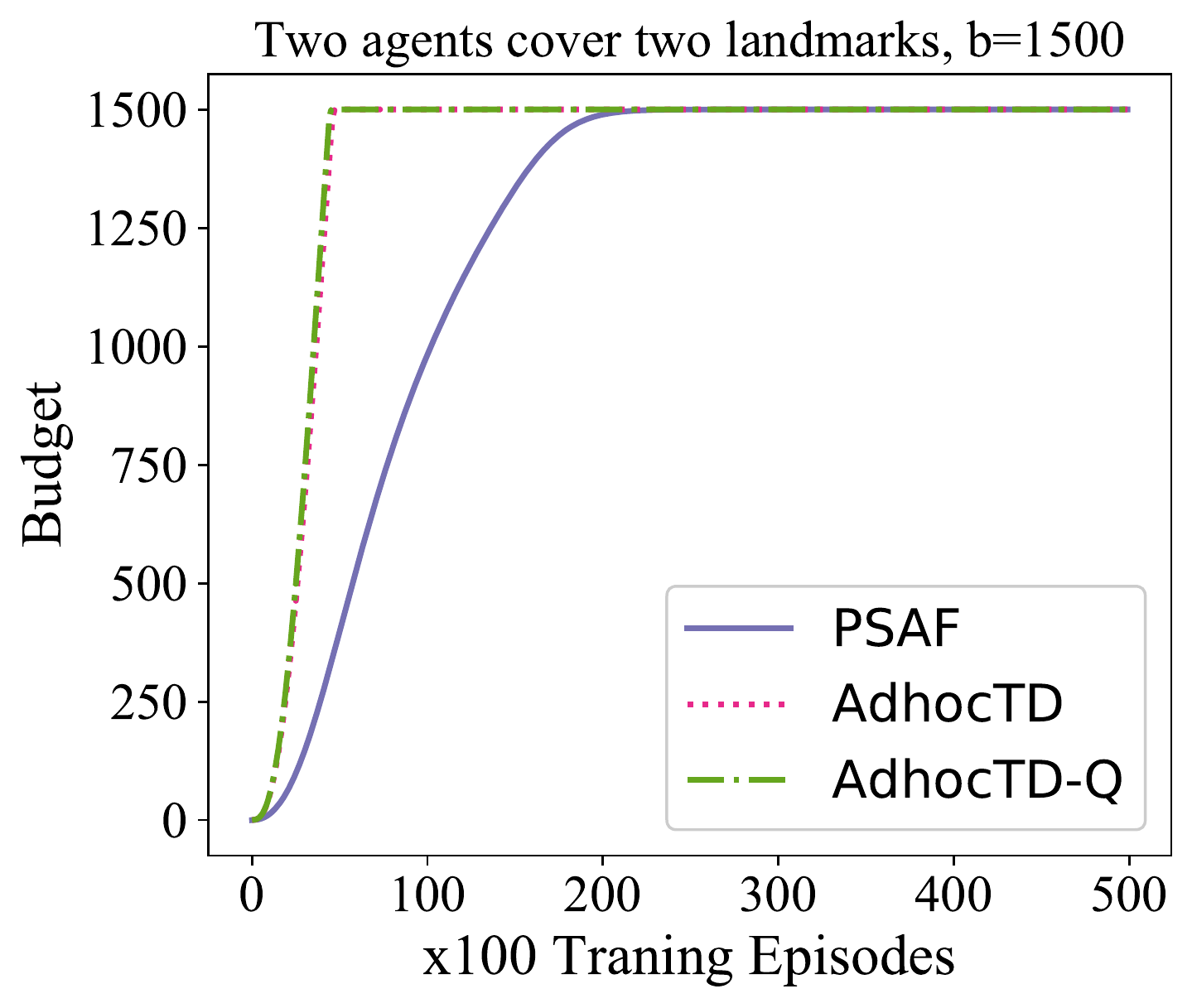}
  }
    \caption{The used budget of PSAF, AdhocTD and AdhocTD-Q in Spread for cases 1, 2 and 3.}
\end{figure}

\begin{figure}[htbp]
  \centering
  \subfigure[The visited times of states]{
    \includegraphics[width=2.5in, height=2in]{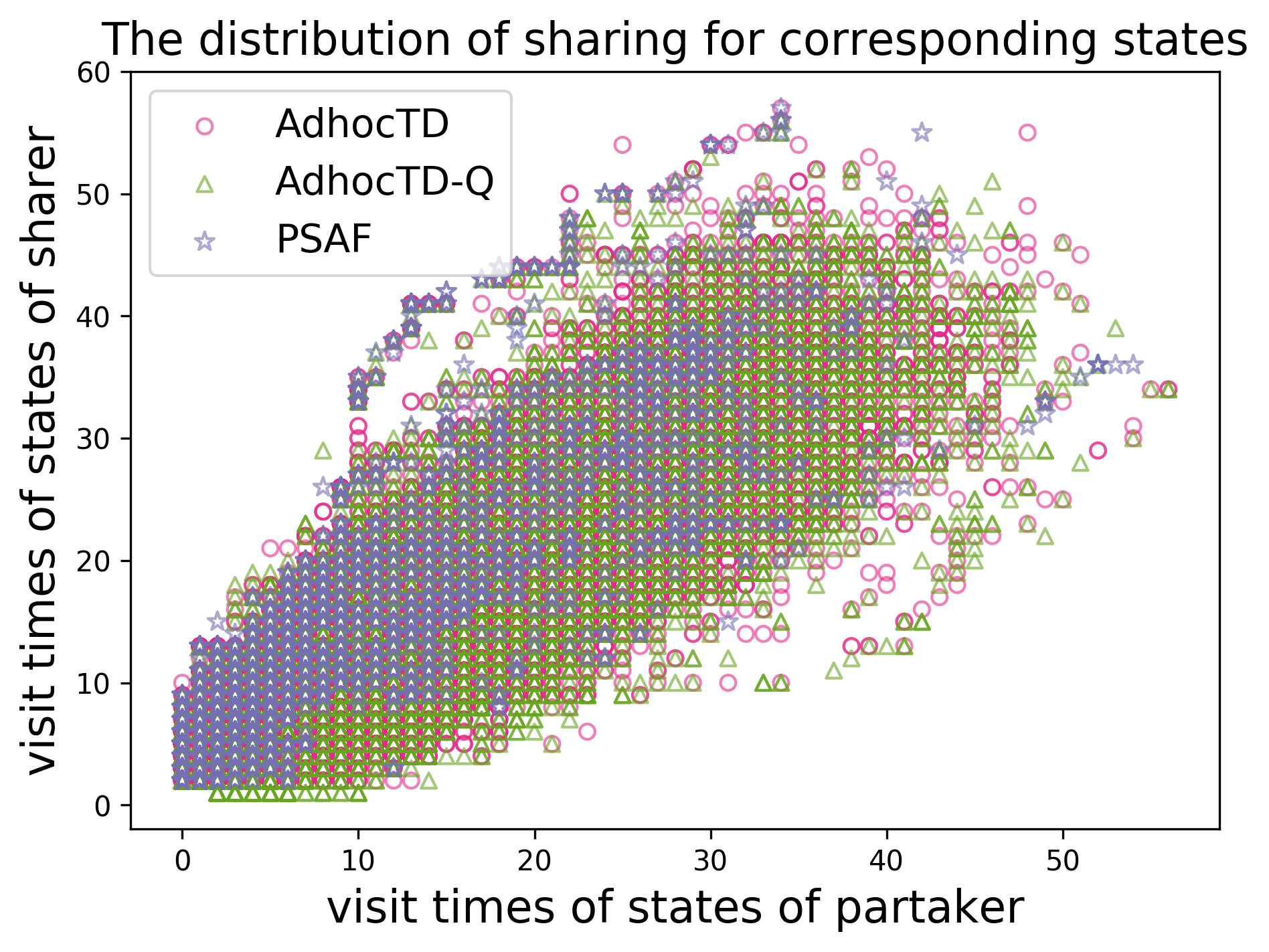}
  }
  \subfigure[The visited times of state-action pairs]{
    \includegraphics[width=2.5in, height=2in]{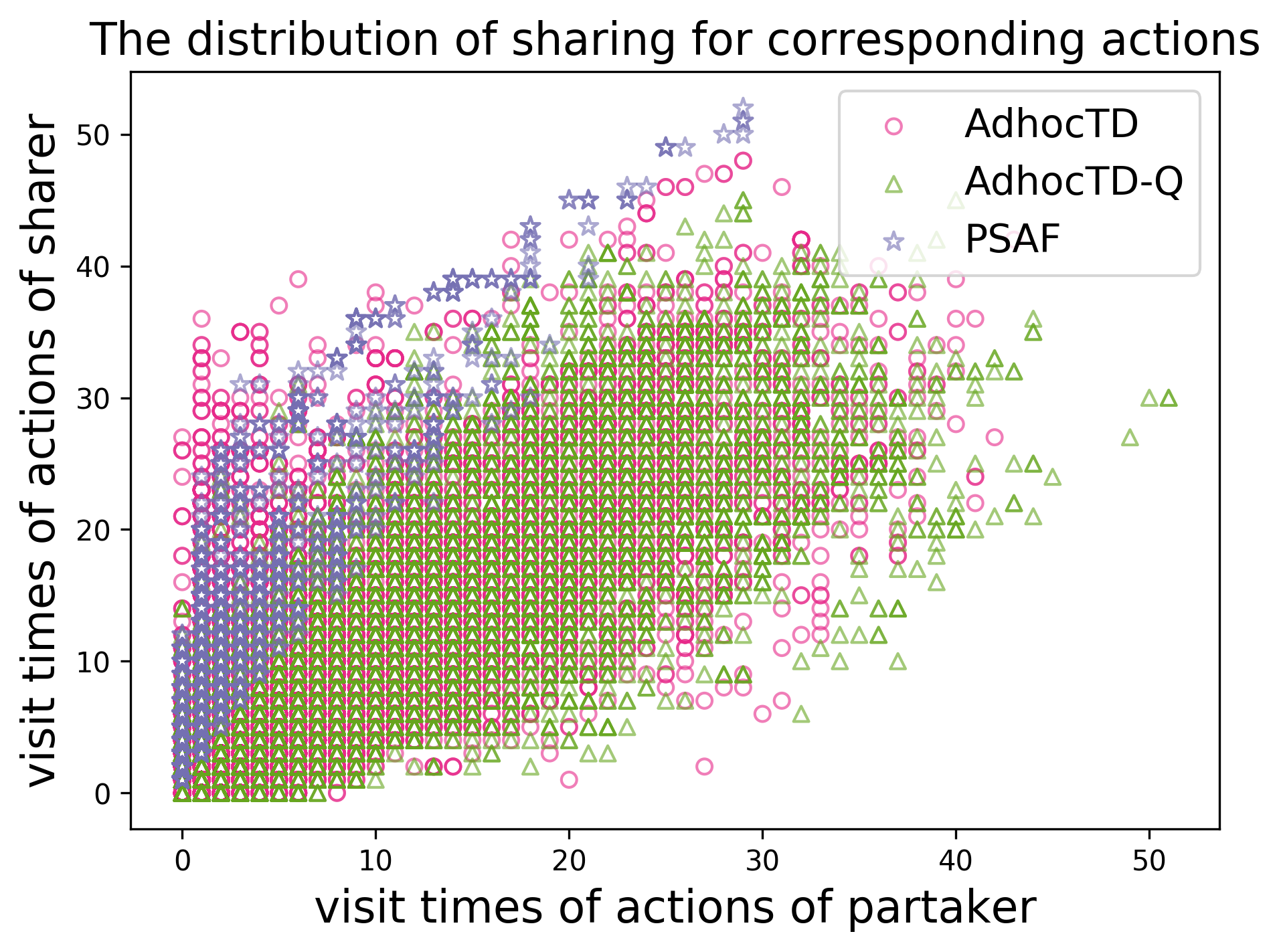}
  }
  \caption{The number of times that a partaker (student) and a sharer (teacher) visit the advised state (or state-action pair) for AdhocTD, AdhocTD-Q and PSAF in Spread.}
\end{figure}

The performance of the agents is assessed by another classical evaluation metric \emph{Average Reward per Step} (ARS). ARS is an agent's average reward of each time step in an episode. The two agents are trained for 50,000 episodes. At every 100 training episodes, a set of 100 evaluation episodes is performed. During evaluation, learning and sharing process are disabled for agents. They only use currently learned optimal policy. The reward received by an agent at every time step is discounted by $\gamma$=0.9, and averaged for one evaluation episode. Then we obtain ARS for the agent by averaging the discounted rewards in the 100 evaluation episodes. The overall process is repeated 1,000 times to smooth the curve of performance. When no Q-value (or action) is shared to agents during learning, they use SARSA($\lambda$) with $\lambda$=0.9, $\alpha$=0.1, and $\gamma$=0.9. In this paper, we use $v_a=$1 and $v_b$=0.5 for AdhocTD, AdhocTD-Q and PSAF. Each agent uses $\epsilon$-greedy strategy as exploration strategy with $\epsilon$=0.1. Similar to the PP domain and HFO, three cases are tested for AdhocTD, AdhocTD-Q, PSAF and Multi-IQL: (1)  $b_{ask}=b_{give}=+\infty$; (2) $b_{ask}$=$b_{give}$=3,000; (3)  $b_{ask}=b_{give}=$1,500. The results of all methods are shown in Figures 11 and 12.

We can see that PSAF is significantly better than other methods both in ARS and AUC socre for three cases. When the budget is unlimited, AdhocTD-Q has much higher ARS and AUC value than AdhocTD and Multi-IQL, and the difference between AdhocTD and Multi-IQL is statistically significant regarding ARS. When the budget is limited to 3,000, however, AdhocTD achieves very similar ARS and AUC value to Multi-IQL, while AdhocTD-Q is slightly better than AdhocTD and Multi-IQL. This could further illustrate the amount of available budget has great impact on the performance of advising actions and sharing Q-values based on AdhocTD. When PSAF runs out of  all budget in case 3, we can see that the performance of PSAF declines compared to case 1 and 2, despite the fact that PSAF still achieves better performance than other methods. Moreover, AdhocTD and AdhocTD-Q perform significantly worse than Multi-IQL, which probably means that learning without proper advice or (shared) Q-values even interferes agents' learning. When the communication among agents becomes expensive, PSAF is more appropriate than other methods, since in Spread game, each agent equipped with PSAF only needs to share about 2,500 Q-values in 50,000 training episodes.

Spread game has 1,230 states (roughly). Similar to the PP domain and HFO, we show the distribution of sharing opportunities for AdhocTD, AdhocTD-Q and PSAF with unlimited budget. In Figure 13, we can see that most sharing relations in PSAF are established when a Q-value of a partaker is rarely updated while a sharer has updated many times. As a partaker repeatedly updates a Q-value, a sharer is required to update the Q-value many more times than the partaker.

\section{DISCUSSION}
In this paper, we emphasize the importance of the amount of budget on agents' learning. If the available budget is high, both advising actions and sharing Q-values can benefit from the more experienced teacher since agents have enough opportunities to be guided. When the budget decreases, agents may stop to ask for action advice or Q-values at the early stage of learning. We observe that in the latter case, sharing Q-values still achieve better performance on the evaluated metrics. Using the Q-values from teachers enables the student to choose a better action when it has no budget to ask for help. However, if the budget is even lower, which means that all budgets can be used up at the very beginning, the learning performance will be dropped as the action advice or Q-values are likely to be non-optimal.

\begin{figure}[htbp]
  \centering
  \subfigure[]{
    \includegraphics[width=1.4in, height=1.4in]{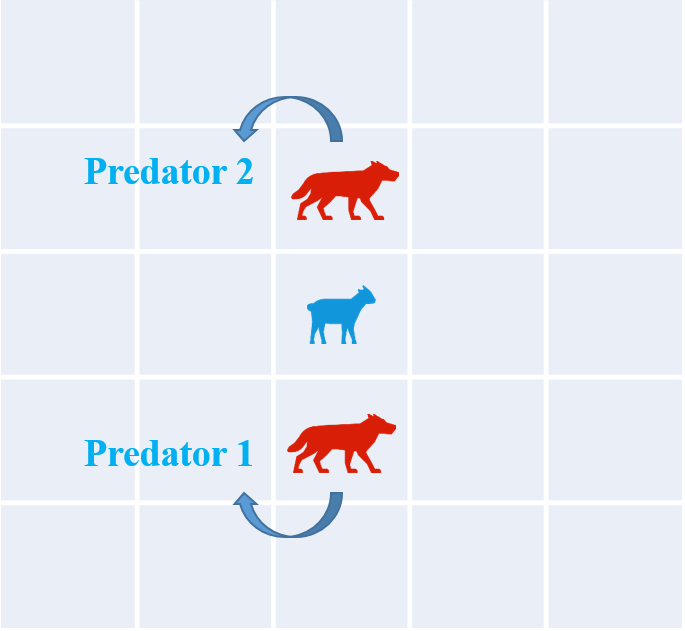}
  }
  \subfigure[]{
    \includegraphics[width=1.4in, height=1.4in]{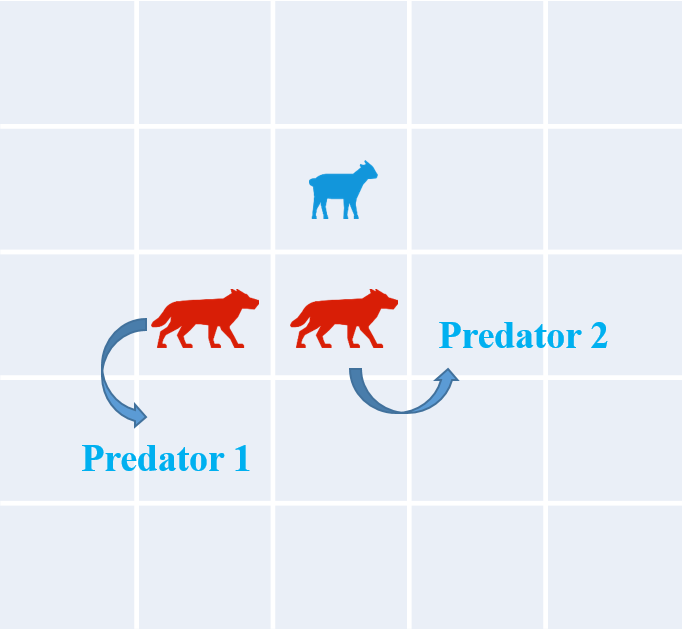}
  }
  \caption{Two examples of the Predator-Prey domain. Left: the state of predator 1 is $\langle 0, 2, 0, 1 \rangle$. Right: the state of predator 1 is $\langle -1, 0, -1, 1 \rangle$}
\end{figure}

We try to understand why sharing Q-values outperforms advising actions and how sharing processes influence agents' learning, which is unexplored in previous works. In this section, we use the PP domain with two predators and one prey, considering that it is quite easy to record and reconstruct the learning process compared to HFO. Now the prey is caught when one predator occupies the same cell as the prey, and another predator is next to the prey in four cardinal directions. This setting makes the predators have more possibilities to achieve the goal. Then we choose one typical run from the PP domain and select two cases where if the predators take right actions, they can successfully catch the prey, as shown in Figures 14a and 14b. The states of predator 1 in two cases are $\langle 0, 2, 0, 1 \rangle$ and $\langle -1, 0, -1, 1 \rangle$ respectively. In Figure 15a, we record the Q-values of predator 1 with AdhocTD in state $\langle 0, 2, 0, 1 \rangle$ after the predator receiving its first action advice at 4,213 step of 1,090 episode. Similarly, the Q-values of predator 1 with PSAF since it firstly receives a Q-value from predator 2 at 1,550 step of 1,216 episode are shown in Figure 15b. We can see that predator 2 always advises predator 1 to go up in state $\langle 0, 2, 0, 1 \rangle$. Intuitively, if predator 1 executes \emph{Up}, the two predators are more likely to catch the prey. After predator 1 has been advised 7 times, its Q-value of action \emph{Up} becomes larger and larger, which may be caused by the dynamic environment. However, in Figure 15b, predator 2 only provides predator 1 with one Q-value, which corresponds to action \emph{Up}. After predator 1 updates the Q-value corresponding to action \emph{Up}, the Q-value simply becomes the maximum one in state $\langle 0, 2, 0, 1 \rangle$. Then, the learning speed of predator 1 with PSAF is accelerated and the budget can be saved.

\begin{figure}[t]
  \centering
  \subfigure[Predator 1 with AdhocTD]{
    \includegraphics[width=2.5in, height=2in]{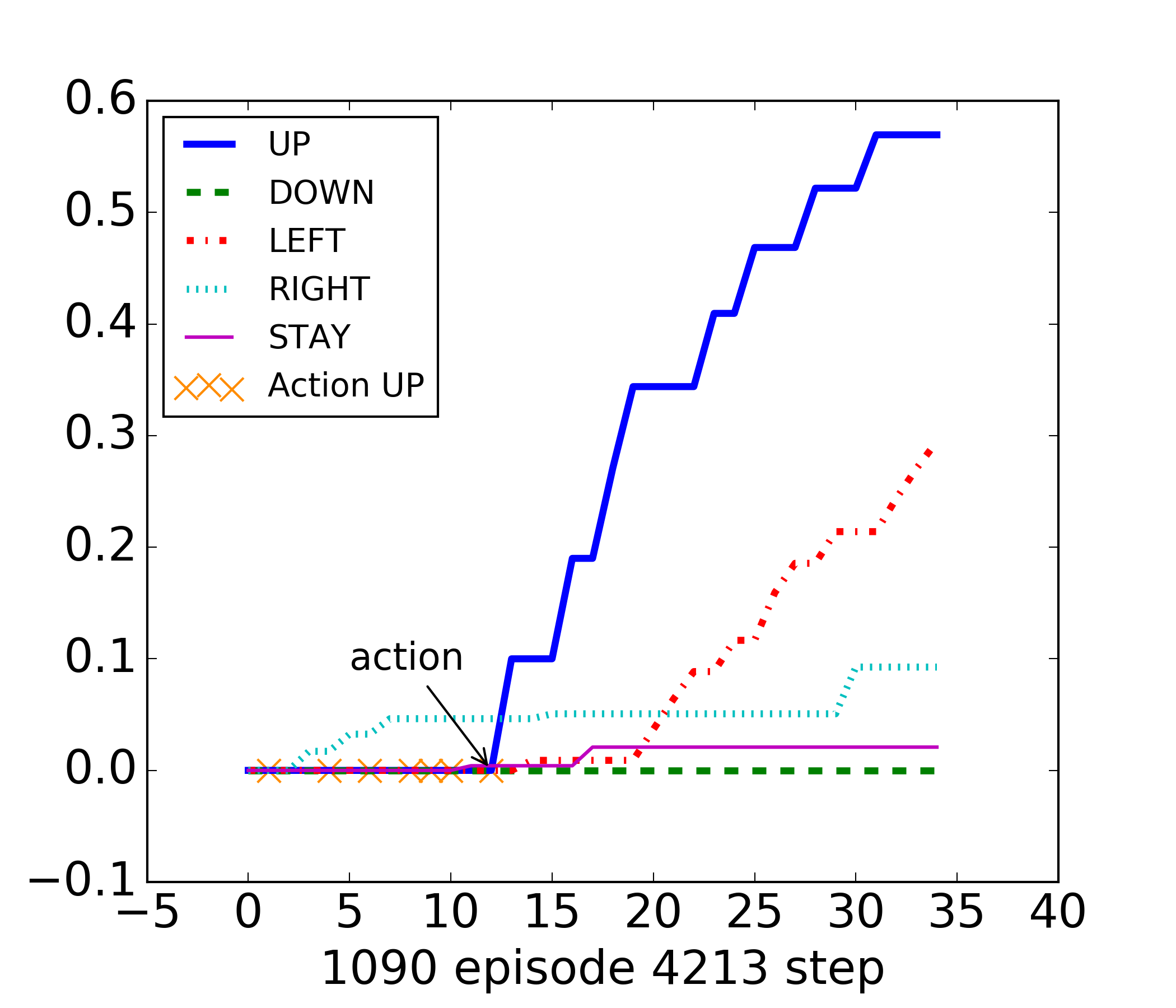}
  }
  \subfigure[Predator 1 with PSAF]{
    \includegraphics[width=2.5in, height=2in]{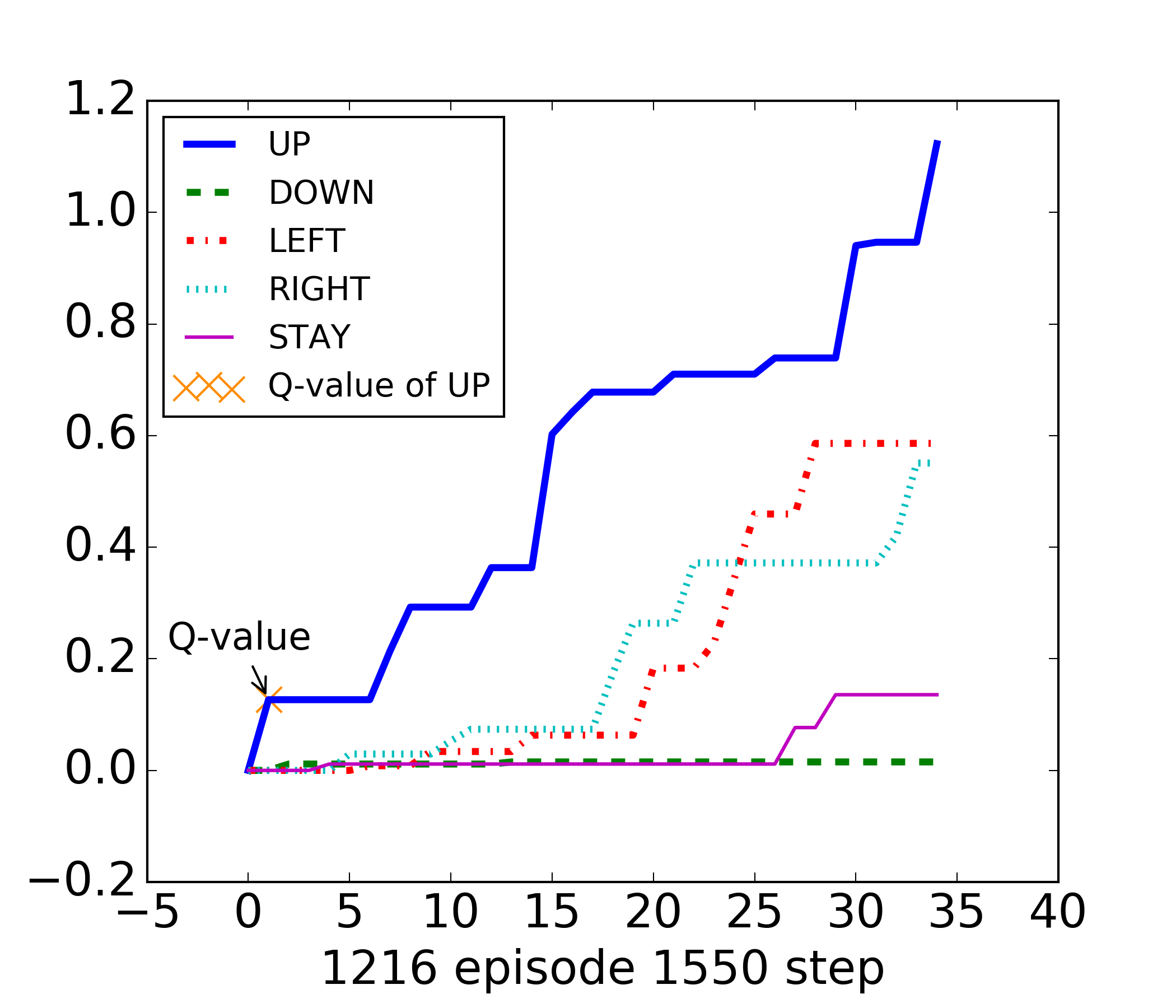}
  }
  \caption{Q-values of predator 1 with AdhocTD (a) and PSAF (b) in state $\langle 0, 2, 0, 1 \rangle$. One solid point in Figure (a) (or (b)) is an advised action (or a shared Q-value) from predator 2.}
\end{figure}

\begin{figure}[t]
  \centering
  \subfigure[Predator 1 with AdhocTD]{
    \includegraphics[width=2.5in, height=2in]{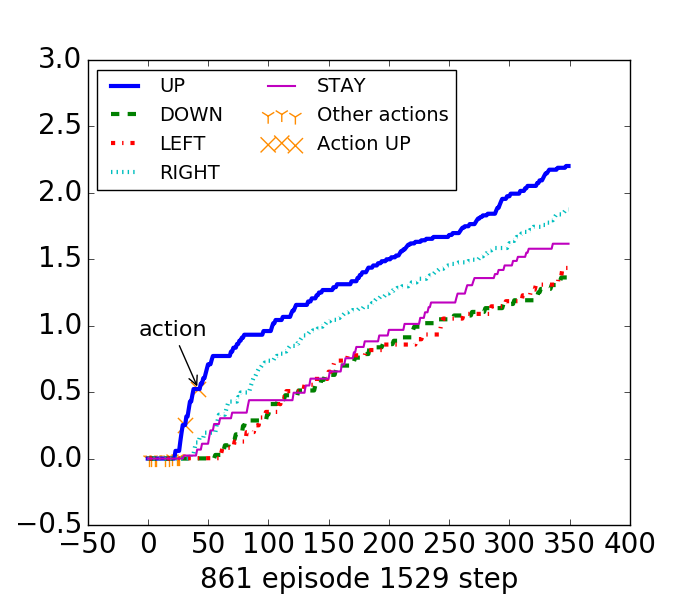}
  }
  \subfigure[Predator 1 with PSAF]{
    \includegraphics[width=2.5in, height=2in]{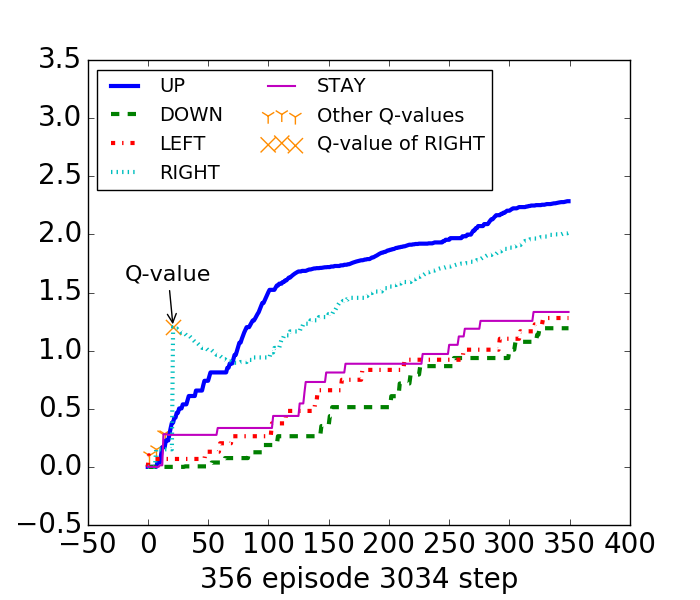}
  }
  \caption{Q-values of predator 1 with AdhocTD (a) and PSAF (b) in state $\langle -1, 0, -1, 1 \rangle$. One solid point in Figure (a) (or (b)) is an advised action (or a shared Q-value) from predator 2.}
\end{figure}

The example in state $\langle -1, 0, -1, 1 \rangle$ is shown in Figure 16. Intuitively, predator 1 should take action \emph{Up} or \emph{Right} so that the two predators are more likely to successfully catch the prey. When predator 1 receives the first action advice or shared Q-value in state $\langle -1, 0, -1, 1 \rangle$, we record the Q-values of AdhocTD and PSAF in Figure 16a and 16b respectively. In Figure 16a, we can see that predator 2 mostly advises action \emph{Left} (corresponding to its maximum Q-value) in the beginning. After that, predator 2 advises action \emph{Up} to predator 1 two times. Then the Q-value corresponding to action \emph{Up} of predator 1 becomes larger and larger. Figure 12b shows that predator 1 only receives 4 Q-values in total. At the early stage, predator 1 may receive some non-optimal Q-values, e.g., the Q-value corresponding to action \emph{Left}. However, predator 1 with PSAF can learn to proceed to the right direction finally.

Experimental results show that sharing Q-values performs better than advising actions, particularly, in the cooperative MARL with budget constraint. We highlight the necessity of cooperative setting and budget constraint due to two reasons. Firstly, in the cooperative MARL, as agents learn to complete the same goal in a shared environment, they are more likely to exchange their knowledge without specific requirements for transforming internal representations, e.g., states, actions and value functions. The second reason is that if the communication is unconstrained, agents can exchange their learned knowledge at every time step. Then, the multiagent learning would be reduced to the single-agent learning, which is less effective at scaling to multiagent systems with a great number of agents, and decreases computational speed. Generally, sharing Q-values is more effective when the Q-values are mainly shared in states where a partaker explores very few times, while a sharer has learned many times. The assumption behinds this view is that the sharer's Q-values rarely hinder the learning of a partaker. Although some Q-values from the sharer may be non-optimal at the very early stage of learning, the partaker still can adjust them as the updated times of the Q-values increase. However, some tasks may have stochastic rewards, making it become much more difficult for each agent to learn the optimal policy. Furthermore, the Q-values in some states can drastically change during learning. In the circumstances, sharing non-optimal Q-values to a partaker may take it a lot of time to adjust its policy and stabilize the performance. PSAF is likely to be used in two situations: (1) an agent arrives a state that the agent rarely visits so that it nearly has no information about the state, while its teammates who have visited the state many times; (2) an agent joins a system in which some agents have already learned a period of time, and their performance becomes stable.

\section{CONCLUSION AND FURTHER WORKS}
We here propose a Q-values sharing framework PSAF for multiple decentralized Q-learners learning with budget constraint. In PSAF, if a learning agent visits current state very few times, it is more likely to take the role of partaker. Then the other agents share their Q-values when they have more confidence than the partaker. The relation of a partaker and a sharer is established only when the sharer's Q-values are expected to be useful for the partaker. For this purpose, we utilize the asking function $P_{ask}$ defined by AdhocTD to decide when a partaker can ask for Q-values, and propose two confidence functions for a partaker and a sharer, respectively, evaluating their own confidence in a particular Q-value. Concretely, a sharer is required to update its maximum Q-value many more times than a partaker, in order to provide a more beneficial Q-value for the partaker, and use the budget more effectively. We conduct experiments in Predator-Prey domain, Half Field Offense and Spread game for three cases: (1) the budget is unlimited; (2) the budget is limited for the other advising frameworks; (3) the budget is even lower for all methods including PSAF. Evaluation results show that the schemes sharing Q-values, such as PSAF and AdhocTD-Q, are significantly better than sharing actions in several metrics, yet PSAF spends much less budget than all other methods. When the budget is unlimited, AdhocTD-Q can achieve similar performance to PSAF in Half Fielf Offense, while has worse results than PSAF in three tasks when the available budget decreases. Moreover, the performance of advising actions heavily depends on the amount of budget. In contrast, PSAF only needs to share a small number of Q-values.

In the present work we randomly select one Q-value for a partaker when it receives several Q-values from multiple knowledge resources. Ongoing work is devoted to study how to select the most effective one, especially when there are hundreds of sharers to response the partaker's request. In particular, the partaker can ask for Q-values from a small set of potential sharers. In this paper, we focus on a tabular representation of Q-function. Even so, PSAF is able to be adapted to some problems with continuous state spaces, where a function approximator is used. One possible way is to incorporate the shared Q-values for a partaker in a Q-learning like rule. Then PSAF can be combined with deep Q-network \cite{Lample2017PlayingFG} to handle the high-dimensional sensory inputs. Furthermore, PSAF can be linked to many learning algorithm, such as double Q-learning \cite{DoubleQlearning}, weighted double Q-learning \cite{WeightedDoubleQ}, and an extension Multi Q-Learning \cite{Duryea2016Exploring}. In our work, all agents are learning from scratch in a shared environment. Another branch of future works can discover sharing Q-values among RL agents with different levels of expertise.

\section*{ACKNOWLEDGEMENT}
This work presented in this paper is partially supported by a CUHK Direct Grant for Research (Project Code EE16963), the Fundamental Research Funds for the Central Universities, SCUT (Nos. 2017ZD048, D2182480), the Tiptop Scientific and Technical Innovative Youth Talents of Guangdong special support program (No. 2015TQ01X633), the Science and Technology Planning Project of Guangdong Province (No. 2017B050506004), the Science and Technology Program of Guangzhou (Nos. 201704030076, 201802010027).

\bibliographystyle{unsrt}

\bibliography{reference}

\end{document}